\documentclass[pre,twocolumn,showpacs,showkeys,amsmath,amssymb,floatfix]{revtex4}
\usepackage{amsfonts} 
\usepackage{graphicx}% Include figure files
\usepackage{dcolumn}% Align table columns on decimal point
\usepackage{bm}% bold math
\usepackage{subfigure}
\usepackage{comment}
\usepackage{times}
\usepackage{color}   
\usepackage{color}
\usepackage{rotating,graphics,psfrag,epsfig}
\begin{document}
%%%%%%%%%%%%%%%%%%%%%%%%%%%%%%%%%%%%%%%%%%%%%%%%%%%
\title{Cosmological particle production in emergent rainbow spacetimes}
%%%%%%%%%%%%%%%%%%%%%%%%%%%%%%%%%%%%%%%%%%%%%%%%%%%
%
\author{Silke Weinfurtner$^{1,2\; \footnote{silke@phas.ubc.ca}}$, Piyush Jain$^{3,4\;\footnote{piyushnz@gmail.com}}$, Matt Visser$^{2\;\footnote{matt.visser@mcs.vuw.ac.nz}}$, and C.~W. Gardiner$^{4\;\footnote{gardiner@physics.otago.ac.nz}}$}
\affiliation{$^1$Department of Physics and Astronomy, University of British Columbia, Vancouver, Canada}
\affiliation{$^2$School of Mathematics, Statistics and Computer Science, Victoria University of Wellington, New Zealand}
\affiliation{$^3$School of Chemical and Physical Sciences, Victoria University of Wellington, New Zealand}
\affiliation{$^4$Jack Dodd Centre of Photonics and Ultra-Cold Atoms, Department of Physics, Otago University, Dunedin, New Zealand}
%------------------------------------------------------------------------------------------------------------------------------------------
\date{21 December 2007; \LaTeX-ed \today}
%------------------------------------------------------------------------------------------------------------------------------------------
%%%%%%%%%%%%%%%%%%%%%%%%%%%%%%%%%%%%%%%%%%%%%%%%%%%
% editing mode
\definecolor{myOrange}{rgb}{1,0.5,0.1}
\definecolor{myRed}{rgb}{0.8,0.1,0.1}
\definecolor{myGreen}{rgb}{0.5,0.7,0.1}
\definecolor{myGray}{rgb}{0.6,0.6,0.6}
%100.0	48.2	7.8
%\renewcommand{\comment}[1]{\par\noindent{\color{myOrange}\begin{center} $\star \, \star \, \star \, \star \, \star$ \end{center}  #1 \begin{center} $\star \, \star \, \star \, \star \, \star$ \end{center} \par\noindent}}
\renewcommand{\comment}[1]{\par\noindent{\em \color{blue} \begin{center} $\star \, \star \, \star \, \star \, \star$ \end{center}  #1 \begin{center} $\star \, \star \, \star \, \star \, \star$ \end{center} \par\noindent}}
\newcommand{\tentative}[1]{{\color{myGreen}#1}}
\newcommand{\silkycomment}[1]{ { \color{myRed} #1}}
\newcommand{\piyushcomment}[1]{ { \color{green} #1}}
\newcommand{\deletecomment}[1]{ { \color{myGray} #1}}
\newcommand{\addcomment}[1]{ { \color{myRed} #1}}
%------------------------------------------------------------------------------------------------------------------------------------------
% some mathematics
\def\e{\mathrm{e}}
\def\exp{\mathrm{exp}}
\newcommand{\half}{\textstyle{\frac{1}{2}}}
\newcommand{\quarter}{\textstyle \frac{1}{4}}
\newcommand{\oneover}[1]{\textstyle \frac{1}{#1}}
%------------------------------------------------------------------------------------------------------------------------------------------
% saving typing time
\newcommand{\refb}[1]{(\ref{#1})}
\newcommand{\bracket}[1]{\left( {#1} \right)}
%------------------------------------------------------------------------------------------------------------------------------------------
% Latin
\def\ie{\emph{i.e.}}
\def\eg{\emph{e.g.}}
\def\ea{\emph{et al.}}
\def\element{\in}
%%%%%%%%%%%%%%%%%%%%%%%%%%%%%%%%%%%%%%%%%%%%%%%%%%%
\begin{abstract}
We investigate cosmological particle production in spacetimes where Lorentz invariance 
emerges in the infrared limit, but is explicitly broken in the ultraviolet regime. Thus these 
models are similar to many (but not all) models of quantum gravity, where a breakdown 
of Lorentz invariance is expected for ultraviolet physics around the Planck\,/\,string scale. 
Our specific model focuses on the boost subgroup that supports CPT invariance and 
results in a momentum-dependent dispersion relation. Motivated by previous studies on 
spacetimes emerging from a microscopic substrate, we show how these modifications 
naturally lead to momentum-dependent rainbow metrics.

Firstly, we investigate the possibility of reproducing cosmological particle production in 
spacetimes emerging from real Bose gases. Several papers have been written on the 
analogy between the kinematics of linearized perturbations in Bose--Einstein condensates 
and effective curved-spacetime quantum field theory. Recently, in  [Phys. Rev. A {\bf 76} (2007)
043625] we have studied the influence of non-perturbative ultraviolet corrections in time-dependent analogue 
spacetimes, leading to momentum-dependent emergent rainbow spacetimes. 
We show that models involving a time-dependent microscopic interaction are 
suitable for mimicking quantum effects in FRW spacetimes. Within certain limits the 
analogy is sufficiently good to simulate relativistic quantum field theory in time-dependent 
classical backgrounds, and the quantum effects are approximately robust against the
model-dependent modifications.
 
Secondly, we analyze how significantly the particle production process deviates from the common 
picture. While very low-energy modes do not see the difference at all, some modes ``re-enter the Hubble horizon'' during the inflationary epoch, and extreme ultraviolet modes 
are completely insensitive to the expansion. The analysis outlined here, because it is 
nonperturbative in the rainbow metric, exhibits features that cannot be extracted simply 
from the standard perturbative modification of particle dispersion relations. However, we 
also show how the final result, after many e-foldings, will approach a time-independent 
exponentially decaying particle spectrum.
\end{abstract}
%------------------------------------------------------------------------------------------------------------------------------------------
\pacs{Valid PACS appear here}
%------------------------------------------------------------------------------------------------------------------------------------------
\maketitle
%------------------------------------------------------------------------------------------------------------------------------------------
%%%%%%%%%%%%%%%%%%%%%%%%%%%%%%%%%%%%%%%%%%%%%%%%%%%
%%%%%%%%%%%%%%%%%%%%%%%%%%%%%%%%%%%%%%%%%%%%%%%%%%%
%\tableofcontents
%%%%%%%%%%%%%%%%%%%%%%%%%%%%%%%%%%%%%%%%%%%%%%%%%%%
%%%%%%%%%%%%%%%%%%%%%%%%%%%%%%%%%%%%%%%%%%%%%%%%%%%
%
%
%
%
%%%%%%%%%%%%%%%%%%%%%%%%%%%%%%%%%%%%%%%%%%%%%%%%%%%
%
\section{Introduction}
%
%%%%%%%%%%%%%%%%%%%%%%%%%%%%%%%%%%%%%%%%%%%%%%%%%%%
The use of Bose--Einstein condensates (BECs) as analogue models for quantum field theory in spatially flat $k=0$ Friedmann--Robertson--Walker (FRW) geometries
\begin{equation}
\label{Eq:FRW.GR}
ds^{2} = g_{ab} \; dx^{a} \, dx^{b}= -d\tau^{2} + a(\tau)^{2} \sum_{i=1}^{d} (dx^{i})^{2} \, 
\end{equation}
(in $d$ spatial dimensions) has recently been extensively explored in \cite{Jain:2006ki}. The proposal~\cite{Barcelo:2003ia,Barcelo:2003yk,Fedichev:2004on,Fedichev:2004fi,Fischer:2004iy,Uhlmann:2005rx,Unruh:1981bi} is based on the analogy between the equation of motion for collective excitations around a macroscopically occupied ground state of an ultra-cold weakly interacting gas of Bosons (\ie, the novel state of matter referred to as the Bose--Einstein condensate), and
\begin{equation}
\label{Eq:KGE}
\frac{1}{\sqrt{\vert g \vert}} \; \partial_{a} \, \left(\sqrt{\vert g \vert} \, g^{ab} \, \partial_{b} \hat \theta \right)=0\, ,  
\end{equation}
the covariant free-field equation (Klein--Gordon equation) for spin-$0$ massless particles. (The indices $a$, $b$ run from $0$ to $d$ whereas $i$, $j$ run from $1$ to $d$ for the spatial coordinates only). Here $g^{ab}$ is a symmetric covariant rank two tensor whose entries include purely collective (mean-field) variables $c=c(t,\mathbf x)$, the speed of sound, and $\mathbf{v}=\mathbf v(t,\mathbf x)$, the background velocity. Using $g^{ab}\,g_{bc}=\delta^{a}{}_{c}$ we are able to define an effective line-element for the mean-field,
\begin{equation}
ds^{2} = \left(\frac{c}{U}\right)^{\frac{2}{d-1}} \left[ -(c^{2}-v^{2}) \, dt^{2} - dt \, \mathbf v^{\mathrm{T}}\, d\mathbf x + d\mathbf{x}^{\mathrm{T}} \, d\mathbf x   \right] \, .
\end{equation}
Here the quantity $U$ arises from the microscopic description of a Bose--Einstein condensate, and represents the inter-atomic potential.

It has previously been shown, see~\cite{Weinfurtner:2007aa,Unruh:2003ln}, that Eq.~\refb{Eq:KGE} is an equation for perturbations in the collective variables, the phase $\hat \theta$ and its conjugate momentum $\hat \Pi_{\hat \theta}$ (see also \cite{Barcelo:2003ia,Barcelo:2003yk,Fedichev:2004on,Fedichev:2004fi,Fischer:2004iy,Uhlmann:2005rx,Unruh:1981bi}).
This kind of behavior is not restricted to Bose--Einstein condensates. Indeed the first modern paper on analogue models for gravity focussed on ordinary fluid mechanics; see \cite{Unruh:1981bi}. Since then numerous media have been suggested as substrates to develop analogue models for gravity~\cite{Barcelo:2005ln,Schutzhold:2007aa,Visser:1993tk,Visser:1998gn}. 
Nevertheless it is sometimes difficult to isolate the fundamental principles behind the analogy.

Inspired by the Bose gas, where both the microscopic and macroscopic theory is understood to an adequate extent, we suggest that it might be useful to transfer the experience gained from the various known examples for emergent spacetimes into a more general framework:

\begin{verse}
\emph{Emergent spacetimes} involve
$\mathrm{(i)}$ a microscopic system of fundamental objects (\eg, strings, atoms, or molecules); 
$\mathrm{(ii)}$ a dominant mean-field regime, where the microscopic degrees of freedom give way to collective variables;
$\mathrm{(iii)}$ a ``geometrical object'' (\eg, a symmetric tensor) dominating the evolution for infrared classical and quantum excitations around the mean-field.
\end{verse}
Within certain limits we are free to choose our geometrical object to mimic Einstein's theory of gravity, (\eg, a symmetric rank two tensor $g_{ab}$ that is conformal to an exact solution obtained by solving the Einstein equation for a physically reasonable stress-energy tensor). 
\begin{verse}
\emph{Analogue models for gravity} are emergent spacetimes that are specifically adjusted to mimic as closely as possible Einstein's geometrical theory of gravity.
\end{verse}
The fundamental difference between emergent spacetimes and general relativity becomes obvious if we consider the following statement: Emergent spacetimes, as they appear in the analogue model programme, are ``a short and simple way'' to summarize the kinematics of linearized perturbations in a geometrical sense, without requiring the notion of a stress-energy tensor or the Einstein equations.

Thus, the analogy is (currently) restricted to the kinematic behavior of the system, and any back-reaction between the excitations and the mean-field might not be in analogy with Einstein's theory of gravity. Therefore it is advisable to restrict the use of analogue models for gravity as toy models for semi-classical quantum gravity at the level of curved spacetime quantum field theory, where the gravitational field is a purely classical field.

Indeed, analogue models for gravity are very idealized constructions, and require careful treatment with respect to the model-dependence to obtain the desired curved-spacetime quantum-field-theory effects. This has first been pointed out in \cite{Jacobson:1991sd,Unruh:2005aa}, where the authors analyzed the robustness of Hawking radiation against modifications in the excitation spectrum (most relevant to fluid dynamics: super- and sub- luminal dispersion).\\

In a related previous work \cite{Jain:2006ki} we made use of classical phase space methods to numerically study ``cosmological particle production'' in a realistic BEC. In principle, there are two different ways to mimic an expanding universe in a Bose gas. In \cite{Fedichev:2004on,Fedichev:2004fi} the authors studied a freely expanding condensate cloud, such that the density is a function of time. This idea has been further investigated in \cite{Uhlmann:2005rx}. In both --- our previous paper \cite{Jain:2006ki} and in this work we pursue an alternative (more recent) idea introduced in \cite{Barcelo:2003ia,Barcelo:2003yk}, where the authors achieved FRW geometry through a time-dependence in the effective two-body inter-atomic potential. (A change in the inter-atomic potential is related to a change in the speed of sound, and consequently corresponds to a time-dependent acoustic metric.) As expected our simulations verified the analogy (in the hydrodynamic limit) to a good extent. However, it is crucial to include quantum pressure effects to understand the ultraviolet\,/\,trans-phononic part of the quasi-particle spectrum\footnote{Note that in the current context \emph{trans-phononic} modes refer to excitations that have a sufficiently short wavelength so that the Lorentz symmetry breaking can no longer be neglected.}. Thus we conclude that our specific model --- an ultra-cold gas of Bosons with time-varying atomic interactions --- shows a specific model-dependence that affects the high-momentum particle production process. One lesson that can be drawn from this new insight is that a careful choice of condensate parameters is necessary in order to mimic the desired curved-spacetime quantum field theory effect.

The main focus in this article is to understand the differences between ordinary quantum field theory in a ``real'' classical FRW-background and quantum field theory in an emergent FRW-background in a Bose gas. This question has been motivated by many effective field theories, where Lorentz invariance is broken at ultraviolet energies in a similar manner as in the emergent spacetimes we are investigating. 
\begin{verse}
Emergent spacetimes exhibit an \emph{emergent\,/\,effective Lorentz symmetry} for low-energy\,/\,infrared excitations around the macroscopic field. This symmetry will be broken in the high-energy\,/\,ultraviolet regime, that is at scales dominated by the underlying microscopic theory. These corrections are of a non-perturbative nature. 
\end{verse}

For now, we leave general questions aside, and focus on our specific emergent spacetime model, investigating the influence on cosmological particle production due to ultraviolet corrections. We include quantum pressure effects --- which lead to a nonlinear excitation spectrum --- as nonperturbative ultraviolet corrections to the emergent geometry. We will show that this leads to momentum-dependent ``rainbow spacetimes''. Partly motivated by extant literature \cite{Magueijo:2004aa,Weinfurtner:2006eq} and our numerical results from the simulations of quantum effects in realistic Bose--Einstein condensates, we present a coherent interpretation for the quasi-particle spectrum produced in FRW rainbow spacetimes \cite{Weinfurtner:2006nl,Liberati:2006kw,Liberati:2006sj,Weinfurtner:2006eq,Weinfurtner:2006iv,Visser:2001ix}.

The ``FRW rainbow metrics'' recover the ``standard'' FRW type geometries in the infrared (phononic) regime, but exhibit momentum-dependent modifications in the ultraviolet (trans-phononic) regime. We show that in such a geometry the scale factor for the universe $a(t)$ is effectively momentum-dependent $a(t) \to a_{k}(t)$, consequently leading to a momentum-dependent Hubble parameter: $H(t) \to H_{k}(t)$. In addition, we are dealing with emergent spacetime that exhibits a time-dependent effective ``Planck-length''. Ultimately, the particle production process in our specific analogue model will show deviations from the one expected in ``standard'' curved-spacetime quantum field theory.

For the ``analogue gravity programme'' it is important to check the robustness of cosmological particle production against those modifications. Is there a window where the analogy is good enough, for example, to use the Bose gas as a toy model for inflation? \\

In this paper we will show that in the present model the particle production process is \emph{in general not} robust against the model-specific modifications. However, for short time duration expansion scenarios, inflation can be simulated in a realistic Bose--Einstein condensate. \\

We perform a quantitative analysis to describe short time \emph{as well as} long-lasting expansions. We use the ratio between the mode frequency and the Hubble frequency (the inverse of the rate of change in the size of the emergent universe), to show that our specific model indeed shows significant deviations in the ultraviolet regime, \eg, both crossing \emph{and} re-entering of the ``Hubble horizon'' during the inflationary epoch; $a(t)$~$\sim$~$\exp(t)$.\\

By means of our numerical analysis we are able to ``read-off'' the final spectrum after a sufficiently long-lasting expansion. We explain why the final particle spectrum --- after an infinitely long-lasting inflationary epoch --- will be time-independent and finite.

%
%%%%%%%%%%%%%%%%%%%%%%%%%%%%%%%%%%%%%%%%%%%%%%%%%%%
%
\section[Emergent spacetimes]{Emergent spacetimes:\protect\\
Excitations in Bose--Einstein condensates}
\label{Sec:Emergent.Spacetimes} 
%
%%%%%%%%%%%%%%%%%%%%%%%%%%%%%%%%%%%%%%%%%%%%%%%%%%%
In the following we extend the calculations presented in our previous work \cite{Jain:2006ki}. The intent is to derive an emergent geometry that is able to simultaneously represent both the phononic and trans-phononic excitations in a Bose--Einstein condensate. It is a well established result --- both in theory and experiment --- that the energy-momentum relation for excitations in Bose--Einstein condensates is given by the nonlinear Bogoliubov dispersion relation. In \cite{Liberati:2006sj,Weinfurtner:2006nl,Weinfurtner:2006iv,Liberati:2006kw,Weinfurtner:2006eq,Visser:2001ix} it has been shown, (within the analogue model for gravity point of view), that phononic modes are relativistic modes, since they exhibit Lorentz symmetry. The corresponding emergent geometry is a flat spacetime with Lorentzian signature (Minkowski metric). Higher energy (trans-phononic) modes break Lorentz symmetry, the excitations being ``supersonic''. The full dispersion relation for excitations in a realistic Bose--Einstein condensate is similar to that expected to occur in certain effective field theories. It is also known that this kind of modification is non-perturbative, since it originates in density fluctuations visible only at small scales (in the order of the healing\,/\,coherence length). These modes are trans-phononic modes, where quantum pressure effects are no longer negligible. We show that in a Bose--Einstein condensate with time-dependent condensate parameters, the crossover between phononic and trans-phononic modes is also time-dependent, and therefore requires a more accurate analysis at the level of the emergent geometry, leading to the concept of a rainbow metric. In other words we extend the treatment for ultraviolet modes from flat to curved spacetimes.

%+++++++++++++++++++++++++++++++++++++++++++++++++++++++++++++++++
\subsection[Ultraviolet non-perturbative corrections]{Ultraviolet non-perturbative corrections:\protect\\
Quantum pressure effects
\label{Label}}
%+++++++++++++++++++++++++++++++++++++++++++++++++++++++++++++++++
In \cite{Jain:2006ki} we derived a coupled pair of equations, 
\begin{eqnarray} 
\label{Eq:on_dynamics}
&& \partial_t \hat n+\nabla \cdot \left[\left(\frac{n_{0} \hbar}{m}\nabla\hat\theta\right)+(n_{0} \, \mathbf v)
\right] =0\, ,\\
\label{Eq:otheta_dynamics}
&&\partial_t \hat \theta+\mathbf v \cdot \nabla\hat\theta+\frac{\widetilde U}{\hbar}\;\hat n =0 \, ;
\end{eqnarray}
for quantum fluctuations in the condensate density $\hat n$ and phase $\hat \theta$. The background velocity $\mathbf v$ is given by
\begin{equation} 
\label{Eq:Background.Velocity}
\mathbf v = \frac{\hbar}{m} \, \nabla \theta_{0} \, ,
\end{equation}
as the gradient of the condensate phase $\theta_{0}$. Here $m$ is the mass of the fundamental Bosons, $n_{0}$ the number density, and $\hbar$ the Planck constant. The differential operator $\widetilde U$ has been defined as \cite{Weinfurtner:2006nl,Jain:2006ki}
\begin{equation}
\label{Eq.Effective.U}
\widetilde U = U - \frac{\hbar^2}{2 m}\;\widetilde D_{2}   \, ,
\end{equation}
where the differential operator
\begin{equation}
\label{QuantumPressure}
\widetilde D_{2} =\frac{1}{2} \left\{ \frac{(\nabla n_{0})^{2} -(\nabla^{2}n_{0})n_{0}}{n_{0}^{3}} -\frac{\nabla n_{0}}{n_{0}^{2}}\nabla +\frac{1}{n_{0}}\nabla^{2} \right\} \, ,
\end{equation}
accounts for the first-order correction obtained from linearizing the quantum potential term
whereas the inter-atomic potential is explicitly given by
\begin{equation}
U  = {4\pi\hbar^2 a_{\mathrm{scatt}}(t)\over m}. 
\end{equation}
The scattering length $a_{\mathrm{scatt}}(t)$ represents the $s$-wave scattering term. (See, for instance \cite{Barcelo:2001gt,Barcelo:2005ln}, and the discussion above.) 
Note the tilde notation is used to emphasize the fact that we are dealing with a differential operator. 

To develop the analogy between condensed matter physics and curved-spacetime quantum field theory, see Eq.~\refb{Eq:KGE}, it is necessary to combine Eq.~\refb{Eq:on_dynamics} and Eq.~\refb{Eq:otheta_dynamics} into a single equation for the phase fluctuations. For that it is important to note that we can rearrange \refb{Eq:otheta_dynamics} to make $\hat n$ the subject. That is
\begin{equation}
\label{Eq:n_of_theta} 
\hat n = \hbar \, \widetilde U^{-1} \, \left[ \partial_t \hat \theta+\mathbf v\cdot\nabla\hat \theta \right] = \hbar \, \widetilde U^{-1} \,
\frac{D \hat \theta}{Dt} \, ,
\end{equation}  
where $D\hat \theta/Dt$ is the rate of change of $\hat \theta$ following a
small volume of the fluid, the \emph{material derivative} of $\hat \theta$.
If the fluid is at rest, the material derivative reduces to
$D\hat \theta/Dt\rightarrow \partial_{t} \hat \theta$, but for a moving fluid it
represents a ``fluid-following'' derivative.
The integral differential operator $\widetilde U^{-1}$ can \emph{formally} be expanded as
\begin{eqnarray}
\label{Eq:Integral.Differential.Operator}
\widetilde U^{-1} = U^{-1} &+& \left[\frac{\hbar}{2m}\right] \, U^{-1} \widetilde D_{2} \, U^{-1}  \\ \nonumber
 &+& \left[\frac{\hbar}{2m}\right]^{2} \, U^{-1} \widetilde D_{2} \,  U^{-1} \widetilde D_{2} \, U^{-1} \\ \nonumber
&+& \left[\frac{\hbar}{2m}\right]^{3} \, U^{-1} \widetilde D_{2} \, U^{-1} \widetilde D_{2} \,  U^{-1} \widetilde D_{2} \, U^{-1} + ...
 \end{eqnarray}
where the \emph{formal} series converges only on the subspace of functions spanned by the eigenfunctions whose eigenvalues satisfy
\begin{equation}
\label{Eq:Convergence.Radius}
\lambda \left( \frac{\hbar^{2}}{2m} U^{-1} \widetilde D_{2} \right) < 1 \, .
\end{equation}
Since $\widetilde D_{2}$ and $\widetilde U$ are second-order linear differential operators, the inverse $\widetilde U^{-1}$ always exists as an integral operator (that is, in the sense of being a Green function). Expanding this Green function as in Eq.~\refb{Eq:Integral.Differential.Operator} above is a convenience that allows us to interpret $\widetilde U^{-1}$ as a sum of differential operators, but this is not a fundamental limitation on the formalism. 
%
%~~~~~~~~~~~~~~~~~~~~~~~~~~~~~~~~~~~~~~~~~~~~~~~~~~~~
\subsubsection{Acoustic and rainbow metrics \label{Sec:Rainbow.Metrics}}
%~~~~~~~~~~~~~~~~~~~~~~~~~~~~~~~~~~~~~~~~~~~~~~~~~~~~
We use equation \refb{Eq:n_of_theta} to eliminate $\hat n$ completely in our equations of motions, by substituting it into equation \refb{Eq:on_dynamics}. That yields a single equation for the perturbed phase
\begin{equation}
\label{Eq:WaveEquation}
\partial_{a} \left(f^{ab}\; \partial_{b} \hat \theta \right) = 0 \, ,
\end{equation}
where we have introduced the matrix $f^{ab}$ with inverse-differential-operator-valued entries:
\begin{equation}
\label{Eq:f}
f^{ab} = \hbar
\left[ \begin{array}{c|c} \vphantom{\Big|} -\widetilde U^{-1} & -\widetilde U^{-1} v^{j} \\ \hline \vphantom{\Big|} -v^{i} \widetilde U^{-1} & \frac{n_{0}}{m}\delta^{ij}- v^i \widetilde U^{-1} v^j \end{array}\right] \, .
\end{equation}
Note that in general $\widetilde U^{-1}$ is an integral operator so Eq.~\refb{Eq:WaveEquation} is an integro-differential equation.
If we additionally require that there
exists an (inverse) metric tensor $g_{ab}$ such that
\begin{equation}
\label{Eq:EmergentGeometry}
f^{ab} \equiv \sqrt{-g} \, g^{ab} \, ,
\end{equation}
where $g$ is the determinant of $g_{ab}$, then the
connection is formally made to the field equation for a minimally
coupled massless scalar field in a curved spacetime; see Eq.~\refb{Eq:KGE}.
Now in many situations of physical interest, the differential operator $\widetilde U$ can be
usefully approximated by a function --- for instance the hydrodynamic and eikonal limits.\\
%

%~~~~~~~~~~~~~~~~~~~~~~~~~~~~~~~~~~~~~~~~~~~~~~~~~~~~
\paragraph{Hydrodynamic approximation:}
%~~~~~~~~~~~~~~~~~~~~~~~~~~~~~~~~~~~~~~~~~~~~~~~~~~~~
In the hydrodynamic limit where the \emph{quantum pressure} is neglected one
has 
\begin{equation}
\label{Eq:Hydrodynamic.Limit}
\vert U \, n_{0} \vert  \gg \vert (\hbar^{2}/2m) \, \widetilde D_{2} \, n_{0} \vert 
\end{equation}
so that 
\begin{equation}
\widetilde U \to U . 
\end{equation} \\

Within this approximation we obtain
\begin{equation}
\label{f1}
f^{ab} =\frac{\hbar}{U} 
\left[ \begin{array}{c|c} \vphantom{\Big|} -1 & - v^{j} \\ \hline \vphantom{\Big|} -v^{i}  & \frac{n_{0} U}{m}\delta^{ij}- v^i  v^j \end{array}\right]
\, .
\end{equation} 
To mimic curved spacetime classical and quantum field theory effects the acoustic metric has to be formally in agreement with Einstein's theory of gravity.
For this low-momentum approximation we are able to define a common (for all wavelengths $k$) speed of sound in the condensate:
\begin{equation} 
\label{Eq:SoundSpeed}
c_k^{2} \rightarrow c^{2}= \frac{n_{0} U(t)}{m} \, .
\end{equation}
Applying Eq.~\refb{Eq:EmergentGeometry} we obtain,
\begin{equation}
\label{Eq:Acoustic.Metric} 
g_{a b} \equiv \left(
\frac{n_{0} \, \hbar}{c \, m } \right)^{\frac{2}{d-1}} \,
\left[ \begin{array}{c|c} -(c^{2}-v^{2}) & -v^j \\ \hline -v^i & \delta^{ij} \end{array}\right] \, ,
\end{equation}
as the \emph{acoustic metric} or \emph{analogue metric}.
In this case all collective excitations behave as
sound waves with the usual linear dispersion form 
\begin{equation}
\omega_{0} = c \, k
\end{equation}
and the quanta of excitations are thus phonons. \\

%~~~~~~~~~~~~~~~~~~~~~~~~~~~~~~~~~~~~~~~~~~~~~~~~~~~~
\paragraph{Eikonal approximation:}
%~~~~~~~~~~~~~~~~~~~~~~~~~~~~~~~~~~~~~~~~~~~~~~~~~~~~
%
Let us now invoke a different approximation that holds for trans-phononic modes, so that the
elements of the matrix $f^{ab}$ can be treated as (possibly momentum
dependent) functions, rather than differential operators:
Consider the \emph{eikonal limit} where $\widetilde U$ can usefully be approximated by a function
\begin{equation} \label{Eq:U.eikonal}
\widetilde U \to U_{k}(t,\mathbf x) = U(t,\mathbf x) + \frac{\hbar^2 k^2}{ 4 m n_0} \, , 
\end{equation}
which we shall conveniently abbreviate by writing $U_k$.
Beyond the hydrodynamic limit we obtain
\begin{equation}
\label{f2}
f^{ab} = \frac{\hbar}{U_k} 
\left[ \begin{array}{c|c} \vphantom{\Big|} -1 & - v^{j} \\ \hline \vphantom{\Big|} -v^{i}  & \frac{n_{0} U_k}{m}\delta^{ij}- v^i  v^j \end{array}\right]
\, .
\end{equation} 
Note that in the eikonal approximation the $k$ dependence hiding in $U_k$ will make this a momentum-dependent metric, a so-called \emph{rainbow metric}. The metric tensor is explicitly given by
\begin{equation}
\label{Eq:Rainbow.Metric}
g_{a b} \equiv
\left( \frac{n_{0}\, \hbar}{c_k \, m} \right)^{\frac{2}{d-1}} \,
\left[ \begin{array}{c|c} -(c_k^{2}-v^{2}) & -v^j \\ \hline -v^i & \delta^{ij} \end{array}\right] \, ,
\end{equation}
where we have introduced the quantity
\begin{equation}
\label{Eq:Eikonal.Sound.Speed}
c_{k}(t)^{2}=c(t)^{2}+\gamma_{\mathrm{qp}}^{2} k^{2} \, ,
\end{equation}
for a uniform condensate density $n_{0}(t,\mathbf x)=n_{0}$.
It is convenient to define 
\begin{equation}
\label{qp}
\gamma_{\mathrm{qp}}=\frac{\hbar}{2m} \, ,
\end{equation}
where $\gamma_{\mathrm{qp}} \ll \vert c(t)/k \vert$ is a useful indication for the hydrodynamic limit; compare with Eq.~\refb{Eq:Hydrodynamic.Limit}. The dispersion relation in the eikonal limit is
\begin{equation}
\label{Eq:Omega.NonLinear}
\omega_{k}(t)=c_{k}(t)\, k=\sqrt{c(t)^{2}k^{2} + \gamma_{\mathrm{qp}}^{2}k^{4}} \, ,
\end{equation}
and hence violates ``\emph{acoustic Lorentz invariance}''.  This is not surprising at all, since we know the quasi-particles become ``atom-like'', and so \emph{non-relativistic}, at high momentum~\cite{Visser:2001ix}. \\

%~~~~~~~~~~~~~~~~~~~~~~~~~~~~~~~~~~~~~~~~~~~~~~~~~~~~
\paragraph{Acoustic or rainbow spacetimes?}
%~~~~~~~~~~~~~~~~~~~~~~~~~~~~~~~~~~~~~~~~~~~~~~~~~~~~
%
It is only in the acoustic/hydrodynamic limit that the $k$ dependence of $U_k$ and $c_{k}$ vanish, so that the rainbow metric is reduced to an ordinary Lorentzian metric.

An interesting consequence of the Bogoliubov theory in Bose condensates is that in general the excitation spectrum displays nonlinear dispersion (see Eq.~\refb{Eq:Omega.NonLinear}),
being linear (\ie, phononic) for low $|\mathbf{k}|$ and becoming quadratic (\ie, free-particle like) at
large $|\mathbf{k}|$. When the nonlinear dispersion \refb{Eq:Omega.NonLinear} is incorporated into analogue models of gravity it is equivalent to breaking Lorentz invariance
\cite{Barcelo:2001gt,Barcelo:2005ln,Weinfurtner:2006iv,Liberati:2006kw,Liberati:2006sj,Weinfurtner:2006nl}.  

The hydrodynamic approximation is a statement about the \emph{smallness} of the quartic term in the dispersion relation for $\omega_{k}^{2}$~\refb{Eq:Omega.NonLinear}, with respect to the quadratic term.
The usual line of argument is that the smallness of $\gamma_{\mathrm{qp}}$ makes it possible to neglect the second order in Eq.~\refb{Eq:Omega.NonLinear} for low-energy excitations, where $\gamma_{\mathrm{qp}} \vert k \vert \ll c(t)$. Here Lorentz invariance is an emergent symmetry.
However, it is important to realise that the propagation speed can be a function of time, $c=c(t)$, and hence if $c(t)\to 0$ then one is dealing with a system that eventually violates Lorentz invariance at \emph{all} energy scales. In principle, there are no theoretical or experimental restrictions to prevent $U(t)\propto c(t)$ becoming arbitrarily small.  %

In the specific cases we are interested in, we are confronted with exactly this situation, and therefore a more subtle analysis is required as to whether the acoustic metric \refb{Eq:Acoustic.Metric} is a sufficient approximation, or whether we have to use the more sophisticated concept of a rainbow spacetime \refb{Eq:Rainbow.Metric}. We will elaborate on the point in Sec.~\ref{Sec:Inlation.FRW.Rainbow.Spacetimes}.\\
%

%~~~~~~~~~~~~~~~~~~~~~~~~~~~~~~~~~~~~~~~~~~~~~~~~~~~~
\paragraph{Limitations of the rainbow analogy?}
%~~~~~~~~~~~~~~~~~~~~~~~~~~~~~~~~~~~~~~~~~~~~~~~~~~~~
%
It is worth reiterating the assumptions used in making the above analogy.  The theory resulting from linearized quantum fluctuations leads to a \emph{free field theory} --- that is, the modes of the field $\hat \theta$ are non-interacting.  Since we are using a linearized theory, the interactions between quantum fluctuations themselves, and between quantum fluctuations  and the condensate mode are neglected. This is equivalent to assuming the metric tensor is an externally-specified classical quantity. 

Our numerical simulations carried out in \cite{Jain:2006ki} do not require these assumptions; there all modes of the system are included and these modes are able to interact --- albeit weakly --- via the nonlinear interaction term.  Thus
we were able to explore the validity of the assumptions of the free-field theory~\cite{Schutzhold:2005wt}.

We further note that while the present form of the analogy only holds for massless scalar (spin zero) particles, in general it is possible to modify the formalism to include massive minimally coupled scalar fields at the expense of dealing with more complex BEC configurations, \eg, a two-component BEC 
\cite{Weinfurtner:2006nl,Weinfurtner:2006eq,Liberati:2006kw,Liberati:2006sj,Weinfurtner:2006eq,Weinfurtner:2006iv,Visser:2005ai,Visser:2004zn}. In BEC language one would explain this situation in terms of a dispersion relation with a gap.
%
%~~~~~~~~~~~~~~~~~~~~~~~~~~~~~~~~~~~~~~~~~~~~~~~~~~~~
\subsubsection{Commutation relations \label{Sec:Commutation.Relations}}
%~~~~~~~~~~~~~~~~~~~~~~~~~~~~~~~~~~~~~~~~~~~~~~~~~~~~
To derive the analogy presented above, we approximated and transformed our field operators several times, see~\cite{Weinfurtner:2007aa} and~\cite{Jain:2006ki}. These are canonical transformations preserving commutation relations:
\begin{enumerate}
\item{$\hat\psi(t,\mathbf x)$ and $\hat\psi^{\dagger}(t,\mathbf x)$: The single Boson annihilation and creation operators; where
\begin{eqnarray} 
&&[\hat{\psi} (t,\mathbf{x}) , \hat{\psi} (t,\mathbf x')] =0 \, , \\
&&[\hat\psi^{\dag}(t,\mathbf{x}), \hat{\psi^{\dag}}(t,\mathbf x')] = 0 \, , \\
&&[\hat\psi^{\dag}(t,\mathbf{x}), \hat{\psi^{\dag}}(t,\mathbf x')] = \delta(\mathbf x  - \mathbf x') \, ;
\end{eqnarray}
}
\item{$\delta\hat\psi(t,\mathbf x)$ and $\delta\hat\psi^{\dagger}(t,\mathbf x)$: Decomposition into a single coherent mode $\psi(t,\mathbf x) = \langle \hat{\psi}(t,\mathbf x)\rangle$, and the quantum excitations $\delta\hat\psi(t,\mathbf x)$ around it. Altogether,
$\hat\psi(t,\mathbf x) = \psi(t,\mathbf x) + \delta\hat\psi(t,\mathbf x)$ and 
$\hat\psi^{\dag}(t,\mathbf x) = \psi^{*}(t,\mathbf x) + \delta\hat\psi^{\dag}(t,\mathbf x)$, where
\begin{eqnarray} 
&&[\delta\hat{\psi} (t,\mathbf{x}) , \delta\hat{\psi} (t,\mathbf x')] =0 \, , \\
&&[\delta\hat\psi^{\dag}(t,\mathbf{x}), \delta\hat{\psi^{\dag}}(t,\mathbf x')] = 0 \, \\
&&[\delta\hat\psi^{\dag}(t,\mathbf{x}), \delta\hat{\psi^{\dag}}(t,\mathbf x')] = \delta(\mathbf x  - \mathbf x') \, ;
\end{eqnarray}
} 
\item{$\hat n$ and $\hat\theta$: Mapping onto Hermitian phase and density fluctuation operators, as studied (for example) in \cite{Barcelo:2003ia,Barcelo:2003yk}.  Here we made use of the fact that the macroscopic
field $\psi(\mathbf x)$ is complex and so for topologically trivial regions ---
without zeros or singularities --- one can always express it as
$\psi(t,\mathbf x) = \sqrt{n(t,\mathbf x) } \; \exp(i\theta(t,\mathbf x) )$. Linearizing around the two parameters of the complex-valued field, $\theta \to \theta_{0} + \hat\theta$ and $n \to n_{0} + \hat n$, we can write
$\hat\psi \simeq \psi + \sqrt{n_{0}} \left( \frac{\hat n}{2 n_{0}} + i \, \hat\theta  \right)$, and its Hermitian conjugate, such that
\begin{eqnarray} 
&&\left[\hat n(t,\mathbf x),\hat n(t,\mathbf x')\right]=0 \, , \\
&&\left[ \hat\theta(t,\mathbf x),\hat\theta(t,\mathbf x')\right]=0 \, , \\
&&\left[\hat n(t,\mathbf x),\hat\theta(t,\mathbf x^{\prime})\right]=i\,\delta(\mathbf x-\mathbf x^{\prime}) \, .
\end{eqnarray}
}
\item{Finally, we are able to use the equation of motion \refb{Eq:n_of_theta} to formally express $\hat n$ in terms of $\hat \theta$:
\begin{eqnarray} 
&&\left[\frac{1}{\widetilde U} \, \frac{D\hat\theta(t,\mathbf x)}{Dt} ,\frac{1}{\widetilde U} \, \frac{D\hat\theta(t,\mathbf x')}{Dt} \right]=0 \, , \\
&&\left[ \hat\theta(t,\mathbf x),\hat\theta(t,\mathbf x')\right]=0 \, , \\
&&\left[ \frac{1}{\widetilde U} \, \frac{D\hat\theta(t,\mathbf x)}{Dt} ,\hat\theta(t,\mathbf x^{\prime})\right]=-\frac{i}{\hbar}\,\delta(\mathbf x-\mathbf x^{\prime}) \, ;
\end{eqnarray}
}
\end{enumerate}
There are two different ways to view the resulting commutator relations in the hydrodynamic and eikonal limit: A condensed matter point of view in terms of time-dependent commutators, or a commutator relationship for the phase perturbation $\hat\theta$ and its conjugate momentum $\hat\Pi_{\hat\theta}$ on the emergent spacetime. \\

%~~~~~~~~~~~~~~~~~~~~~~~~~~~~~~~~~~~~~~~~~~~~~~~~~~~~
\paragraph{Condensed matter point of view: \label{Sec:Condensed.Matter.Commutators}}
%~~~~~~~~~~~~~~~~~~~~~~~~~~~~~~~~~~~~~~~~~~~~~~~~~~~~
First we face the problem of how to deal with the differential operator $\widetilde D_{2}$, which involves studying two interesting limits where the commutation relation takes a simpler form. \\

In the \emph{hydrodynamic approximation} we get 
\begin{equation}
\left[ \frac{D\hat\theta(t,\mathbf x)}{Dt},\hat\theta(t,\mathbf x^{\prime})\right]=\frac{U(t)}{i\hbar}\,\delta(\mathbf x-\mathbf x^{\prime}),
\end{equation}
which is now a time-dependent commutation relation. \\

For $U(t) \rightarrow 0$ the
hydrodynamic commutator vanishes completely and we are left with
purely classical statements for $\hat\theta$.  This situation changes
significantly if one instead considers the eikonal approximation. \\

In the \emph{eikonal approximation} we get (in momentum space) 
\begin{equation}
\left[ \frac{D\hat\theta(t,\mathbf{k})}{Dt},\hat\theta(t,\mathbf{k}^{\prime})\right]
=\frac{U(t)+{\hbar^{2} k^2 \over{4m n_{0}}}}{i\hbar}\,\delta_{kk^{\prime}} \, ,
\end{equation}
and the commutator does not vanish for $U(t)\rightarrow 0$, though it can
vanish if $U(t)$ becomes negative.  \\

This suggests that the
presence of $\widetilde D_{2}$ cannot in general be neglected for a time-dependent atomic interaction $U(t)$. \\
%

%~~~~~~~~~~~~~~~~~~~~~~~~~~~~~~~~~~~~~~~~~~~~~~~~~~~~
\paragraph{Emergent spacetime point of view: \label{Sec:Emergent.Spacetime.Commutators}}
%~~~~~~~~~~~~~~~~~~~~~~~~~~~~~~~~~~~~~~~~~~~~~~~~~~~~
An alternative insight can be gained if we define an emergent Lagrange density,
\begin{equation} \label{Eq:emergent.Lagrange.density}
\mathcal{L}=-\frac{1}{2} f^{ab} \; \partial_{a}\hat\theta \, \partial_{b}\hat\theta\, ,
\end{equation}
in correspondence with Eq.~\refb{Eq:KGE}. 
The momentum conjugate to $\hat \theta$ is given by
\begin{equation}
\hat\Pi_{\hat\theta} :=\frac{ \partial \mathcal{L}}{\partial (\partial_{t} \hat\theta)}=-f^{tb} \, \partial_{b}\hat\theta,
\end{equation}
and hence we evaluate the conjugate momentum to $\hat\theta$ as,
\begin{equation}
\hat\Pi_{\hat\theta}=\frac{\hbar}{\widetilde U} \, \frac{D\hat\theta}{Dt} \, .
\end{equation}
With this new insight we are able to add another set of commutation relations, one that makes only sense after having introduced the emergent spacetime:
\begin{itemize}
\item[5.]{The phase and density operators are a canonical set of quantum field operators and conjugate field operators,
\begin{eqnarray}
\label{Eq:C1_5}&& \left[ \hat \theta(t,\mathbf x),\hat\theta(t,\mathbf x')\right] = 0 \, , \\ 
\label{Eq:C2_5}&& \left[ \hat\Pi_{\hat\theta}(t,\mathbf x),\hat\Pi_{\hat\theta}(t,\mathbf x')\right] =0 \, ,\\ 
\label{Eq:C3_5}&& \left[ \hat \theta(t,\mathbf x),\hat\Pi_{\hat\theta}(t,\mathbf x')\right] = i \delta(\mathbf x - \mathbf x') \, ; 
\end{eqnarray}
in an effective curved spacetime represented by Eq.~\refb{Eq:Acoustic.Metric}, for a massless spin-zero scalar field.}
\end{itemize}

In the \emph{hydrodynamic limit} $\widetilde U \to U$ we recover
\begin{equation}
\label{Eq:Conjugate.Momentum.Acoustic.Limit}
\hat\Pi_{\hat\theta}=\frac{\hbar}{U} \, \frac{D\hat\theta}{Dt} \, ,
\end{equation}
the standard result for the conjugate momentum in curved spacetime; for more details see \cite{Birrell:1984aa,Parker:1969aa}. \\

The conjugate momentum \emph{for ultraviolet modes} $\tilde U \to U_{k}$ is given by
\begin{equation}
\label{Eq:Conjugate.Momentum.Eikonal.Limit}
\hat\Pi_{\hat\theta} \to \hat\Pi_{\hat\theta,k}=\frac{\hbar}{U_{k}} \, \frac{D\hat\theta}{Dt} \, ,
\end{equation}
as the conjugate momentum in our rainbow geometry.\\

This is a significant result, since it shows that knowledge of the emergent spacetime picture provides a deeper insight into the full dynamics for the density and phase perturbations, and explains the explicit time-dependence in their commutation relations.

%+++++++++++++++++++++++++++++++++++++++++++++++++++++++++++++++++
\subsection[FRW-rainbow geometries]{FRW-rainbow geometries:\protect\\ 
Specific time-dependence for atom-atom scattering
\label{Sec:FRW.type.geometries}}
%+++++++++++++++++++++++++++++++++++++++++++++++++++++++++++++++++
Clearly time dependence can enter in any of the parameters $n_{0}$, $c$, and $\mathbf v$. We now focus on the case where $\mathbf{v}= 0$, and $n_{0}$ is constant throughout space and time, so that the system is homogeneous. This choice of parameters leads to the specific class of $k=0$ spatially flat FRW spacetimes \cite{Barcelo:2003yk,Duine:2002aa}.  Such geometries are always conformally flat and at any particular time the spatial geometry is simply that of flat Euclidean space. \emph{All} the time dependence is contained entirely in the speed of sound given by Eq.~\refb{Eq:SoundSpeed}. In a homogeneous condensate the speed of sound and the scale factor are position independent, and within the acoustic\,/\,eikonal limit a separation of the field operators into time and position dependent parts is possible. \\

We introduce the dimensionless scale function $b(t)$ so that the interaction strength (or equivalently the scattering length) becomes time-dependent, 
\begin{equation}
U(t) = U_{0} \, b(t) \, ,
\end{equation}
and under consideration of Eq.~\refb{Eq.Effective.U} we get
\begin{equation}
\label{Eq.Effective.U.time}
\widetilde U = U_{0} \, \widetilde b(t)   \, ,
\end{equation}
and a time-dependent differential scaling operator $\widetilde b(t)$;
\begin{equation}
\label{Eq.Effective.b.time}
\widetilde b(t) = b(t) - \frac{\hbar^2}{2 m \, U_{0}}\;\widetilde D_{2} \, .
\end{equation}
It is important to realize that a change in the interaction strength, $U \to U(t)$, inevitably involves a shift in the crossover between the phononic and trans-phononic regime, and hence the nature of  collective excitations in the condensate. To see this relationship more clearly we write down the eikonal approximation of Eq.~\refb{Eq.Effective.b.time},
\begin{equation}
\label{Eq.Effective.b.time.eikonal}
U_{k}(t) =U_{0} \; b_{k}(t) = U_{0} \; \left(b(t)+ (k/K)^{2} \right) \, ,
\end{equation} 
where for further convenience it is useful to introduce $K$, 
\begin{equation}
\label{Eq:K}
K = \frac{c_{0}}{\gamma_{\mathrm{qp}}} \, ,
\end{equation}
which represents the crossover between the phononic and trans-phononic regimes. Here we defined $c_0=c(t_0)$, as the initial speed of sound.
We would like to stress the importance of the last two equations for the understanding of everything that follows below. Motivated by the ultraviolet deviations in our results obtained by the numerical simulation of time-dependent spacetimes in a realistic Bose--Einstein condensate \cite{Jain:2006ki}, we set out to find a description that includes trans-phononic modes into the spacetime picture. This forced us to generalize acoustic metrics to momentum-dependent (rainbow) metrics, that cover a larger $k$-range. These rainbow metrics, see Eq.~\refb{f2}, show a time- and  momentum-dependent scaling in the effective scale function for the interaction strength. To show this we take $b(t_0)=1$ at some arbitrary initial time $t_0$. 
From the convergence constraint \refb{Eq:Convergence.Radius} for a uniform condensate at rest at $t=0$, and working in the eikonal approximation,
\begin{equation}
\label{Eq:Convergence.Constraint.Uniform.Condensate}
\left\vert k \right\vert< \left\vert K \right\vert \, .
\end{equation}
This implies that for modes $k>K$ the spacetime picture begins to break down. This strongly suggests that one should consider $1/K$ as the analogue Planck length,
$\ell_{\mathrm{Planck}} = 1/K$. For modes with wavelength $k \ll K$ we recover the ``standard'' geometry (\ie, momentum-independent spacetimes). These modes are phononic modes. Making use of our rainbow geometries we extended the emergent spacetime picture for higher energetic (ultraviolet) modes with wavelengths $0< k < K$. For rainbow metrics the equation of motion for modes with wavelengths $k>K$ are most easily derived in momentum space. This can immediately be seen when one Fourier transforms the initial quantum fluctuations $\delta \hat \psi$ and $\delta \hat \psi^{\dag}$, since then it is not necessary to introduce $g_{ab}$ at all.
The situation is more complicated if we involve time-dependent atomic interactions. At any later time $t>0$ we get a different convergence constraint,
\begin{equation}
\left\vert k \right\vert< \left\vert  \sqrt{b(t)} \, K \right\vert \, ,
\end{equation}
and with it a time-dependent limit on the breakdown of the validity of the emergent geometry, 
\begin{equation}
\label{Eq.Time.Dependent.Effective.Planck.Length}
\ell_{\mathrm{Planck}}(t) = 1/(K \sqrt{b(t)}) \, . 
\end{equation}
In other words, the analogue Planck volume $\ell_{\mathrm{Planck}}(t)^{d}$ is --- depending on the form of $b(t)$ --- shrinking or expanding. We will return to this point in Secs.~\ref{Sec:Qualitative.Behavior.Quantum.Fluctuations} and \ref{Sec:Conclusions.Outlook}, when we explain particle production in a de Sitter-like universe, with a growing Planck volume. Note that for a time dependent condensate density the problem is different, and has been studied in \cite{Fedichev:2004on,Fedichev:2004fi,Fischer:2004iy,Uhlmann:2005rx}.
\\

It should be noted that in the acoustic approximation the scale factor,
\begin{equation} 
\label{Eq:U.acoustic.time}
U_{k}(t) \to U_0 \; b(t) \,
\end{equation}
is indeed momentum-independent as expected. In practice a variation in the interaction strength is possible by using a Feshbach resonance \cite{Vogels:1997aa,Inouye:1998aa,Barcelo:2003ia,Barcelo:2003yk}. \\

From Eq.~\refb{Eq:Eikonal.Sound.Speed} we obtain the time-dependent speed of sound in the eikonal approximation,
\begin{equation}
\label{Eq:Eiknoal.Sound.Speed.time}
c_{k}(t) =  c_0  \; \sqrt{b(t)+ k^{2}/K^{2}} \, ,
\end{equation}
and thus
\begin{equation}
\label{Eq:Omega.FRW}
\omega_{k}(t) = \omega_{0}  \; \sqrt{b(t)+ k^{2}/K^{2}} = \omega_{0}  \; \sqrt{b_{k}(t)} \, ,
\end{equation} 
where
\begin{equation}
\label{Eq:bkt}
b_{k}(t) = b(t)+ k^{2}/K^{2} \, .
\end{equation} 
Their acoustic counterparts are obtained in the limit $b_{k}(t) \to b(t)$. \\

The equation of motion \refb{Eq:WaveEquation}, which depends on $f^{ab}$, therefore shows no explicit dependence on the spatial dimensions $d$ of the condensate. 
In contrast, the emergent spacetime 
\begin{equation} 
\label{Eq:FRW.Acoustic}
d s^2 =  \left( \frac{n_0}{c_0}\right)^{\frac{2}{d-1}}
\left[ -c_0^2 \, b_{k}(t)^{\alpha} \, dt^2 + b_{k}(t)^{\alpha- 1} \, d\mathbf x^2 \right] \, .
\end{equation}
is explicitly dependent on $d$ through $g_{ab}$, see Eqs.~\refb{Eq:Acoustic.Metric} and~\refb{Eq:Rainbow.Metric}.
Here the exponent $\alpha$ is dimension-dependent and given as
\begin{equation}
\label{Eq:Acoustic.Lineelement.Exponent}
\alpha = \frac{d-2}{d-1} \, .
\end{equation}
In order to use a Bose--Einstein condensate (in the infrared limit) as an analogue model for Einstein's theory of gravity, we have to equate the Friedmann--Roberton--Walker universe line element given in Eq.~\refb{Eq:FRW.GR}, with the line element~\refb{Eq:FRW.Acoustic}.
This explains why the choice of $b(t)$ --- to mimic a specific FRW-type universe scale factor $a(\tau)$ --- depends on the spatial dimensions $d$. 

%
%%%%%%%%%%%%%%%%%%%%%%%%%%%%%%%%%%%%%%%%%%%%%%%%%%%
%
\section[Quantum field theory in rainbow geometries]{Quantum field theory in rainbow geometries:\protect\\
Parametric excitations}
\label{Sec:Cosmo.Particle.Production}			
%
%%%%%%%%%%%%%%%%%%%%%%%%%%%%%%%%%%%%%%%%%%%%%%%%%%%
Before we address problems of interest to the programme of cosmological particle production in a specific FRW-type universe, we prepare the necessary mathematical and physical formalism. Quantum field theory in rainbow geometries requires a careful treatment of the equation of motion and the commutation relations for the quantum field operators. 
There are three major differences between ordinary massless spin-$0$ particles in real curved spacetimes and BEC quasi-particles in the particle production process. These are: (1) The numerical finiteness of the particle production in all circumstances due to the non-perturbative ultraviolet corrections, as we will show below. (2) We are taking the point of view that any Bose--Einstein condensate experiment will have a finite time duration, and therefore well-defined initial and final vacuum states. (3) Finally, there exists a preferred frame, the laboratory frame in which the experiment (or at this stage gedanken-experiment) is implemented. The laboratory time therefore can be viewed as the most relevant coordinate choice. \\

Conditions (1) and (2) allow us to employ the method of instantaneous Hamiltonian diagonalization to adequately calculate the particle production.
The condition (3), that our particle detector is bound to the laboratory-frame simplifies the decision as to which time coordinate to choose. However, this limitation is also very disappointing. A major awareness resulting from ``conventional'' curved-spacetime quantum field theory has been the observer dependence of the particle spectrum~\cite{Unruh:1976aa}: Different coordinatizations motivate different vacuum choices. One might stretch the good nature of any condensed matter experimentalist, asking for a co-moving particle detector --- adjusted to a suitable vacuum in the chosen coordinate system --- in an infinitely long-lasting expanding emergent spacetime. Any unambiguous measurement requires one to stop the expansion, and to project onto a positive and negative plane wave basis for the final value of $U_{k}(t)$.

%
%+++++++++++++++++++++++++++++++++++++++++++++++++++++++++++++++++
\subsection[Particle production]{Particle production\protect\\
Linearized quantum excitations in condensate  
 \label{Sec:ParticleProduction}}
%+++++++++++++++++++++++++++++++++++++++++++++++++++++++++++++++++
The equation of motion for quantum fluctuations in such a $d$-dimensional condensate are given by
\begin{equation}
\partial_{t} \left( \frac{\hbar}{\widetilde U(t)} \, \partial_{t} \hat \theta(t,\mathbf{x}) \right) - \frac{n_{0}\hbar}{m} \nabla^{2} \hat\theta(t,\mathbf{x}) =0 \, ;
\end{equation}
compare with Eq.~\refb{f2}.
In the following we are going to rewrite the equation of motion in a more suitable form, applying an auxiliary field, and simultaneously transform to momentum space.

%~~~~~~~~~~~~~~~~~~~~~~~~~~~~~~~~~~~~~~~~~~~~~~~~~~~~
\subsubsection{Auxiliary field operators in Fourier space \label{Sec:Auxiliary.Field.Fourier.space}}
%~~~~~~~~~~~~~~~~~~~~~~~~~~~~~~~~~~~~~~~~~~~~~~~~~~~~
We use the differential operator $\widetilde R = \widetilde U(t) / \hbar$ to write the field operators in terms of auxiliary field operators $\hat \chi$, where $\hat\theta = (\widetilde R)^{1/2} \,\hat\chi$. As long as $\widetilde U=\widetilde U(t)$, and with it $\hat \theta=\hat \theta(t)$ and $\hat \chi=\hat \chi(t)$, is position independent, we can always write
\begin{eqnarray}
\nonumber
&&\hat\theta(t,\mathbf x)= \int \frac{d^{d}k}{(2\pi)^{d/2}}  \sqrt{\frac{\widetilde U(t)}{\hbar}} \,
\hat\chi_{k}(t) \, \e^{i \, \mathbf{k} \cdot \mathbf{x}}  \\
\nonumber
&\,&= \frac{1}{\sqrt{\hbar}} \int \frac{d^{d}k}{(2\pi)^{d/2}} \hat\chi_{k}(t) \, \sum_{s=0}^{\infty} \frac{(-1)^{s} (2s)!}{(1-2s)(s!)^{2} 4^{s}} \, (\widetilde D_{2})^{s}  \e^{i \, \mathbf{k} \cdot \mathbf{x}}  \\
&\,&= \sqrt{\frac{U_{k}(t)}{\hbar}} \, \int \frac{d^{d}k}{(2\pi)^{d/2}}  
\hat\chi_{k}(t) \, \e^{i \, \mathbf{k} \cdot \mathbf{x}} \, .
\end{eqnarray}
The last statement is naively based on a Taylor series which only converges for modes with wavenumbers $\vert k \vert < \vert K \vert$. Within this radius of convergence the transformation is exact, but can be extended to arbitrary $k$-values in the eikonal approximation, where $\widetilde U(t) \to U_{k}(t)$.
The equation of motion for the mode operators $\hat \chi_{k}$,
\begin{equation}
\label{Eq:Equation.Motion.Chi}
\sqrt{\frac{\hbar}{U_{k}(t)}} \,
\left(
\ddot{\hat{\chi}}_{k}(t) +\Omega_{k}(t)^{2} \, \hat\chi_{k}(t) \right) = 0
\, , 
\end{equation}
and the equal time commutation relations,
\begin{eqnarray}
\label{Eq:C1_6}&& \left[ \hat \chi_{k}(t),\hat\chi_{k'}(t)\right] = 0 \, , \\ 
\label{Eq:C2_6}&& \left[ \partial_{t}\hat \chi_{k}(t),\partial_{t}\hat\chi_{k'}(t)\right] =0 \, ,\\ 
\label{Eq:C3_6}&& \left[ \hat \chi_{k}(t),\partial_{t}\hat\chi_{k'}(t)\right] = i \delta_{k,k'} \, ; 
\end{eqnarray}
are now slightly more convenient than~\refb{Eq:C1_5}-\refb{Eq:C3_5}.
The function $\Omega_{k}(t)$ is defined as
\begin{equation}
\label{Eq:Harmonic.Oscillator.Frequency}
\Omega_{k}(t)^{2} = c_{0}^{2}k^{2}\,b_{k}(t) -\frac{3}{4} \left(\frac{\dot b_{k}(t)}{b(t)}\right)^{2} +\frac{1}{2}\frac{\ddot b_{k}(t)}{b_{k}(t)} \, , 
\end{equation}
and therefore we have the connection between quantum field theory in FRW-type spacetimes and a parametrically excited harmonic oscillator. The notation $\Omega_{k}(t)$ will become quite obvious for time-independent cases, where it reduces to the usual dispersion relation $\Omega_{k} \to \omega_{k}$;  see Eq.~\refb{Eq:Omega.FRW}.
Notice the overall factor, $\sqrt{\hbar/U_{k}(t)}$, in Eq.~\refb{Eq:Equation.Motion.Chi}. This is of importance for the particle production process in cases of discontinuous, and continuous (but not differentiable) changes in $U(t)$. We will revisit this issue shortly in Sec.~\ref{Sec:Bogoliubov.Transformation.Time.Dependent.Emergent.Spacetime}.
%~~~~~~~~~~~~~~~~~~~~~~~~~~~~~~~~~~~~~~~~~~~~~~~~~~~~
\subsubsection{Mode expansion\label{Sec:Mode.Expansion}}
%~~~~~~~~~~~~~~~~~~~~~~~~~~~~~~~~~~~~~~~~~~~~~~~~~~~~
The equation of motion \refb{Eq:Equation.Motion.Chi} is a homogeneous differential equation, which can be written as
\begin{equation}
\label{Eq:Equation.Motion.Chi.L}
\widetilde L \, \hat \chi_{k} = 0 \, ,
\end{equation}
where
\begin{equation}
\widetilde L = \partial_{t}^{2} + \Omega_{k}(t)^{2} 
\end{equation}
is a linear second order differential operator with a $2$-dimensional solution space. 

A common tool in quantum mechanics is to describe quantum states --- with unknown or variable number of particles --- with respect to an orthonormal occupancy number basis (\ie, the Fock space basis in the infinite-dimensional function space, the Hilbert space of state).
A Fock state is a quantum state consisting of an ensemble of excited non-interacting particles,
\begin{equation}
\vert n_{\mathbf k1}, n_{\mathbf k2}, ... \rangle = \frac{1}{\sqrt{n_{\mathbf k1}! \, n_{\mathbf k2}! \, ...}} \left[ (\hat a_{\mathbf k1}^{\dag})^{n_{\mathbf k1}}  (\hat a_{\mathbf k2}^{\dag})^{n_{\mathbf k2}} ... \right]  \vert 0 \rangle \, ,
\end{equation}
with definite occupation numbers $(n_{\mathbf k1},n_{\mathbf k2},...)$ in the modes $(\hat\chi_{\mathbf k1},\hat\chi_{\mathbf k2},...)$.
The creation, $\hat a_{k}^{\dag}$, and annihilation, $\hat a_{k}$ operators create or destroy a single-particle in the mode $\chi_{\mathbf k}$;
\begin{eqnarray}
\hat a_{\mathbf ki}^{\dag} \vert ..., n_{\mathbf ki}-1, ... \rangle &=& \sqrt{n_{\mathbf ki}} \, \vert ..., n_{\mathbf ki}, ... \rangle \, , \\
\hat a_{\mathbf ki} \vert ..., n_{\mathbf ki}, ... \rangle &=& \sqrt{n_{\mathbf ki}} \, \vert ..., n_{\mathbf ki}-1, ... \rangle \, .
\end{eqnarray}
The state $\vert 0 \rangle$ --- short for $\vert 0_{\mathbf k1},0_{\mathbf k2},... \rangle$ --- is a special state, the vacuum state. It is defined as the eigenstate of all annihilation operators $\hat a_{k}$ with eigenvalue $0$, such that $\hat a_{k} \vert 0 \rangle = 0$. Thus any arbitrary quantum state $\vert \varphi \rangle $ in the Fock spaces is a linear combination of all excited states,
\begin{equation}
\vert \varphi \rangle = \sum_{n_{\mathbf k1}, n_{\mathbf k2}, ...} P_{{n_{\mathbf k1}, n_{\mathbf k2}, ...}} \, \vert n_{\mathbf k1}, n_{\mathbf k2}, ... \rangle \, ,
\end{equation}
that can be created out of the vacuum. Here $P_{{n_{\mathbf k1}, n_{\mathbf k2}, ...}} $ is the probability to measure the single Fock state with the mode occupation $(n_{\mathbf k1}, n_{\mathbf k2}, ...)$.

Within this framework we expand the mode operators in terms of destruction and creation operators,
\begin{equation}
\label{Eq:Mode.Expansion.1}
\hat \chi_{k}(t) = \frac{1}{\sqrt{2}} \,
\left[ v_{k}^{*}(t) \, \hat a_{k} + v_{k}(t) \, \hat a_{-k}^{\dag}\right] \, .
\end{equation}
The coefficients $v_{k}$ and $v_{k}^{*}$ are a set of linearly independent mode functions and any linear combination is a complete solution of
\begin{equation}
\widetilde L v_{k}(t) = 0 \, .
\end{equation}
(Due to the isotropy of the $s$-wave scattering amplitude of the atomic interactions, the mode functions are also isotropic, \ie, $v_{\mathbf k} = v_{k}$ and $v_{\mathbf k}^{*}=v_{k}^{*}$. It therefore seems unlikely that we would need to extend the specific analogue set-up to mimic anisotropic expansion scenarios. The case is different for analogue models involving a changing condensate density, $n_{0}=n_{0}(t)$; where a non-uniform density expansion can easily be achieved, for example see \cite{Fedichev:2004on,Fedichev:2004fi,Fischer:2004iy,Uhlmann:2005rx}.)

In order to obtain a ``nice'' canonical set of operators, which obey the common equal time commutation relations (see Eq.~\refb{Eq:C1_5}-\refb{Eq:C2_5}),
\begin{eqnarray}
\label{Eq:C1_7}&& \Big[ \hat a_{k}^{},\hat a_{k'}^{}\Big] = 0 \, , \\ 
\label{Eq:C2_7}&& \left[ \hat a_{k}^{\dag},\hat a_{k'}^{\dag}\right] =0 \, ,\\ 
\label{Eq:C3_7}&& \left[ \hat a_{k},\hat a_{k'}^{\dag} \right] = \delta_{k,-k'} \, ,
\end{eqnarray}
the Wronskian $W$ of the mode functions has to be normalized as follows,
\begin{equation}
\label{Eq:Wronskian}
W[v_{k},v_{k}^{*}] = \dot v_{k} \, v_{k}^{*} - v_{k} \, \dot v_{k}^{*} = 2 i \,.
\end{equation}
It is easy to see that --- under the application of the equation of motion for the mode functions --- the Wronskian is always time-independent. 
Note, a spin-zero scalar particle is its own anti-particle, where $(\hat a_{k})^{\dag}=\hat a_{-k}^{\dag}$. We would also like to draw attention to the fact that with a particular choice of Fock space, we select a particular subspace of the Hilbert space of states (\ie, the subspace that can be created out of the vacuum state, determined by our choice of $\hat a_{k}$). This will be of further interest when we discuss the validity and limitations of the Bogoliubov transformation.
%
%~~~~~~~~~~~~~~~~~~~~~~~~~~~~~~~~~~~~~~~~~~~~~~~~~~~~
\subsubsection{Bogoliubov transformation\label{Sec:Bog.Trafo}}
%~~~~~~~~~~~~~~~~~~~~~~~~~~~~~~~~~~~~~~~~~~~~~~~~~~~~
In Eq.~\refb{Eq:Mode.Expansion.1} we can in principle pick some particular mode expansion,
$(\vec v_{k})^{\dag} \cdot \vec{A}_{k}$, with a specific set of mode functions $\vec v_{k}^{T}=(v_{k}(t),v_{k}^{*}(t))$, and set of mode operators $\vec{A}_{k}^{T}=(\hat a_{k},\hat a_{-k}^{\dag})$. This choice is not unique and any other mode expansion $(\vec u_{k})^{\dag} \cdot \vec{B}_{k}$, with mode functions $\vec u_{k}^{T}=(u_{k}(t),u_{k}^{*}(t))$, and mode operators $\vec{B}_{k}^{T}=(\hat b_{k},\hat b_{-k}^{\dag})$ would have been possible. The relation between these two different representations is referred to as Bogoliubov transformation.

As mentioned above, the mode expansion $(\vec v_{k})^{\dag} \cdot \vec{A}$ corresponds to an orthonormal basis in the infinite-dimensional function space, provided the $\vec v_{k}$ are normalized; that is $W[v_{k}(t),v_{k}^{*}(t)]=2i$ for all times $t$. If we restrict ourselves to exclusively mapping between orthonormal frames --- then the mode functions fulfill the normalization constraint $W[u_{k}(t),u_{k}^{*}(t)]=2i$, and the mode operators obey the commutation relations given in Eqs.~(\ref{Eq:C1_7}-\ref{Eq:C3_7}) ---
the Bogoliubov transformation is given by $\vec{v}_{k} = M \cdot \vec{u}_{k}$, where $M$ must be a $2\times 2$ matrix,
\begin{equation}
\label{Eq:M.Transformation.Matrix}
M =
\left(
\begin{array}{cc}
\alpha_{k}^{*} & \beta_{k}^{*} \\ \beta_{k} & \alpha_{k}
\end{array}
\right) \, .
\end{equation}
The complex-valued coefficients $\alpha_{k}$ and $\beta_{k}$ are called Bogoliubov coefficients, and the transformation preserves the normalization condition if $\det(M)=1$, that is for
\begin{equation}
\label{Eq:Bogoliubov.Normalization.Alpha.Beta}
\vert \alpha_{k} \vert^{2} - \vert \beta_{k} \vert^{2} = 1 \, .
\end{equation}
At this stage we notice a minor technical subtlety when we look at summing over discreet $k$ modes versus integrating over continuous $k$ modes. Equation \refb{Eq:Bogoliubov.Normalization.Alpha.Beta} is appropriate for summing over discreet modes whereas for continuous modes we should strictly speaking use $\vert \alpha_{k} \vert^{2} - \vert \beta_{k}\vert^{2} =\delta^{(d)}(\vec{0})=V/(2\pi)^{d}$. Here $\delta^{(d)}(\vec{0})$ is the $d$-dimensional momentum space Dirac function. We will work with continuous integrals in momentum space but will suppress unnecessary occurrence of the volume V, where including it would lead to unnecessary clutter that might not aid understanding. \\

The relationship between the ``old'' and ``new'' mode operators, where $(\vec v_{k})^{\dag} \cdot \vec{A}=(\vec u_{k})^{\dag} \cdot \vec{B}$ and $(\vec v_{k})^{\dag} = (\vec u_{k})^{\dag} \cdot M^{\dag}$, is given as $\vec{A}_{k}=(M^{\dag})^{-1} \cdot \vec{B}_{k}$ and $\vec{B}_{k}=(M^{\dag}) \cdot \vec{A}_{k}$.
Here
\begin{equation}
M^{\dag} =
\left(
\begin{array}{cc}
\alpha_{k} & \beta_{k}^{*} \\ \beta_{k} & \alpha_{k}^{*}
\end{array}
\right)  
~~\mbox{and}~~ 
(M^{\dag})^{-1} =
\left(
\begin{array}{cc}
\alpha_{k}^{*} & -\beta_{k}^{*} \\ -\beta_{k} & \alpha_{k}
\end{array}
\right)  \, ,
\end{equation}
and therefore
\begin{eqnarray}
\label{Eq:operator.ak}
\hat a_{k} &=& \alpha_{k}^{*} \hat b_{k} - \beta_{k}^{*} \hat b_{-k}^{\dag},~~\mbox{and}~~
\hat a_{-k}^{\dag} = \alpha_{k}\hat b_{-k}^{\dag} - \beta_{k} \hat b_{k}^{\dag}  \,  ;  \; \;\\
\label{Eq:operator.bk}
\hat b_{k} &=& \alpha_{k} \hat a_{k}       + \beta_{k}^{*} \hat a_{-k}^{\dag},~~\mbox{and}~~
\hat b_{-k}^{\dag} = \alpha_{k}^{*}\hat a_{-k}^{\dag} - \beta_{k} \hat a_{k}^{\dag}  \, . \; \;
\end{eqnarray}
Obviously, the two vacuum states, $\hat a_{k} \, \vert {}_{(a)} 0 \rangle = 0$ and $\hat b_{k} \, \vert {}_{(b)} 0 \rangle = 0$, are different, as $\hat b_{k} \, \vert {}_{(a)} 0 \rangle \neq 0$ for $\beta_{k}\neq 0$. 

%~~~~~~~~~~~~~~~~~~~~~~~~~~~~~~~~~~~~~~~~~~~~~~~~~~~~
\subsubsection{Particles and lowest energy eigenstate\label{Sec:Meaning.Particles}}
%~~~~~~~~~~~~~~~~~~~~~~~~~~~~~~~~~~~~~~~~~~~~~~~~~~~~
\emph{Quasi-particles} are excited states, and therefore they depend on the choice of (mode functions and the associated) vacuum state. 
Under the application of Eq.~\refb{Eq:operator.bk} we can \emph{formally} calculate the mean density of $b$-particles, $\hat n^{(b)}$, the occupation number for the mode $\hat \chi_{k}$, with respect to the $a$-vacuum $\vert {}_{(a)} 0 \rangle$ as follows:
\begin{equation}
\label{Eq:Occupation.Number.Desity}
n_{k}^{(b)} = \frac{\langle {}_{(a)} 0 \vert \, \hat b_{k}^{\dag} \hat b_{k} \, \vert {}_{(a)}0 \rangle}{\mbox{Volume}} = \vert \beta_{k} \vert^{2}  \,.
\end{equation}
The challenge is to find particular specified mode functions that represent physically meaningful vacuum states, that minimize the expectation value of the Hamiltonian. Only then does the chosen vacuum correspond to the ``actual'' physical vacuum, and its excitations describe ``real'' quasi-particles. (That is, we are looking for an unambiguous measure of particles relative to some vacuum state.)
The problem of finding the ``right'' vacuum state (``best'' orthonormal frame) has been controversial from the outset. However, as we shall see, it is straightforward to find the zero-particle vacuum state for Minkowski spacetimes, and the whole controversy arises only for time-dependent cases, where the Hamiltonian $\hat H(t)=\int_{\mathbf k} \hat{\mathcal{H}}_{k}(t)$ is explicitly time-dependent;
\begin{eqnarray}
\nonumber
\hat{\mathcal{H}}_{k}(t) &=&  \frac{   
\{ [\partial_{t} ( \sqrt{U_{k}(t)} \hat \chi_{k}(t) ) ] 
[\partial_{t} ( \sqrt{U_{k}(t)} \hat \chi_{-k}(t) ) ]\}
}{2 \, U_{k}(t)} \\
\label{Eq:Hamiltonian.Density}
&\;& +\frac{1}{2}\, \omega_{k}(t)^{2} \, \chi_{k}(t) \, \chi_{-k}(t) \, .
\end{eqnarray}
In combination with the mode expansion \refb{Eq:Mode.Expansion.1} the Hamiltonian in momentum space is given by
\begin{eqnarray}
\nonumber
\hat H(t) &=& \frac{1}{4} \, \int d^{d}\mathbf k \, F_{k}(t) \, \hat a_{k}^{\dag} \hat a_{-k} + F_{k}^{*}(t) \, \hat a_{k} \hat a_{-k}\\ 
\label{Eq:Hamiltonian.Mode.Expansion}
&\;& \quad \quad \quad \quad  + E_{k}(t) \, (2 \hat a_{k}^{\dag} \hat a_{k} + \delta^{(d)}(0) ) \, ,
\end{eqnarray}
where the factor $E_{k}(t)$ for the diagonal terms is,
\begin{eqnarray}
\nonumber
E_{k}(t) &=& \vert \dot{v}_{k}(t) \vert^{2} + \{ \omega_{k}^{2}(t) + 1/4 \, [ \dot U_{k}(t)/U_{k}(t)]^{2} \} \, \vert v_{k} \vert^{2} \\  
\label{Eq:Ekt}
&+& 1/2 \,  \dot U_{k}(t)/U_{k}(t)  \, \partial_{t} \vert v_{k}(t) \vert^{2} \, ,
\end{eqnarray}
and the off-diagonal term $F_{k}(t)$ is given by
\begin{eqnarray}
\nonumber
F_{k}(t) &=& ( \dot{v}_{k}(t) )^{2} + \{ \omega_{k}^{2}(t) + 1/4 \, [ \dot U_{k}(t)/U_{k}(t)]^{2} \} \, (v_{k})^{2} \\  
\label{Eq:Fkt}
&+& 1/2 \,  \dot U_{k}(t)/U_{k}(t)  \, \partial_{t} (v_{k}(t))^{2}  \, .
\end{eqnarray}
Therefore there are no time-independent eigenstates that represent the physical vacuum state, one that minimizes the expectation value for the Hamiltonian,
\begin{equation}
\langle {}_{(t)} 0 \vert \, \hat H_{k}(t) \, \vert {}_{(t)} 0 \rangle = \frac{1}{4} \, \int d^{d}\mathbf{k} \, E_{k}(t)  \, ,
\end{equation}
for all times.

Yet somehow, in order to obtain a meaningful statement about the particle production in our effective curved spacetime, we need to deal with or circumvent this problem. \\

%~~~~~~~~~~~~~~~~~~~~~~~~~~~~~~~~~~~~~~~~~~~~~~~~~~~~
\paragraph{Instantaneous Hamiltonian diagonalization:
\label{Sec:Instantaneous.Hamiltonian.Diagonalization}}
%~~~~~~~~~~~~~~~~~~~~~~~~~~~~~~~~~~~~~~~~~~~~~~~~~~~~
One possibility to find an approximate vacuum at an instant of time --- say $t=t_{0}$ ---  is to define the vacuum state $\vert {}_{(t_{0})} 0 \rangle$ of the \emph{instantaneous} Hamiltonian $\hat H(t_{0})$, where we project our mode functions instantaneously onto a plane-wave basis with the dispersion relation $\omega_{k}(t_{0})$. Within this approximation we find for the coefficients in Eq.~\refb{Eq:Hamiltonian.Mode.Expansion}:
\begin{eqnarray}
\label{Eq:Ekt.Minkowski}
E_{k} &=&  \vert \dot{v}_{k}(t) \vert^{2} +  \omega_{k}^{2}(t_{0})  \, \vert v_{k} \vert^{2}  \, ; \\
F_{k} &=&  ( \dot{v}_{k}(t) )^{2} + \omega_{k}^{2}(t_{0})  \, (v_{k})^{2} \, .
\end{eqnarray}
It can be shown that the expectation value for the instantaneous Hamiltonian --- up to an arbitrary phase --- at $t=t_{0}$ is minimal for mode functions $v_{k} = 1/\sqrt{\omega_{k}(t_{0})}$ and $\dot v_{k}=i \, \sqrt{\omega}=i\, \omega_{k} \, v_{k}(t_{0})$. 

Note that at $t_{0}$ the two coefficients $F_{k}(t_{0})=0$ and $E_{k}(t_{0})=2\omega_{k}(t_{0})$, and that under this condition the instantaneous Hamiltonian (see Eq.~\refb{Eq:Hamiltonian.Mode.Expansion}) is diagonal. Therefore it is referred to as the \emph{vacuum state of instantaneous Hamiltonian diagonalization}.

In a time-dependent problem the vacuum state changes, and therefore one has to give up on this particular definition of vacuum for times $t>t_{0}$. The association between the $\vert \beta_{k} \vert^{2}$ Bogoliubov coefficients (as outlined above) and physical quasi-particles requires that we approach --- either asymptotically or abruptly  --- flat initial and final state spacetimes. Only a time-independent spacetime perpetuates its vacuum state, where $\omega_{k}(t) = \omega_{k}$, and thus the normalized mode functions
\begin{eqnarray}
\label{Eq:Mode.Functions.Minkowski.v}
v_{k}(t) &=& \frac{1}{\sqrt{\omega_{k}(t)}} \, \exp(i\,\omega_{k} \, t) \, , \\
\label{Eq:Mode.Functions.Minkowski.vc}
v_{k}^{*}(t) &=& \frac{1}{\sqrt{\omega_{k}(t)}} \, \exp(-i\,\omega_{k} \, t) \, ,
\end{eqnarray}
in this case represent a physically meaningful vacuum state for all times $t$.  
%
%~~~~~~~~~~~~~~~~~~~~~~~~~~~~~~~~~~~~~~~~~~~~~~~~~~~~
\subsubsection{Bogoliubov transformation in emergent spacetimes \label{Sec:Bogoliubov.Tranformation.Emergent.Spacetimes}}
%~~~~~~~~~~~~~~~~~~~~~~~~~~~~~~~~~~~~~~~~~~~~~~~~~~~~
The Bogoliubov transformation outlined in Sec.~\ref{Sec:Bog.Trafo} can be applied to time-dependent problems, as long as we avoid making any statement about the presence of ``real'' quasi-particles unless the expansion has started from an initial time-independent spacetime, and eventually approaches a time-independent spacetime. We also require $\omega_{k}^{2}>0$, such that at the beginning and the end the mode functions are simple (normalized) in- and out-going plane waves. For a constant, but imaginary frequency, $\omega_{k}^{2}<0$, the instantaneous Hamiltonian is still diagonal, but its expectation value no longer has a minimum.
(A detailed treatment of this problem in spacetimes emerging from a Bose gas can be found in Ref.~\cite{Weinfurtner:2007aa}. There the underlying physical model exhibits a negative scattering length.)

To apply the Bogoliubov transformation (see, Sec.~\ref{Sec:Bog.Trafo}, and see Eqs.~(\ref{Eq:operator.ak}-\ref{Eq:operator.bk})) to our specific problem it is necessary to consider that the original problem was defined in terms of two linearly independent field variables, the field operator $\hat \theta$, and its conjugate momentum $\hat \Pi_{\hat \theta}$.
The connection conditions arise from the necessity that the field operator,
\begin{equation}
\label{Eq:C1}
\left[ \hat \theta \right] = \lim_{\epsilon \to 0} \left\{ \hat \theta(t-\epsilon)  -  \hat \theta(t+\epsilon) \right\}  = 0 \, ,
\end{equation}
and its conjugate momentum (on the emergent spacetime),
\begin{equation}
\label{Eq:C2}
\left[ \hat \Pi_{\hat\theta}  \right] =  \lim_{\epsilon \to 0} \left\{   \hat \Pi_{\hat\theta}(t-\epsilon) -  \hat \Pi_{\hat\theta}(t+\epsilon) \right\} = 0 \, ,
\end{equation}
have to be continuous at all times. (A more detailed treatment of this problem in the context of emergent spacetimes from Bose gases can be found in \cite{Weinfurtner:2007aa}.)
Note, that these two conditions have to be fulfilled for any arbitrary (albeit physically reasonable) change in the contact potential $U(t)$.
In Sec.~\ref{Sec:Auxiliary.Field.Fourier.space} we have already pointed out that our emergent spacetime is special in the sense that all time-dependence (in our model) has to be implemented via $U(t)$, which also shows up in the overall conformal factor in the equation of motion, see 
Eq.~\refb{Eq:Equation.Motion.Chi}. \\

%~~~~~~~~~~~~~~~~~~~~~~~~~~~~~~~~~~~~~~~~~~~~~~~~~~~~
\paragraph{Bogoliubov coefficients for a time-dependent emergent spacetime: 
\label{Sec:Bogoliubov.Transformation.Time.Dependent.Emergent.Spacetime}}
%~~~~~~~~~~~~~~~~~~~~~~~~~~~~~~~~~~~~~~~~~~~~~~~~~~~~
Let us consider a contact potential defined as follows,
\begin{equation}
U_{k}(t) =  g_{k}(t) \, \Theta_{\mathrm{HS}}(t-t_{0}) +  h_{k}(t) \, \Theta_{\mathrm{HS}}(t_{0}-t)\, .
\end{equation}
We are dealing with different sets of mode functions in each region; $v_{k}^{T}=(v_{k},v_{k}^{*})$ for $t < t_{0}$, and $u_{k}^{T}=(u_{k},u_{k}^{*})$ for $t > t_{0}$. (Here $\Theta_{\mathrm{HS}}$ is the symbol for the Heaviside step function.) For the time being we do not make any further assumption except that the mode functions have to be a solution of the equation of motion~\refb{Eq:Equation.Motion.Chi} in their respective region.
In particular, we do \emph{not yet} assume the modes are normalized. The connection conditions, given in Eqs.~\refb{Eq:C1} and \refb{Eq:C2}, in combination with the Bogoliubov transformation --- see transformation matrix \refb{Eq:M.Transformation.Matrix}, mapping between two pairs of complex-conjugate mode functions (it is always possible to choose such solutions) --- provide us with two matrix equations (that is, four component equations), 
\begin{eqnarray}
\sqrt{\frac{g_{k}}{\hbar}} \, \vec{v}_{k} &=& M \, \sqrt{\frac{h_{k}}{\hbar}} \, \vec{u}_{k} \, , \\
\frac{\hbar}{g_{k}} \,  \partial_{t}\left(\sqrt{\frac{g_{k}}{\hbar}} \, \vec{v}_{k}\right) &=& \frac{\hbar}{h_{k}} \,  M \, \partial_{t}\left( \sqrt{\frac{h_{k}}{\hbar}} \, \vec{u}_{k} \right) \, ,
\end{eqnarray}
for the four unknown transmission coefficients $\alpha_{k}$, $\alpha_{k}^{*}$, $\beta_{k}$, and $\beta_{k}^{*}$ --- which are contained in the matrix $M$. Fortunately, due to our specific choice of complex mode functions, we only need to calculate $\alpha_{k}$ and $\beta_{k}$, and get the others by calculating their complex conjugate. We obtain
\begin{equation}
\label{Eq.Alpha.General}
\alpha_{k}(t) = \frac{2 \, \left( g \, \dot u_{k}  v_{k}^{*} - h \, u_{k} \dot v_{k}^{*} \right) + v_{k}^{*} u_{k} \left( g \, \frac{\dot h}{h} - \frac{\dot g}{g} \, h \right)}{2 \sqrt{g\,h} \; W[u_{k},u_{k}^{*}]} \, ,
\end{equation}
and
\begin{equation}
\label{Eq.Beta.General}
\beta_{k}(t) = \frac{2 \, \left( h \, \dot v_{k}^{*} u_{k}^{*} - g \,  v_{k}^{*} \dot u_{k}^{*} \right) - v_{k}^{*} u_{k}^{*} \left( g \, \frac{\dot h}{h} - \frac{\dot g}{g} \, h \right)}{2 \sqrt{g\,h} \; W[u_{k},u_{k}^{*}]} \, .
\end{equation}
A somewhat time-consuming, but trivial calculation shows that further conditions are necessary to obtain a Bogoliubov transformation in the desired normalized form~\refb{Eq:Bogoliubov.Normalization.Alpha.Beta}, since in general~\refb{Eq.Alpha.General} and~\refb{Eq.Beta.General} lead to
\begin{equation}
\vert \alpha_{k}(t)\vert^{2} - \vert \beta_{k}(t)\vert^{2} = \frac{W[v_{k},v_{k}^{*}]}{W[u_{k},u_{k}^{*}]} \neq 1 \, .
\end{equation}
Hence it is necessary to choose a consistent normalization condition, $W[v_{k},v_{k}^{*}]=W[u_{k},u_{k}^{*}]=\mathrm{constant}$, so that $\vert \alpha_{k}(t)\vert^{2} - \vert \beta_{k}(t)\vert^{2} =1$. Also note that for multiple-step events, meaning multiple Bogoliubov transformations described by the compound matrix $M_{n} \cdot M_{n-1} \cdots M_{2} \cdot M_{1} $, the physics is independent of the particular choice for the normalization of the intermediate mode functions, since
\begin{equation}
\det(M_{n}  \cdots M_{1} ) =\det(M_{n})  \cdots \det(M_{1}) 
= \frac{W[v_{k}^{1},(v_{k}^{1})^{*}]}{W[v_{k}^{n},(v_{k}^{n})^{*}]} \, .
\end{equation}

Our main focus in this article is to calculate the particle production in a FRW-rainbow metric, and therefore we wish to restrict our problems to cases where we start and end in a physically meaningful vacuum state. This way we are able to connect $\vert \beta_{k} \vert^{2}$ with the occupation number for real condensate excitations in an effectively expanding spacetime. In our companion paper, see \cite{Jain:2006ki}, we investigated various expansion scenarios. We will now concentrate on the most relevant expansion scenario in terms of cosmology, \ie, de Sitter like inflation.

There are three principal different cases, in terms of $U(t)$, that are of interest when addressing the problem of inflation in emergent spacetimes:

\begin{description}
\item[(i)]{\emph{$U(t)$ discontinuously, connects two flat spacetime regions:} The sudden case is indirectly interesting for inflation, since the de Sitter expansion approaches the sudden case for infinitely fast expansion~\cite{Visser:1999aa}. But, we will show shortly that in emergent spacetimes --- due to the nonperturbative ultraviolet corrections --- the particle production for the extreme limit is finite, thus the Bogoliubov transformation for the de Sitter case is always well-defined. We discuss this in Sec.~\ref{Sec:Finiteness.Particle.Production.Emergent.Spacetimes}.} 
\item[(ii)]{\emph{$U(t)$ continuously, but not continuously differentiably, connects two flat spacetime regions with a finite de Sitter like phase in between:} This calculation can be carried out in the hydrodynamic limit, but fails to be a good approximation for infinitely long-lasting inflation in emergent spacetimes.
Due to an unsolvable second order differential equation in the eikonal limit, we present a qualitative analysis, that should be compared with the numerics presented in our companion paper \cite{Jain:2006ki}. We discuss this in Sec.~\ref{Sec:Toy.Model.Conventional.Inflation}.}
\item[(iii)]{\emph{$U(t)$ is a smooth function everywhere, such that we are left with one de Sitter like region:} The qualitative analysis shows that in our particular emergent spacetime the de Sitter expansion has, in the infinite past and in the infinite future, two distinct physical vacua, and therefore we are able to predict the existence of time-independent real (unambiguous) quasi-particles created during an infinitely long-lasting expansion. This might at first seem of less interest for the condensed matter community, but as our simulations show, this can be realized for a sufficiently long expansion time. We will see that for the right parameter choice the particle spectrum approaches a characteristic final shape, for which we can numerically determine the form of the final particle spectrum. We discuss this in Sec.~\ref{Sec:Long.Lasting.Rainbow.Inflation}.}
\end{description}

%+++++++++++++++++++++++++++++++++++++++++++++++++++++++++++++++++
\subsection[Finiteness of particle production in emergent spacetimes]{Finiteness of particle production in emergent spacetimes \label{Sec:Finiteness.Particle.Production.Emergent.Spacetimes}}
%+++++++++++++++++++++++++++++++++++++++++++++++++++++++++++++++++
In a universe that is subjected to a (finite-size) expansion, and in particular a sudden variation in the size of the universe~\cite{Visser:1999aa}, there is a relatively simple way to calculate an upper bound on the particle production.

The case of a particle production for a sudden transition has been previously explored by Jacobson for a parametric oscillator \cite{Jacobson:2003ac}. (Though the underlying physics is rather different, there is also a model for \emph{sonoluminescence} that is based on a rapid change in refractive index --- that model shares many of the mathematical features encountered in the present calculation~\cite{Liberati:1998aa,Liberati:2000aa,Liberati:2000ab,Liberati:2000ac,Belgiorno:1999ha,Liberati:1999aa,Liberati:2006sj}.)
A more closely related work on sudden changes --- between Lorentzian\,/\,Euclidean signatures --- in emergent spacetime  has recently been carried out in \cite{Weinfurtner:2007aa}.

To calculate the particle production in the limiting case of a sudden expansion, we consider the situation where the atom-atom interaction is ``instantaneously'' switched from $U$ to $U/X$ at some time $t_0$. 
The scale function~\refb{Eq:U.acoustic.time} is given by
\begin{eqnarray}
\label{bsudden}
b(t) = 1 - \left(1 - \frac{1}{X}\right) \Theta_{\mathrm{HS}}(t-t_{0}) \, .
\end{eqnarray}
This corresponds to a change between two regions with distinct dispersion relations, connected at $t_{0}$. We choose as positive and negative frequency mode functions for $t<t_{0}$,
\begin{eqnarray}
\label{Eq:Mode.Functions.Minkowski.in}
v_{k}(t)=\frac{\e^{+i\omega_{k}^{\mathrm{in}}t}}{\sqrt{\omega_{k}^{\mathrm{in}}}} 
~~\mbox{and}~~
v_{k}^{*}(t)=\frac{\e^{-i\omega_{k}^{\mathrm{in}}t}}{\sqrt{\omega_{k}^{\mathrm{in}}}} \, , 
\end{eqnarray}
and for  $t>t_{0}$,
\begin{eqnarray}
\label{Eq:Mode.Functions.Minkowski.out}
u_{k}(t)=\frac{\e^{+i\omega_{k}^{\mathrm{out}}t}}{\sqrt{\omega_{k}^{\mathrm{out}}}} 
~~\mbox{and}~~
u_{k}^{*}(t)=\frac{\e^{-i\omega_{k}^{\mathrm{out}}t}}{\sqrt{\omega_{k}^{\mathrm{out}}}} \, ; 
\end{eqnarray}
these are normalized using Eq.~\refb{Eq:Wronskian}. As explained in Sec.~\ref{Sec:Instantaneous.Hamiltonian.Diagonalization}, these mode functions only represent physically meaningful vacua in flat spacetime regions. 

Straightforwardly, we can apply Eqs.~\refb{Eq.Alpha.General} and \refb{Eq.Beta.General} to calculate $\alpha$ and $\beta$. We get for the Bogoliubov coefficients: 
\begin{eqnarray}
\label{Eq:Alpha.Sudden}
\alpha_{k}(t_{0}) &=& \frac{1}{2} \left[ \sqrt{\frac{\omega_{k}^{\mathrm{out}}}{\omega_{k}^{\mathrm{in}}}} +\sqrt{\frac{\omega_{k}^{\mathrm{in}}}{\omega_{k}^{\mathrm{out}}}} \right]  \e^{i\,(\omega_{k}^{\mathrm{out}} - \omega_{k}^{\mathrm{in}}) \, t_{0}}  ; \quad \; \; \\
\label{Eq:Beta.Sudden}
\beta_{k}(t_{0}) &=& \frac{1}{2} \, \left[ \sqrt{\frac{\omega_{k}^{\mathrm{out}}}{\omega_{k}^{\mathrm{in}}}} -\sqrt{\frac{\omega_{k}^{\mathrm{in}}}{\omega_{k}^{\mathrm{out}}}} \right] \e^{-i\,(\omega_{k}^{\mathrm{out}} + \omega_{k}^{\mathrm{in}}) \, t_{0}}  . \quad \; \;
\end{eqnarray}
Using the transformation laws between the $\mathrm{in}$ and $\mathrm{out}$ Fock state operators we obtain the occupation number density~\refb{Eq:Occupation.Number.Desity} of $\mathrm{out}$-particles in the $\mathrm{in}$-vacuum, 
\begin{equation}
\label{Eq:nk.sudden}
n_{k}^{\mathrm{out}} =\frac{1}{4} \, \left\vert \frac{(\omega_{k}^{\mathrm{out}}-\omega_{k}^{\mathrm{in}})^{2}}{\omega_{k}^{\mathrm{out}}\omega_{k}^{\mathrm{in}}} \right\vert \, ,
\end{equation}
where we assumed that the eigenfrequencies are real. Note that this relation is valid beyond the eikonal limit, such that we are dealing with a nonlinear dispersion, see Eq.~\refb{Eq:Eiknoal.Sound.Speed.time}, in both regions;
\begin{eqnarray}
\omega_{k}^{\mathrm{in}} &=& \omega_{0}^{\mathrm{in}} \sqrt{1 + k^{2}/K^{2}} \, , \\
\omega_{k}^{\mathrm{out}} &=& \omega_{0}^{\mathrm{in}} \sqrt{1/X + k^{2}/K^{2}} \, . 
\end{eqnarray}
Thus we get
\begin{equation}
n_{k}^{\mathrm{out}} = \frac{ 
\left(\sqrt{X} \sqrt{K^{2} + k^{2}} -  \sqrt{K^{2} + X \, k^{2}} \right)^{2}}{4\, \sqrt{X} \sqrt{K^{2} + k^{2}} \, \sqrt{K^{2} + X \, k^{2}}} \, .
\end{equation}
It is perhaps surprising how simple the final result is, despite the additional technical machinery required to derive it.  Of course, particle production is momentum-independent only within the hydrodynamic
limit, when 
\begin{equation}
\omega_{k}^{\mathrm{in}/\mathrm{out}}\to \omega_{0}^{\mathrm{in} /\mathrm{out}} \, . 
\end{equation}
The number density in the infrared limit, when $k < K$, can be simplified to
\begin{equation}
n_{\mathrm{hydro}} = \frac{1}{4} \, \left\vert X^{1/4} - X^{-1/4} \right\vert^{2} \, ,
\end{equation}
and is in agreement with the results obtained in ``ordinary'' momentum independent spacetimes with sudden jumps in the dispersion relation, instead of the scale factor; for example see \cite{Jacobson:2003ac}.
Bogoliubov transformations with infinite total particle production per unit volume are ill-defined, as
the $\mathrm{out}$-vacuum state $\vert {}_{\mathrm{out}} 0 \rangle$ can be written in terms of a \emph{normalizable} linear-combination of $\mathrm{in}$-particle states, only if $\vert \beta_{k} \vert^{2} \to 0$ faster $k^{-d}$ for large $k$. To see an explicit derivation consult, for example, Ref.~\cite{Mukhanov:2007aa}.
We shall now show how due to non-perturbative corrections in our FRW-rainbow spacetimes, this problem does not appear. Here the total number of particles produced in the ultraviolet limit remains finite under any circumstances. 
At large momenta
\begin{equation}
\lim_{k\gg K} n_{k} =   \frac{(X-1)^{2}}{X^{2}} \, \left( \frac{K}{2 \, k} \right)^{4} + O(1/k^6)  \, ,
\end{equation}
and separately we see that for an infinitely large expansion that
\begin{equation}
\label{Eq:sudden.X.Infinity}
\lim_{X\to \infty} n_{k} = \frac{1}{4} \, \frac{(\sqrt{K^{2}+k^{2}}-k)^{2}}{k \, \sqrt{K^{2}+k^{2}}} \, . 
\end{equation}
We can also combine these two limits, and obtain $n_{k} \to [K/(2 \, k)]^{4}$.
The quantum pressure term supresses particle production at high momenta --- effectively because at high momenta the quasi-particles are free and do not ``see'' the evolving spacetime. (The high-momentum limit of the rainbow metric is ``static''.)

As a consequence the number particle production per spacetime volume,
\begin{equation}
N_{\leqslant k} \sim 2^{d-1} \pi \int  d k \; k^{d-5} \sim k^{d-4} \, ,
\end{equation}
for the here relevant cases of $2$ or $3$ spatial dimensions, is finite.

It is also possible, using Eq.~\refb{Eq:sudden.X.Infinity}, to exactly calculate the \emph{total} number of
particles produced during, $c_{0} \to c_{0}/\sqrt{X}$, for $X \to \infty$, an \emph{infinite} expansion:
\begin{equation}
N_{\infty} = \frac{2^{d-3} \, \pi}{d} \, K^{d} \, ,
\end{equation}
which is also finite.

Particle production in our emergent spacetime is a good example as to how corrections --- at the level of effective field theories --- from the underlying microscopic structure can circumvent the problem of an infinite number density. On the other hand, it seems naive to assume the validity of common quantum field theory on all scales. If spacetime is indeed the infrared limit of a more fundamental theory, the presence of ultraviolet deviations seems to be unavoidable.
%
%%%%%%%%%%%%%%%%%%%%%%%%%%%%%%%%%%%%%%%%%%%%%%%%%%%
%
\section[FRW-rainbow spacetimes in $(2+1)$ dimensions]{FRW-rainbow spacetimes in $(2+1)$ dimensions:\protect\\
Controlled parametric excitations in $2$-dimensional condensate}
\label{Sec:Inlation.FRW.Rainbow.Spacetimes}			
%
%%%%%%%%%%%%%%%%%%%%%%%%%%%%%%%%%%%%%%%%%%%%%%%%%%%
In the following we investigate the robustness of the cosmological particle production process against model-dependent ultraviolet corrections in the dispersion relation. The modifications arise from the microscopic substructure, therefore are of non-perturbative nature and not the result of some perturbative loop-calculations. Related work on Hawking radiation from acoustic black holes has been carried out, where the various authors mainly studied the problem in media with subsonic and supersonic dispersion relations \cite{Jacobson:1991sd,Unruh:2005aa}. In addition, similar efforts have been made in the field of ``conventional'' cosmology, by implementing \emph{ad hoc} trans-Planckian modifications to the system, for example see~\cite{Martin:2003rp,Mersini-Houghton:2001aa,Bastero-Gil:2003aa}. This motivated us to extend our main intention --- testing the analogy --- to a detailed study of the deviations (from standard cosmology) in our specific model. We demonstrate this with a particular physically relevant model, a rapidly exponentially growing effective universe, how ``conventional'' and ``modified'' models naturally merge in the infrared limit at early times. Whereas, they show significant deviations at late times, yet for a sufficiently short-time expansion, the resulting particle spectra --- in systems with linear and nonlinear dispersion relations --- are very similar to each other.
%
%+++++++++++++++++++++++++++++++++++++++++++++++++++++++++++++++++
\subsection[$(2+1)$ dimensional FRW-rainbow spacetimes]{$(2+1)$ dimensional FRW-rainbow spacetimes:\protect\\
$2$ dimensional condensates with time-dependent atomic interactions
 \label{Sec:d=2}}
%+++++++++++++++++++++++++++++++++++++++++++++++++++++++++++++++++
In Eq.~\refb{Eq:FRW.Acoustic} we established the concept of a FRW-type rainbow universe to describe the $(2+1)$ or $(3+1)$ dimensional spacetime (that is, for $d=2$ and $d=3$ space dimensions and $1$ time dimension) as experienced by small quantum fluctuations in a $2$ or $3$ dimensional condensate. As pointed out in Sec.~\ref{Sec:FRW.type.geometries}, besides the dimensionality, the scale factor $b(t)$ is also different for $2$ and $3$ dimensions. For example, if we wish to mimic a de Sitter geometry, where the scale factor in the FRW line element, see Eq.~\refb{Eq:FRW.GR}, takes the form 
\begin{equation}
\label{Eq:a.FRW.deSitter.hydrodynamic}
a(\tau) = \exp(H\, \tau) \, ,
\end{equation}
we have to choose the scale factor for the atomic interactions,
\begin{equation}
\label{SwitchedDeSitter}
b(t) = 
\left\{ 
\begin{array}{ll}
\exp(-t/t_{s})  & ~~\mbox{for}~~ d=2,\\
t_{s}/(t+t_{s}) & ~~\mbox{for}~~ d=3. \\ 
\end{array} \right.
\end{equation}
A more detailed treatment of this problem --- that originates in the discrepancy of laboratory $t$ and proper time $\tau$ for the $3$-dimensional case ---  can be found in \cite{Barcelo:2003yk}. In $2$ spatial dimensions --- where  $\alpha(2)=0$ --- the effective line-element~\refb{Eq:FRW.Acoustic} is then given by
\begin{equation}
\label{met2Dconstrho}
\mathrm{d} s_{d=2}^2 
= \left( \frac{n_0}{c_0}\right)^{2} \left[ -c_0^2 \,  d t^2+ b_{k}(t)^{-1} d\mathbf x^2 \right] \, ,
\end{equation}
here laboratory time $t$ and proper time $\tau$ are of the same form, and we obtain the very simple relationship,
\begin{equation}
\label{Eq:FRWsaclefatcor_2d}
a_{k}(t) = b_{k}(t)^{-1/2}=\frac{1}{\sqrt{b(t) + (k/K)^{2}}} \, ,
\end{equation}
between the two scale factors. To connect $t_{s}$, the scale unit in the laboratory, to the Hubble parameter $H$, we need to choose
\begin{equation}
H = \frac{1}{2 \, t_{s}} \, .
\end{equation}
We will soon revisit the Hubble parameter, when we discuss its physical motivation, and how the Hubble parameter applies to our de Sitter-rainbow spacetime.
For practical reasons we shall be mainly interested in $(2+1)$ spacetime dimensions, simply because the numerical simulations we wish to compare the theory with are more easily carried out in 2 space dimensions.

We will always use $a(t)$ to refer to the \emph{FRW scale factor}; if we ever really need the scattering length  we will refer to it as $a_{\mathrm{scatt}}(t)$.

In what physical respect do we really have an ``expanding universe'', given
that the condensate is physically contained in a fixed volume $V$? A
decrease in the scattering length corresponds to a decrease in the
speed of sound propagating in the condensate; therefore any quasi-particle
excitation will propagate with decreasing speed in the condensate as
time passes. To an ``internal'' observer at rest in the effective
spacetime, and communicating by means of acoustic signals, a decrease
of the speed of sound is indistinguishable from an isotropic expansion
of the spatial dimensions. In contrast ``external'' observer made out
of real particles belongs to the microscopic world, and thus is inevitably bound to
laboratory frame and \emph{not} part of the analogy~\cite{Volovik:2003jn}.
%
%+++++++++++++++++++++++++++++++++++++++++++++++++++++++++++++++++
\subsection[Inflation in emerging universe]{Inflation in emerging universe:\protect\\ 
Exponentially decreasing atomic interactions
\label{Sec:Exponential.Universe}}
%+++++++++++++++++++++++++++++++++++++++++++++++++++++++++++++++++
Maybe the most interesting cosmological case to study in our emergent spacetime is the \emph{de Sitter} universe, where the scale factor is given by an exponentially expanding (or contracting) universe.
The concept of inflation was introduced simultaneously around 1981 and 1982 by Guth~\cite{Guth:1981aa}, Linde~\cite{Linde:1990aa}, and Albrecht and Steinhardt~\cite{Albrecht:1982aa} to explain the homogeneity of the temperature observed in our universe, beyond casually disconnected areas. Not long after (\eg, see Guth, Hawking~\cite{Hawking:1982aa}, Bardeen~\cite{PhysRevD.22.1882}, Turner~\cite{Turner:1993aa} and Brandenburger~\cite{Brandenberger:1983aa}) it has been realized that inflation also accounts for the existence of the perturbations in our universe today. These perturbations in the form of a slight deviation from a uniform temperature in our cosmological microwave background (CMB) has been measured in 2001 by Netterfield \emph{et al}.
To date, inflation --- a phase of a rapid de Sitter-like expansion between a predating radiation $a(t)\sim t^{2/3}$ and postdating matter $a(t)\sim t^{1/2}$ dominated area --- seems to be the most plausible explanation for the CMB map. For more details see \cite{Rothman:1993aa,Dodelson:2003aa}.
While we expect deviations from the standard picture in our emergent rainbow spacetime, it seems desirable to hold on to the concept of a thermal spectrum resulting from an exponentially expanding universe. We will come back to this point at the end of this section and show that this hope might not be in vain.

Before we focus on the particle spectrum in our de Sitter-rainbow metric, we would like to introduce some physically important parameters. They will be of great value throughout the remaining part of this article, and are necessary to understand the particle production process for a $2d$ analogue FRW-type universe. Let us start with the obvious question:
%~~~~~~~~~~~~~~~~~~~~~~~~~~~~~~~~~~~~~~~~~~~~~~~~~~~~
\subsubsection{Why rainbow spacetimes?\label{Sec:Why.Rainbow.Spacetime}}
%~~~~~~~~~~~~~~~~~~~~~~~~~~~~~~~~~~~~~~~~~~~~~~~~~~~~
The modified rainbow spacetimes owe their names to their momentum-dependence. This kind of modification can be absorbed into the time and (now also) momentum-dependent scale factor. For a de Sitter like universe (see Eqs~\refb{Eq:a.FRW.deSitter.hydrodynamic} and \refb{SwitchedDeSitter}) in the hydrodynamic limit, we get
\begin{equation}
\label{Eq:Rainbow.scale.factor.deSitter.eikonal}
a_{k}(t) =\frac{1}{ \sqrt{\exp(-2H \, t)+ (k/K)^{2}} }\, ,
\end{equation}
for the modified scale factor~\refb{Eq:a.FRW.deSitter.hydrodynamic}; in $2$ spatial dimensions.
Thus the hydrodynamic,
\begin{equation}
\exp(-2H \, t) \gg \vert k/K \vert^{2} \, ,
\end{equation}
crossover,
\begin{equation}
\exp(-2H \, t) \sim  \vert k/K \vert^{2} \, ,
\end{equation}
and free particle,
\begin{equation}
\exp(-2H \, t) \ll  \vert k/K \vert^{2} \, ,
\end{equation}
limits are a matter of dividing the spectrum into appropriate energy regimes \emph{at a particular time} $t$. 
It is interesting that for early times --- when the interactions between the atoms are strong --- we naturally approach the hydrodynamic case,
\begin{equation}
\lim_{t \to - \infty} a_{k}(t) \to a(t) \, ,
\end{equation} 
in the sense that most modes are phononic, and therefore larger and larger $k$-values are covered by ``conventional'' FRW-type quantum-field-theory. 

Quite the contrary occurs after an infinitely long-lasting expansion, where all modes behave as free particles,
\begin{equation}
\lim_{t \to + \infty} a_{k}(t) \to \vert K/k \vert \, ,
\end{equation} 
and the universe, as seen by a mode with the wavelength $k$, will approach a final finite fixed size.

\begin{figure*}[!htb]
\begin{center}
\mbox{
\subfigure[$\,$ Scale factor without quantum pressure effects ; $t_{s}=1 \times 10^{-5}$. 
\label{Fig:FIG_desitter_ts1e-05_4000000_200000_SF_hydro_-55_25}]{\includegraphics{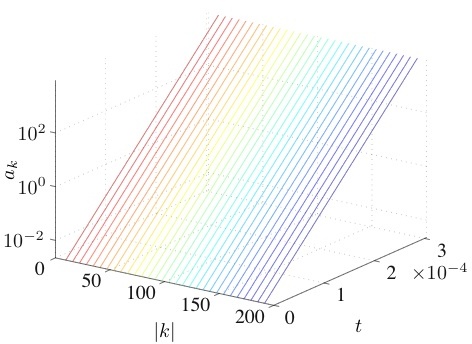}}
\hspace{0mm}
\subfigure[$\,$ Scale factor without quantum pressure effects; $t_{s}=1 \times 10^{-5}$. 
\label{Fig:FIG_desitter_ts1e-05_4000000_200000_SF_hydro_0_0}]{\includegraphics{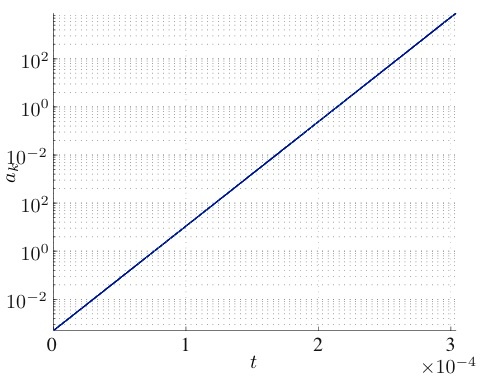}}     
}
\mbox{
\subfigure[$\,$ Scale factor quantum pressure effects ; $t_{s}=1 \times 10^{-5}$. 
\label{Fig:FIG_desitter_ts1e-05_4000000_200000_SF_-55_25}]{\includegraphics{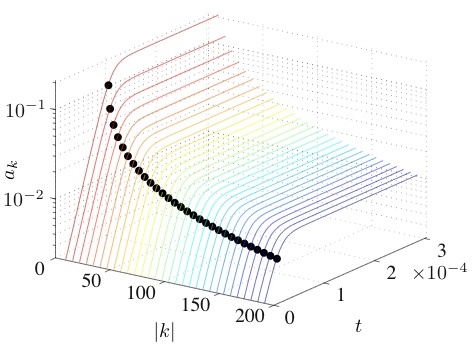}}
\hspace{0mm}
\subfigure[$\,$ Scale factor quantum pressure effects; $t_{s}=1 \times 10^{-5}$. 
\label{Fig:FIG_desitter_ts1e-05_4000000_200000_SF_-55_25}]{\includegraphics{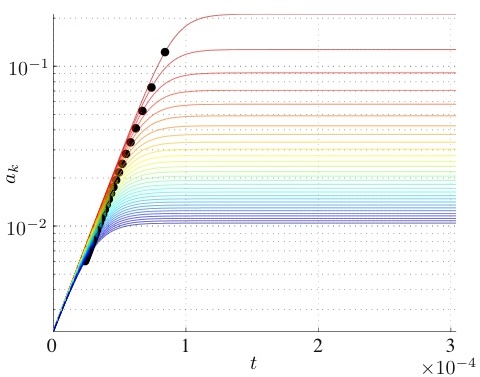}}     
}
\caption[Rainbow scale factor for effective inflation in Bose gas.]{(Colors online only.) In this figure we plot the logarithm of the scale function $a_{k}(t)$ for each $k$-value --- for $k \in [9,191]$ --- in a different color. The different colors encode the energy of the modes: Gradually changing from low-energy\,/\,infrared (dark red) to high-energy\,/\,ultraviolet (dark blue). 
The upper row shows the behavior of the scale function in the hydrodynamic limit.
While the rainbow-scale function --- shown in the lower row --- approaches the hydrodynamic limit for low-energy modes, the ultraviolet modes show strong deviations. Note, that in the infinite past all modes are phononic, and therefore $a_{k}(t)\to a(t)$. The black dots indicate the time-dependent crossover (phononic to trans-phononic) in every quantum mode. Parameters are $C_{NL}(\bar{t}=0)=2 \times 10^{5}$, $N_{0}=10^{7}$ and $X=4 \times 10^{6}$.}
\label{Fig:deSitter.Rainbow.ScaleFactor}
\end{center}
\end{figure*}

Due to this fundamental difference between our analogue model and the ``theory'' we wish to mimic, we know already that there will only be a finite time-period --- its length depends on $K$, and therefore on the tunable initial interaction strength $U(0)=U_{0}$ --- beyond which the analogy breaks down. Note that the particle production process must naturally come to an end, when the expansion rate slows down to zero. That the effective expansion rate of the FRW-rainbow universe approaches zero will be shown next, but before we would like to answer the question posed in the headline of this subsection, with a simple illustration for the de Sitter-rainbow scale factor in Fig.~\ref{Fig:deSitter.Rainbow.ScaleFactor}. There we plot the emergent rainbow scale factor $a_{k}(t)$ for each $k$ mode using different colors --- gradually changing from dark red for infrared modes to dark blue for ultraviolet modes. The resulting color-spectrum is reminiscent of on the color spectrum obtained from real rainbows.

%
%_figure__figure__figure__figure__figure__figure__figure__figure__figure__figure__figure__figure_
%_figure__figure__figure__figure__figure__figure__figure__figure__figure__figure__figure__figure_
%

%~~~~~~~~~~~~~~~~~~~~~~~~~~~~~~~~~~~~~~~~~~~~~~~~~~~~
\subsubsection{What about the Hubble parameter?\label{Sec:What.About.Hubble.Paramter}}
%~~~~~~~~~~~~~~~~~~~~~~~~~~~~~~~~~~~~~~~~~~~~~~~~~~~~
The de Sitter universe is special in the sense that its Hubble parameter, the ``rate of expansion''
\begin{equation}
H := \frac{\dot a(t)}{a(t)} \, ,
\end{equation}
is constant. The de Sitter universe is a solution of the Einstein equations with a positive cosmological constant, $\Lambda$; \ie, $H \propto \sqrt{\Lambda}$. If the acceleration of our universe can be put down to this cosmological constant, the universe will expand forever --- and further dilute the matter and radiation distribution in our universe --- until it approaches the de Sitter spacetime~\cite{Cattoen:2007aa,Cattoen:2007sk}.

What is the situation in our emergent de Sitter universe? The rate of size change in the emergent de Sitter universe using~\ref{Eq:Rainbow.scale.factor.deSitter.eikonal} is given by,
\begin{equation}
\label{Eq:Hubble.general.2d.deSitter}
H_{k} = H \; \frac{\exp(-2H \, t)}{\exp(-2H \, t)+ (k/K)^{2}}  \, , 
\end{equation}
a momentum-dependent \emph{rainbow Hubble parameter}. At early times, or for phononic modes, when $a_{k}(t) \to a(t)$, the rainbow Hubble parameter,
\begin{equation}
\lim_{t \to - \infty} H_{k}(t) \to H \, ,
\end{equation} 
reduces to the conventional Hubble parameter, while for late times,
\begin{equation}
\lim_{t \to + \infty} H_{k}(t) \to H \, K^{2} \, (\exp(-H\,t)/ k)^{2} \to 0 \, .
\end{equation} 
The universe gradually --- mode by mode --- stops expanding as modes leave the phononic regime.

Therefore, in our particular scenario the long-time-kinematics of the particle production process is determined by the non-perturbative corrections from the substructure, and not by the emergent spacetime picture. Any speculations with respect to an everlasting expanding universe are, (within our model), in vain.
%~~~~~~~~~~~~~~~~~~~~~~~~~~~~~~~~~~~~~~~~~~~~~~~~~~~~
\subsubsection{Characteristic value for quantum process?\label{Sec:Characteristic.Value.Quantum.Process}}
%~~~~~~~~~~~~~~~~~~~~~~~~~~~~~~~~~~~~~~~~~~~~~~~~~~~~
To get a grasp on the cosmological particle production process, these two parameters (the effective scale function $a_{k}(t)$, and the effective Hubble parameter $H_{k}(t)$) are not quite enough. They are a good measure to describe the kinematics of the emergent gravitational field, but one also needs to know how the microscopic corrections affect the energy of the modes. Only the ratio between the mode frequency,
\begin{equation}
\label{Eq:Omega.FRW.2d.deSitter.eikonal}
\omega_{k}(t) = \omega_{0} \;  \sqrt{\exp(-2H \, t)+  (k/K)^{2}}   \, ,
\end{equation}
and the Hubble frequency can tell us whether the mode will be disturbed by the classical background or not. This result has been established for quantum field theory in conventional de Sitter spacetimes, and we review this point in the following section.

For now, we simply transfer the qualitative description known from conventional cosmological particle production, to our rainbow spacetimes. In this spirit we define \emph{the ratio} between the modified dispersion relation and the effective Hubble parameter as follows:
\begin{equation}
\label{Eq:Ratio.deSitter.eikonal}
\mathcal R_{k}(t) =\frac{\omega_{k}(t)}{H_{k}(t)}= \frac{\omega_{0}}{H}  \; \frac{(\exp(-2H \, t) +  (k/K)^{2} )^{3/2} }{\exp(-2H \, t)} \, .
\end{equation}
Within the hydrodynamic limit this ratio simplifies to,
\begin{equation}
\label{Eq:Ratio.deSitter.hydro}
R_{k}(t) = \frac{\omega_{0}}{H}  \; \exp(-H \, t)  \, ,
\end{equation}
a monotonically decreasing function with time.
Again, we calculate limits for very early, 
\begin{equation}
\lim_{t \to - \infty} \mathcal R_{k}(t) \to \frac{\omega_{0}}{H}  \; \exp( - H \, t) = R_{k}(t) \, ,
\end{equation} 
and very late times, 
\begin{equation}
\lim_{t \to + \infty} \mathcal R_{k}(t) \to \frac{\omega_{0}}{H} \;  (k/K)^{3}  \; \exp( + 2H \, t)  \, ,
\end{equation} 
to see once again, that for late times\,/\,trans-phononic modes the deviations in our emergent spacetime play 
an important rule. \\

We will show that this ratio is sufficient to understand, and therefore predict qualitatively, the particle production process in our analogue spacetimes.  
%+++++++++++++++++++++++++++++++++++++++++++++++++++++++++++++++++
\subsection[Quantum field theory and rainbow inflation]{Quantum field theory and rainbow inflation:\protect\\ 
Excitations from an exponentially changing interaction strength
\label{Sec:Quantum.Field.Theory.Rainbow.Inflation}}
%+++++++++++++++++++++++++++++++++++++++++++++++++++++++++++++++++
The equation of motion in the presence of time-dependent atomic interactions is given in Eq.~\refb{Eq:Equation.Motion.Chi}. In the de Sitter rainbow universe, with the scale factor \refb{Eq:Rainbow.scale.factor.deSitter.eikonal}, and the dispersion relation \refb{Eq:Omega.FRW.2d.deSitter.eikonal}, the harmonic oscillator frequency \refb{Eq:Harmonic.Oscillator.Frequency} simplifies to
\begin{equation}
\label{Eq:Harmonic.Oscillator.Frequency.Eikanol.General}
\Omega_{k}(t)^{2} = \omega_{0}^{2} \, b_{k}(t) - H^{2} + \Delta_{k}(t)^{2} \, ,
\end{equation}
where the effective (or rainbow) scale factor is given by
\begin{equation}
\label{Eq:Rainbow.Scale.Factor.deSitter}
b_{k}(t) = \exp[-2Ht] + (k/K)^{2} \, ,
\end{equation}
and the last term is given as
\begin{equation}
\Delta_{k}(t)^{2} = H^{2} \, \, \frac{[4 \, \exp(-2Ht) +(k/K)^{2}] \, (k/K)^{2} }{[\exp(-2Ht) +(k/K)^{2}]^{2}} ,
\end{equation}
which can be neglected in hydrodynamic approximation. Therefore the equation of motion for our rainbow de Sitter spacetime is given by a rather complicated differential equation, where common techniques for solving second-order differential equations fail. 

However, this is not the end of the story, since at early times, or for phononic modes we get,
\begin{equation}
\label{Eq:Harmonic.Oscillator.deSitter.Hydrodynamic}
\lim_{t \to  -\infty} \;
\left\{ \begin{array}{rcl}
b_{k}(t) & \to & b(t) \\ &\,& \\
\Delta_{k}(t)& \to & 0
\end{array} \right\} \quad
\Omega_{k}(t)^{2} \to \omega_{0}^{2} \, b(t) - H^{2} \, ,
\end{equation}
where 
\begin{equation}
\label{Eq:Atomic.Scale.Factor.deSitter.H}
b(t)=\exp(-2Ht) \,  ,
\end{equation}
and within this limit the equation of motion simplifies to a differential equation solvable by Bessel functions. Within this regime, our emergent spacetimes can be regarded as a good toy model for quantum field theory in conventional de Sitter spacetimes. However, we need to apply special boundary conditions, that is to set limits on the time-duration and the $k$-range. Only then will the cosmological particle production process be robust against the model-specific non-perturbative corrections that are present in a spacetime emerging from a gas of Bosons. 

As the expansion continues, the more modes will cross-over from phononic to trans-phononic, and finally for an infinitely long-lasting expansion, we get
\begin{equation}
\label{Eq:Harmonic.Oscillator.Frequency.Early.Times}
\lim_{t \to  +\infty} \;
\left\{ \begin{array}{rcl}
b_{k}(t) & \to & (k/K)^{2} \\ &\,& \\
\Delta_{k}(t)& \to & H
\end{array} \right\} \quad
\Omega_{k}(t)^{2} \to \omega_{0}^{2} \, (k/K)^{2} \, ,
\end{equation}
and therefore are left with no phononic regime whatsoever. That is, in the infinite future all excitations of the fluctuations of the system are free-particle like. In this regime the quantum fluctuations are decoupled from the collective behavior, the emergent spacetime picture, and thus are static.

Therefore, without any further considerations, we deduce that in the infinite future we will always end up with well-defined $\mathrm{out}$-states, that represent the vacuum state of instantaneous Hamiltonian diagonalization, see Sec.~\ref{Sec:Instantaneous.Hamiltonian.Diagonalization}. The normalized mode functions in the infinite future are given by Eq.~\refb{Eq:Mode.Functions.Minkowski.out}, where $\omega_{k}^{\mathrm{out}}=\omega_{0} \vert k \vert / K$. \\

We will subsequently revisit this point in Sec.~\ref{Sec:Qualitative.Behavior.Quantum.Fluctuations}, but for now our concern will be to focus on a regime will be to focus on a regime --- the hydrodynamic limit --- where the problem of particle production is tractable, and there is a straightforward description of the physics.
%
%~~~~~~~~~~~~~~~~~~~~~~~~~~~~~~~~~~~~~~~~~~~~~~~~~~~~
\subsubsection{Toy-model for conventional inflation\label{Sec:Toy.Model.Conventional.Inflation}}
%~~~~~~~~~~~~~~~~~~~~~~~~~~~~~~~~~~~~~~~~~~~~~~~~~~~~
%
%_figure__figure__figure__figure__figure__figure__figure__figure__figure__figure__figure__figure_
\begin{figure}[htb]
 \begin{center}
 \includegraphics{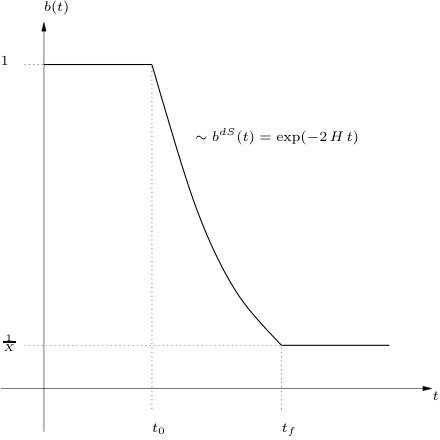}
 \caption[Fnite de Sitter universe.]  {\label{deSitterlikeconcept}Time-dependence for atom-atom interactions to mimic a de~Sitter universe in a $2$ dimensional Bose--Einstein condensate.}
 \end{center} 
\index{de~Sitter}      
\end{figure}
%_figure__figure__figure__figure__figure__figure__figure__figure__figure__figure__figure__figure_
%
%
%_figure__figure__figure__figure__figure__figure__figure__figure__figure__figure__figure__figure_
\begin{figure*}[!htb]
\begin{center}
\mbox{
\subfigure[$\,$ $t=0.25 \times t_{f}$ ; $t_{s}=5 \times 10^{-5}$. \label{Fig:timeslice_dsa_t_0.25tf_ts5e-5}]{\includegraphics[width=0.45\textwidth]{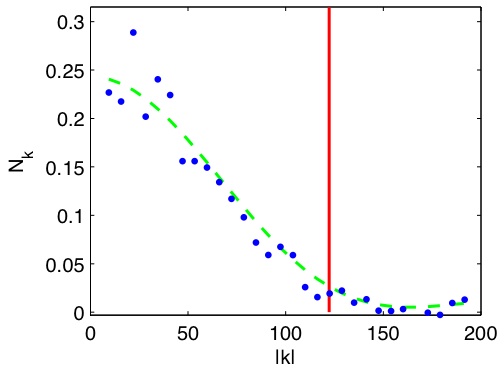}}
\hspace{3mm}
\subfigure[$\,$ $t=0.50 \times t_{f}$ ; $t_{s}=5 \times 10^{-5}$. \label{Fig:timeslice_dsa_t_0.50tf_ts5e-5}]{\includegraphics[width=0.45\textwidth]{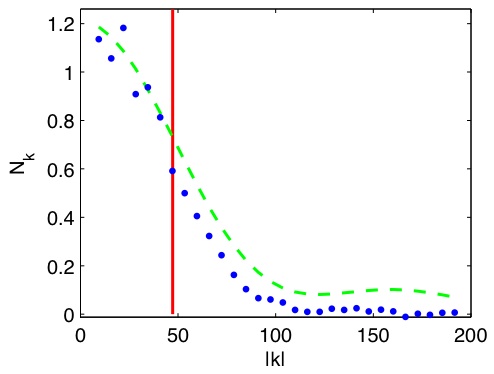}}
}
\vspace{3mm}
\mbox{
\subfigure[$\,$ $t=0.75 \times t_{f}$ ; $t_{s}=5 \times 10^{-5}$. \label{Fig:timeslice_dsa_t_0.75tf_ts5e-5}]{\includegraphics[width=0.45\textwidth]{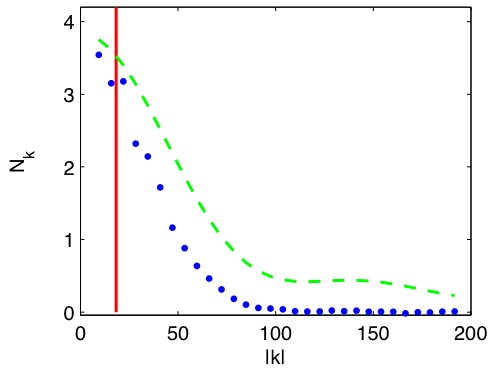}}
\hspace{3mm}
\subfigure[$\,$ $t=1.00 \times t_{f}$ ; $t_{s}=5 \times 10^{-5}$. \label{Fig:timeslice_dsa_t_1.00tf_ts5e-5}]{\includegraphics[width=0.45\textwidth]{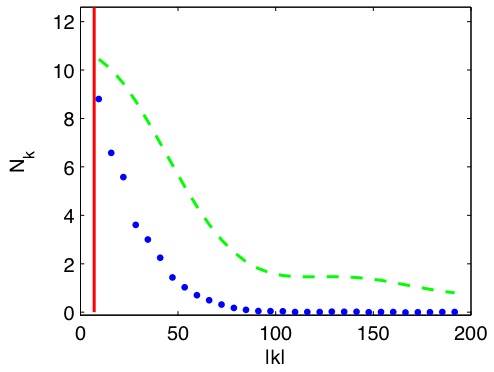}}     }
\caption[Quasi-particle production during rainbow inflation --- I.]{The four figures show time-slices of the  ``quasi-particle'' production $N_{k}$ in a de Sitter like spacetime, for $t_{s}=5\times 10^{-5}$. The blue dots represent the actual data obtained from the simulations for a realistic Bose gas with time-dependent interaction-strength~\cite{Jain:2006ki}. The dashed green line shows the theoretical results for a finite de Sitter calculation obtained in the hydrodynamic limit, as presented in Sec.~\ref{Sec:Toy.Model.Conventional.Inflation}. The vertical (red) line indicates the crossover between the two relevant regimes, the excitations being phononic on the left and trans-phononic\,/\,free-particle like on the right.}\label{Fig:timeslice_ts5e-5}.
\end{center}
\end{figure*}
%_figure__figure__figure__figure__figure__figure__figure__figure__figure__figure__figure__figure_
%

To apply the method of instantaneous Hamiltonian diagonalization to particle production from a finite de Sitter phase, we assume a continuous function for all $t$, but at the times $t_{0}$ and $t_{f}$ the function is not differentiable:
In the hydrodynamic limit 
\begin{equation}
b_{k}(t) \to b(t)
\end{equation}
we assume
\begin{equation}
b(t) = 1 +[b^{\mathrm{dS}}(t)-1]  \, \Theta_{\mathrm{HS}}(t-t_{0}) - [b^{\mathrm{dS}}(t)-1/X] \, \Theta_{\mathrm{HS}}(t-t_{f});
\end{equation}
compare with Fig.~\refb{deSitterlikeconcept}.

Here $b^{\mathrm{dS}}(t)=\exp[-2H(t-t_{0})]$, such that $b(t_{0})=1$, and $b(t_{f})=1/X$. 
In addition, we define $\omega_{k}(t_{0})=\omega_{k}^{\mathrm{in}}$, and $\omega_{k}(t_{f})=\omega_{k}^{\mathrm{out}}$, such that for $t<t_{0}$ we choose the $\mathrm{in}$-mode functions, given in Eq.~\refb{Eq:Mode.Functions.Minkowski.in}, and similarly for $t>t_{f}$ we choose the $\mathrm{out}$-mode functions, given in Eq.~\refb{Eq:Mode.Functions.Minkowski.out}, for the initial and final flat regions.

Note that for any laboratory set-up the time-scales are finite, but given that this calculation is only valid for modes that are of phononic nature both before \emph{and} after the expansion, we are not free to choose the expansion time arbitrarily. \\

The Bogoliubov transformation is slightly more difficult here, where two single-step processes have to be carried out:
\begin{equation}
\vec{v}_{k}(t) = M(t_{f}) \cdot \vec{v}_{k}^{dS}(t) = M(t_{f}) \cdot M(t_{0}) \cdot \vec{u}_{k}(t) \, . 
\end{equation}

For the mode functions during the de Sitter phase, which are $v_{k}^{\mathrm{dS}}$ and ($v_{k}^{\mathrm{dS}})^{*}$ for $t \in [t_{0},t_{f}]$,  we solve the harmonic oscillator equation with the time-dependent frequency given in Eq.~\refb{Eq:Harmonic.Oscillator.Frequency.Early.Times}. The solution is a linear combination of first order Hankel functions of the first $\mathrm H_{1}^{(1)}$ and second $\mathrm H_{1}^{(2)}$ kind. The normalized mode functions --- using the Wronskian condition~\refb{Eq:Wronskian} --- are given by
\begin{eqnarray}
\label{Eq:Mode.Functions.deSitter.v}
v_{k}^{\mathrm{dS}} &=& \sqrt{\frac{\pi}{2H}} \, \mathrm H_{1}^{(1)} (R_{k}(t)) \, ,  \\
\label{Eq:Mode.Functions.deSitter.vc}
(v_{k}^{\mathrm{dS}})^{*} &=& \sqrt{\frac{\pi}{2H}} \,\mathrm H_{1}^{(2)} (R_{k}(t))  \, .
 \end{eqnarray}
Note that in this representation the mode functions are indeed a set of complex conjugate functions, since $(\mathrm H^{(1)})^{*}=\mathrm H^{(2)}$; see for example \cite{Abramowitz:1964aa}. The argument of the mode functions is $R_{k}(t)$, the ratio between the mode frequency and the Hubble frequency, see Eq.~\refb{Eq:Ratio.deSitter.hydro}. These mode functions are only valid within the hydrodynamic limit, \ie~for modes that satisfy $\vert k/K \vert \ll \exp(-H \, t)$.
 
To calculate the Bogoliubov coefficients at each step we again apply Eqs.~\refb{Eq.Alpha.General} and \refb{Eq.Beta.General}. 
Below we will briefly outline the necessary steps to calculate the final Bogoliubov coefficients for the whole process.
To begin with we write down the single-step
\begin{equation}
U(t) =  g(t) \, \Theta_{\mathrm{HS}}(t-t_{0}) +  h(t) \, \Theta_{\mathrm{HS}}(t_{0}-t)\, .
\end{equation}
Bogoliubov coefficients, 
\begin{eqnarray}
\label{Eq.Alpha.General.Continous.Normalized}
\alpha_{k}(t) &=&\frac{1}{4i} \left( 2\, W[ u_{k}, v_{k}^{*}]  + v_{k}^{*} u_{k}  (\dot h/h - \dot g/g)  \right), \quad \\
\label{Eq.Beta.General.Continous.Normalized}
\beta_{k}(t) &=&\frac{1}{4i} \left(  2\, W[ v_{k}^{*},u_{k}^{*}]  - v_{k}^{*} u_{k}^{*}  (\dot h/h - \dot g/g)  \right) , \quad
\end{eqnarray}
appropriate to a continuous, but not continuously differentiable scale function at the time $t$.  
At the first step, where we map from plane waves onto Bessel functions, and for our specific scale factor before the step, $g(t)=1$, and after the step, $h(t)=\exp[-2H(t-t_{0})]$, we get
\begin{eqnarray}
\label{Eq.Alpha.General.Continous.Normalized.deSitter.temp1}
\alpha_{k}(t) &=&\frac{1}{2i} \left\{  W[ v_{k}, (v_{k}^{\mathrm{dS}})^{*}]  -H \, v_{k}^{*} v_{k}^{\mathrm{dS}}  \right\}  \, , \\
\label{Eq.Beta.General.Continous.Normalized.deSitter.temp1}
\beta_{k}(t) &=&\frac{1}{2i} \left\{  W[ v_{k}^{*},(v_{k}^{\mathrm{dS}})^{*}]  + H\, v_{k}^{*} (v_{k}^{\mathrm{dS}})^{*}  \right\}  \, . 
\end{eqnarray}
The final coefficients for our specific mode functions are given by
\begin{equation}
\label{Eq.Alpha.deSitter.1step}
\alpha_{k}^{\mathrm{in}}(t_{0}) =
\frac{ \sqrt{\pi \, R_{k}^{\mathrm{in}}}}{-i\, 2}  \e^{i \, \omega^{\mathrm{in}} \,t_{0}}  \left\{\mathrm H_{0}^{(1)}(R_{k}^{\mathrm{in}}) + i \, \mathrm H_{1}^{(1)} (R_{k}^{\mathrm{in}})  \right\} \, ,
\end{equation}
and
\begin{equation}
\label{Eq.Beta.deSitter.1step}
\beta_{k}^{\mathrm{out}}(t_{0}) =\frac{ \sqrt{\pi \, R_{k}^{\mathrm{in}}}}{i\, 2} \e^{i \, \omega^{\mathrm{in}} \,t_{0}}    \left\{ \mathrm H_{0}^{(2)}(R_{k}^{\mathrm{in}}) - i \, \mathrm H_{1}^{(2)} (R_{k}^{\mathrm{in}})  \right\} \, .
\end{equation}
In the last two equations we have used $R_{k}^{\mathrm{in}}=R_{k}(t_{0})$ and $\omega_{k}^{\mathrm{in}}=\omega_{k}(t_{0})$.
Instead of calculating the Bogoliubov coefficients for the second step explicitly, we suggest a little shortcut. The transformation at the first step was a mapping from plane waves, represented as $v_{k}$ and $v_{k}^{*}$, onto the de Sitter mode functions $v_{k}^{\mathrm{dS}}$ and $(v_{k}^{\mathrm{dS}})^{*}$: $\vec{v}_{k} = M(t) \vec{v}_{k}^{\mathrm{dS}}$. At the second step the calculation is exactly the opposite, a mapping from the de Sitter mode functions onto plane waves: $\vec{v}_{k}^{\mathrm{dS}} = M(t)^{-1} \vec{v}_{k}$. The inverse of the transition matrix displayed in Eq.~\refb{Eq:M.Transformation.Matrix} is
\begin{equation}
\label{Eq:M.Transformation.Matrix.Inverse}
M^{-1} =
\left(
\begin{array}{cc}
\alpha_{k} & -\beta_{k}^{*} \\ -\beta_{k} & \alpha_{k}^{*}
\end{array}
\right) \, ,
\end{equation}
where we have employed $\vert \alpha_{k}\vert^{2} -\vert \beta_{k}\vert^{2} =1$, since we map between normalized mode functions. Altogether we can formally derive the Bogoliubov coefficients at the second step as:
\begin{eqnarray}
\alpha_{k}^{\mathrm{out}}(t_{f})  &:& \quad 
\alpha_{k}^{\mathrm{in}}(t_{0})
\xrightarrow[ \mathrm{in} \to \mathrm{out}]{t_{0} \to t_{f}}
[\alpha_{k}^{\mathrm{out}}(t_{f})]^{*} \, \\
\beta_{k}^{\mathrm{out}}(t_{f})  &:& \quad 
\beta_{k}^{\mathrm{in}}(t_{0})
\xrightarrow[ \mathrm{in} \to \mathrm{out}]{t_{0} \to t_{f}}
-[\beta_{k}^{\mathrm{out}}(t_{f})] \, ,
\end{eqnarray}
Thus we can write down the Bogoliubov coefficients at the second step, at $t=t_{f}$, without any further calculation being required, as
\begin{equation}
\label{Eq.Alpha.deSitter.2step}
\alpha_{k}^{\mathrm{out}}(t_{f}) =
\frac{ \sqrt{\pi R_{k}^{\mathrm{out}}}}{2\, i}  \e^{-i  \omega^{\mathrm{out}} t_{f}}  \left\{ \mathrm H_{0}^{(2)}(R_{k}^{\mathrm{out}}) - i \, \mathrm H_{1}^{(2)} (R_{k}^{\mathrm{out}})  \right\} ,
\end{equation}
and
\begin{equation}
\label{Eq.Beta.deSitter.2step}
\beta_{k}^{\mathrm{out}}(t_{f}) =\frac{ \sqrt{\pi R_{k}^{\mathrm{out}}}}{-2\, i} \e^{-i  \omega^{\mathrm{out}} t_{f}}    \left\{\mathrm H_{0}^{(1)}(R_{k}^{\mathrm{out}}) + i \, \mathrm H_{1}^{(1)} (R_{k}^{\mathrm{out}})  \right\}  .
\end{equation}
Here $R_{k}^{\mathrm{out}}=R_{k}(t_{f})$ and $\omega_{k}^{\mathrm{out}}=\omega_{k}(t_{f})$.

The overall particle production, at later times $t > t_{f}$ can be obtained by simple matrix multiplication, since
\begin{eqnarray}
\label{Eq:M.Transformation.Matrix.deSitter.Overall}
&M_{t>t_{f}} = 
M(t_{0}) \, M(t_{f}) =
\left(
\begin{array}{cc}
\alpha_{k}^{\mathrm{final}} &
\beta_{k}^{\mathrm{final}}\\
(\beta_{k}^{\mathrm{final}})^{*} &
(\alpha_{k}^{\mathrm{final}})^{*} 
\end{array}
\right)  . \; \; \;
\end{eqnarray}
and therefore we get for the final Bogoliubov coefficients
\begin{eqnarray}
\alpha_{k}^{\mathrm{final}} &=&
\alpha_{k}^{\mathrm{in}} \alpha_{k}^{\mathrm{out}}+\beta_{k}^{\mathrm{in}} (\beta_{k}^{\mathrm{out}})^{*} \, ,
\\
\beta_{k}^{\mathrm{final}} &=&
\alpha_{k}^{\mathrm{in}} \beta_{k}^{\mathrm{out}}+\beta_{k}^{\mathrm{in}} (\alpha_{k}^{\mathrm{out}})^{*} \, .
\end{eqnarray}
Note that these coefficients can in some sense be considered as time-dependent Bogoliubov coefficients, where 
\begin{eqnarray}
\alpha_{k}^{\mathrm{final}(t')} &=&
\alpha_{k}^{\mathrm{in}} \alpha_{k}^{\mathrm{out}(t')}+\beta_{k}^{\mathrm{in}} (\beta_{k}^{\mathrm{out}(t')})^{*} \, ,
\\
\beta_{k}^{\mathrm{final}(t')} &=&
\alpha_{k}^{\mathrm{in}} \beta_{k}^{\mathrm{out}(t')}+\beta_{k}^{\mathrm{in}} (\alpha_{k}^{\mathrm{out}(t')})^{*} \, ,
\end{eqnarray}
where we project at any instant of time onto a plane wave basis at a particular $t'$ with the eigen-frequency $\omega_{k}(t')$. But, as we pointed out in Sec.~\ref{Sec:Meaning.Particles}, as long as the expansion is continuing, the corresponding $\mathrm{out}$-frequency modes do not represent a physically meaningful vacuum state. Nevertheless, for $t>t_{f}$, we stop the expansion and force the Hamiltonian of the system to become static, so that we are able to associate $n_{k}^{\mathrm{final}}=\vert \beta_{k}^{\mathrm{final}}\vert^{2}$ with the ``real'' mode occupation of a mode $k$ per unit volume. \\

A lengthy but straightforward calculation now yields the mode occupation number after stopping the de Sitter like expanding phase. The number of particles produced depends only on the initial and final values of the frequency ratio $R_k(t)$, respectively $R_{k}^{\mathrm{in}}$ and $R_{k}^{\mathrm{out}}$, as:
\begin{eqnarray}
\nonumber
n_{k}^{\mathrm{dS}} &=& \frac{\pi^{2}}{64}R_{k}^{\mathrm{in}} R_{k}^{\mathrm{out}} \, \times \Big\vert \\
\nonumber
&-&[ \mathrm H^{(1)}_{1}(R_{k}^{\mathrm{in}}) \, \mathrm H^{(2)}_{1}(R_{k}^{\mathrm{out}}) -\mathrm H^{(2)}_{1}(R_{k}^{\mathrm{in}}) \, \mathrm  H^{(1)}_{1}(R^{\mathrm{out}}) ]^2 \\
\nonumber
&-& [ \mathrm H^{(1)}_{0}(R_{k}^{\mathrm{in}})\, \mathrm H^{(2)}_{0}(R_{k}^{\mathrm{out}})-\mathrm H^{(2)}_{0}(R_{k}^{\mathrm{in}}) \, \mathrm H^{(1)}_{0}(R_{k}^{\mathrm{out}}) ]^2 \\
\nonumber
&-&[ \mathrm H^{(1)}_{0}(R_{k}^{\mathrm{in}}) \, \mathrm H^{(2)}_{1}(R_{k}^{\mathrm{out}})- \mathrm H^{(2)}_{1}(R_{k}^{\mathrm{in}}) \, \mathrm H^{(1)}_{0}(R_{k}^{\mathrm{out}}) ]^2 \\
\nonumber
&-&[ \mathrm H^{(2)}_{0}(R_{k}^{\mathrm{in}}) \, \mathrm H^{(1)}_{1}(R_{k}^{\mathrm{out}})- \mathrm H^{(1)}_{1}(R_{k}^{\mathrm{in}}) \, \mathrm H^{(2)}_{0}(R_{k}^{\mathrm{out}})]^2 \\
\nonumber
&+& 2 \, \mathrm H^{(2)}_{1}(R_{k}^{\mathrm{out}}) \, \mathrm H^{(2)}_{0}(R_{k}^{\mathrm{in}}) \times \\
\nonumber 
&& \;
[\mathrm H^{(1)}_{0}(R_{k}^{\mathrm{in}}) \, \mathrm H^{(1)}_{1}(R_{k}^{\mathrm{out}}) - \mathrm H^{(1)}_{1}(R_{k}^{\mathrm{in}}) \, \mathrm H^{(1)}_{0}(R_{k}^{\mathrm{out}})] \\
\nonumber
&+& 2 \, \mathrm H^{(2)}_{1}(R_{k}^{\mathrm{in}}) \, \mathrm H^{(2)}_{0}(R_{k}^{\mathrm{out}}) \\
\nonumber 
&& \;
[ \mathrm H^{(1)}_{1}(R_{k}^{\mathrm{in}}) \, \mathrm H^{(1)}_{0}(R_{k}^{\mathrm{out}})- \mathrm H^{(1)}_{1}(R_{k}^{\mathrm{out}}) \, \mathrm H^{(1)}_{0}(R_{k}^{\mathrm{in}}) ] \Big\vert \, . \\ 
\label{Eq:Mode.Occupation.Finite.deSitter}
&& 
\end{eqnarray}

At first sight this formula seems to be rather complicated, but there are two relatively simple and feasible consistency checks. 

For example, we expect no particle production in the limit where the universe has not changed at all. It can easily been seen that for
\begin{equation}
 \lim_{R_{k}^{\mathrm{out}}\to R_{k}^{\mathrm{in}}} n_{k}^{\mathrm{dS}} \to 0 \, ,
\end{equation}
the occupation number indeed goes to zero.

Also interesting are limits resulting form an infinitely slow (adiabatic), that is $H\to 0$, and an infinitely fast (sudden) $H\to \infty$ expansion. In either case we replace the ratios with $R_{k}^{\mathrm{in}}=\omega_{k}^{\mathrm{in}}/H$ and $R_{k}^{\mathrm{out}}=\omega_{k}^{\mathrm{out}}/H$, hold $\omega_{k}^{\mathrm{out}}$ and $\omega_{k}^{\mathrm{out}}$ fixed, while we allow the Hubble parameter $H$ to vary. Such a parameter choice requires $H \, t_{exp}= \mathrm{constant}$, and therefore we expect the expansion time $t_{exp}=t_{f}-t_{0}$ to behave inverse proportional to $H$. The leading order, obtained from a Taylor-series-expansion around $H=\infty$,
\begin{equation}
 \lim_{H\to \infty} n_{k}^{\mathrm{dS}} \to  \frac{1}{4} \left\vert \frac{(\omega_{k}^{\mathrm{in}}-\omega_{k}^{\mathrm{out}})^{2}}{\omega_{k}^{\mathrm{in}}\omega_{k}^{\mathrm{out}}} \right\vert + O(1/H^{2}) \, ,
\end{equation}
is as expected --- since $(t_{f}-t_{0})\to 0$ --- in agreement with the result from the sudden calculation within the hydrodynamic limit, see Eq.~\refb{Eq:nk.sudden}. In contrast, a Taylor-series expansion around $H=0$ yields
\begin{equation}
 \lim_{H\to 0} n_{k}^{\mathrm{dS}} \to 0 \, .
\end{equation}
As expected, within the limit of infinitely slow expansion, we produce zero particles, and we recover adiabatic invariance.
Furthermore if we consider the asymptotic expansion where $1 \ll  R_{k}^{\mathrm{in}} \ll \sqrt{X}$, and employ the asymptotic limits of the Hankel functions~\cite{Abramowitz:1964aa}, we get at linear order:
\begin{equation}
n_{k} \to \frac{1}{2 \pi \, R_{k}^{\mathrm{out}}} = \frac{H}{2 \pi \, \omega_{k}^{\mathrm{out}}} \, .
\end{equation}
Of course, we cannot rely on our calculation in the case of an infinitely long-lasting expansion, since it is based on the validity of the hydrodynamic limit, which is completely inappropriate for $t \gg t_{s}$. However, in our previous numerical simulation of a ``realistic'' Bose gas~\cite{Jain:2006ki} we compared our theoretically obtained result with short-time expansion scenarios, and were able to match them to the phononic part of the particle production spectrum, see Figs.~\ref{Fig:timeslice_ts5e-5}.
(In~\cite{Jain:2006ki} we applied the truncated Wigner method to a Bose--Einstein condensate, where we selected a nonlinearity and atom number such that most of the modes were phononic at the start of the simulation, and that for our choice of parameters the effects of back-reaction and mode-mixing were minimal.)

In Figure~\ref{Fig:timeslice_ts5e-5}, for $t_{s}=5\times 10^{-5}$ we see an excellent agreement between our theoretical predictions (dashed green line), and the numerical data (blue dots) for the ``quasi-particle'' production in the phononic regime (left of the vertical red line). As shown in Fig.~\ref{Fig:timeslice_ts5e-5}, at the end of the expansion we see that almost all excitations are trans-phononic, hence the analogy \emph{eventually breaks down}.
But, to answer the question stated at the beginning of this section, where we asked about the applicability of our specific model to mimic cosmological particle production: Yes, the BEC can be used as an analogue model for cosmological particle production \emph{within certain limits}, as our numerical results clearly confirm.

The calculations we have presented are somewhat tedious, but they are more than worth the effort since there is a fundamental lesson to be learnt from Eq.~\refb{Eq:Mode.Occupation.Finite.deSitter}: It establishes our previous intuition (see Sec.~\ref{Sec:Characteristic.Value.Quantum.Process}), that the characteristic value controlling the particle production is the ratio, $R_{k}(t)$ between the mode frequency, and the Hubble frequency. Comparison with Eq.~\refb{Eq:Mode.Occupation.Finite.deSitter} shows explicitly that the final quantity of particle production --- for truly phononic modes before, during, and after the expansion --- only depends on the initial $R_{k}^{\mathrm{in}}$ and final $R_{k}^{\mathrm{out}}$ frequency ratio. This motivated us to extend the role of the frequency ratio beyond the hydrodynamic limit, and compare our predictions with the data obtained from our simulations, where the non-perturbative corrections are included; see \cite{Jain:2006ki} for details of the numerics.

%~~~~~~~~~~~~~~~~~~~~~~~~~~~~~~~~~~~~~~~~~~~~~~~~~~~~
\subsubsection{Qualitative behavior of quantum fluctuations\label{Sec:Qualitative.Behavior.Quantum.Fluctuations}}
%~~~~~~~~~~~~~~~~~~~~~~~~~~~~~~~~~~~~~~~~~~~~~~~~~~~~
%
%_figure__figure__figure__figure__figure__figure__figure__figure__figure__figure__figure__figure_
\begin{figure*}[!htb]
\begin{center}
\mbox{
\subfigure[$\,$ $N_{k}(t)$. 
\label{Fig:FIG_desitter_ts0.001_2000_100000_QP_-55_25}]{\includegraphics{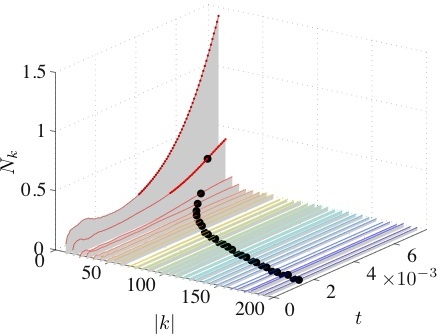}}
\hspace{3mm}
\subfigure[$\,$ $\mathcal R_{k}(t)$. 
\label{Fig:FIG_desitter_ts0.001_2000_100000_FR_-55_25}]{\includegraphics{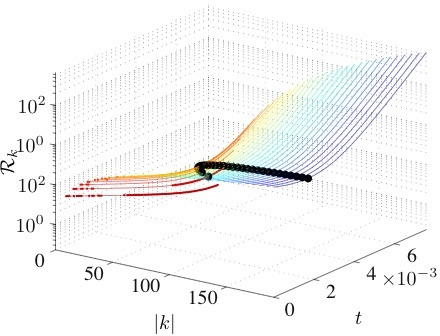}}     
}
\vspace{3mm}
\mbox{
\subfigure[$\,$  $N_{k}(t)$ projected onto the $t$-$N_{k}$ plane.
\label{Fig:FIG_desitter_ts0.001_2000_100000_QP_0_0}]{\includegraphics{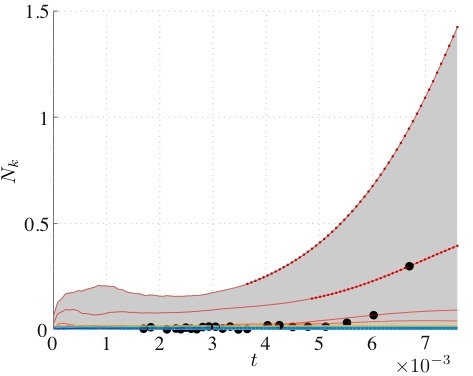}}
\hspace{3mm}
\subfigure[$\,$  $\mathcal R_{k}(t)$ projected onto the $t$-$R_{k}$ plane.
\label{Fig:FIG_desitter_ts0.001_2000_100000_FR_0_0}]{\includegraphics{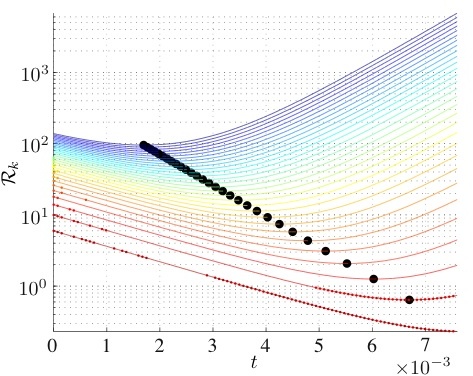}}     
}
\vspace{3mm}
\mbox{
\subfigure[$\,$  $N_{k}(t)$ projected onto the $k$-$N_{k}$ plane. 
\label{Fig:FIG_desitter_ts0.001_2000_100000_QP_-90_0}]{\includegraphics{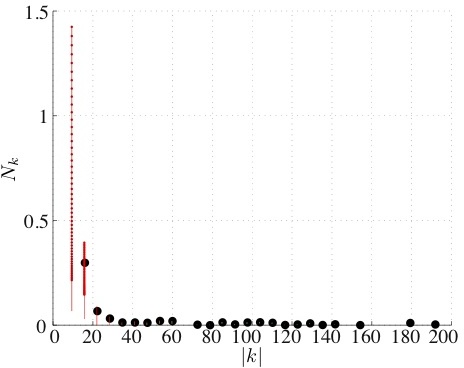}}
\hspace{3mm}
\subfigure[$\,$  $\mathcal R_{k}(t)$ projected onto the $k$-$R_{k}$ plane.
\label{Fig:FIG_desitter_ts0.001_2000_100000_FR_-90_0}]{\includegraphics{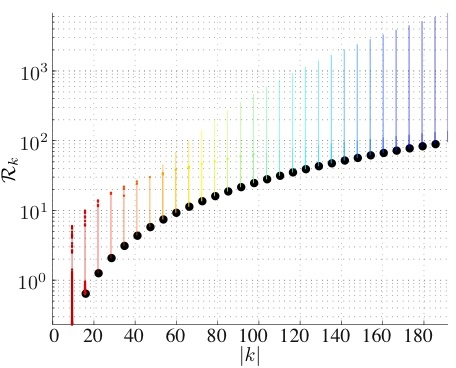}}     
}
\vspace{3mm}
\caption[Relationship between quasiparticle production and frequency ration for quantum modes.]{(Colors online only.) In this figure we compare the quasiparticle production per quantum mode (left column) with its frequency ratio (right column), for $t_{s}=1\times10^{-3}$. Parameters are $C_{NL}(\bar{t}=0)=1 \times 10^{5}$, $N_{0}=10^{7}$ and $X=2 \times 10^{3}$. The bold plotted dots on the left hand side indicate that the frequency ratio is below one, hence the quantum mode corresponds to a super-Hubble horizon mode. While on the right hand side we indicated with the bold dots when a change in the mode occupation number is above a certain threshold --- here $\Delta N_k \geqslant 0.004$ --- to filter out quantum noise fluctuations. }
\label{Fig:desitter_ts0.001_2000_100000_QP_FR}
\end{center}
\end{figure*}
%_figure__figure__figure__figure__figure__figure__figure__figure__figure__figure__figure__figure_
%
%
%_figure__figure__figure__figure__figure__figure__figure__figure__figure__figure__figure__figure_
\begin{figure*}[!htb]
\begin{center}
\mbox{
\subfigure[$\,$ $N_{k}(t)$. 
\label{Fig:FIG_desitter_ts0.0001_2000_100000_QP_-55_25}]{\includegraphics{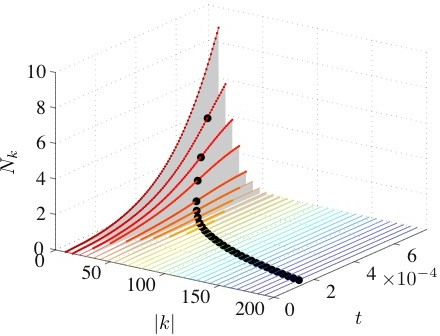}}
\hspace{3mm}
\subfigure[$\,$ $\mathcal R_{k}(t)$. 
\label{Fig:FIG_desitter_ts0.0001_2000_100000_FR_-55_25}]{\includegraphics{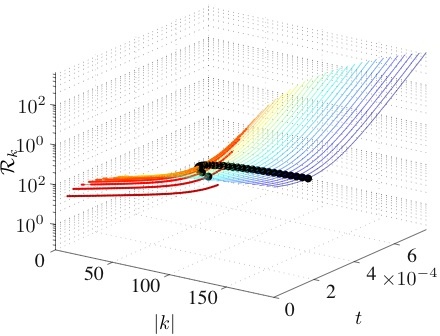}}     
}
\vspace{3mm}
\mbox{
\subfigure[$\,$  $N_{k}(t)$ projected onto the $t$-$N_{k}$ plane.
\label{Fig:FIG_desitter_ts0.0001_2000_100000_QP_0_0}]{\includegraphics{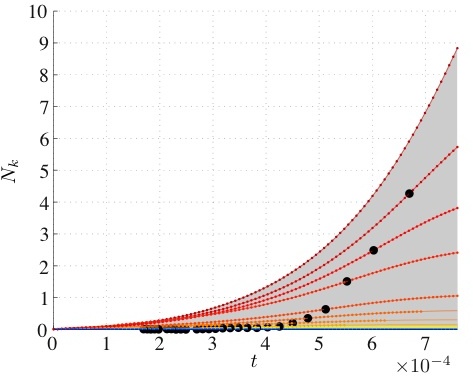}}
\hspace{3mm}
\subfigure[$\,$  $\mathcal R_{k}(t)$ projected onto the $t$-$R_{k}$ plane.
\label{Fig:FIG_desitter_ts0.0001_2000_100000_FR_0_0}]{\includegraphics{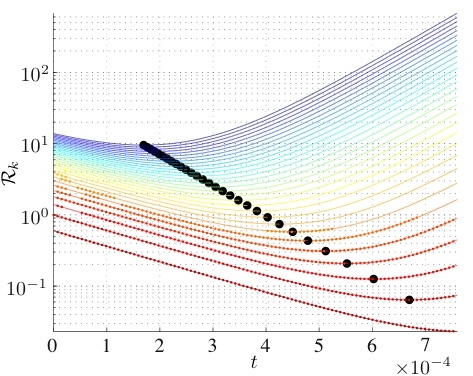}}     
}
\vspace{3mm}
\mbox{
\subfigure[$\,$  $N_{k}(t)$ projected onto the $k$-$N_{k}$ plane. 
\label{Fig:FIG_desitter_ts0.0001_2000_100000_QP_-90_0}]{\includegraphics{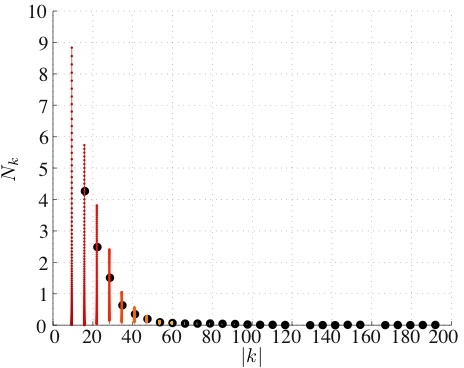}}
\hspace{3mm}
\subfigure[$\,$  $\mathcal R_{k}(t)$ projected onto the $k$-$R_{k}$ plane.
\label{Fig:FIG_desitter_ts0.0001_2000_100000_FR_-90_0}]{\includegraphics{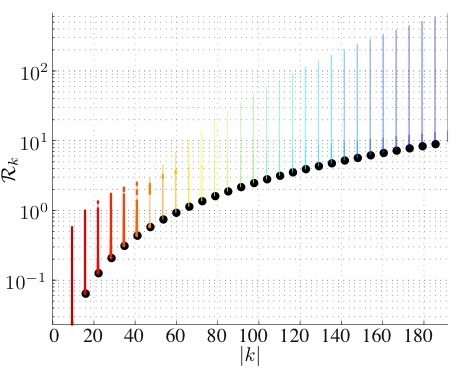}}     
}
\vspace{3mm}
\caption[Relationship between quasiparticle production and frequency ration for quantum modes.]{(Colors online only.) In this figure we compare the quasiparticle production per quantum mode (left column) with its frequency ratio (right column), for $t_{s}=1\times10^{-4}$. Parameters are $C_{NL}(\bar{t}=0)=1 \times 10^{5}$, $N_{0}=10^{7}$ and $X=2 \times 10^{3}$. The bold plotted dots on the left hand side indicate that the frequency ratio is below one, hence the quantum mode corresponds to a super-Hubble horizon mode. While on the right hand side we indicated with the bold dots when a change in the mode occupation number is above a certain threshold --- here $\Delta N_k \geqslant 0.004$ --- to filter out quantum noise fluctuations. }
\label{Fig:desitter_ts0.0001_2000_100000_QP_FR}
\end{center}
\end{figure*}
%_figure__figure__figure__figure__figure__figure__figure__figure__figure__figure__figure__figure_
%
%
%_figure__figure__figure__figure__figure__figure__figure__figure__figure__figure__figure__figure_
\begin{figure*}[!htb]
\begin{center}
\mbox{
\subfigure[$\,$ $N_{k}(t)$. 
\label{Fig:FIG_desitter_ts5e-05_2000_100000_QP_-55_25}]{\includegraphics{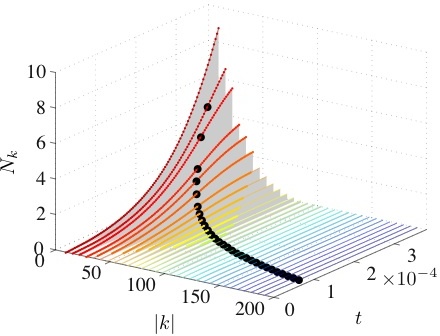}}
\hspace{3mm}
\subfigure[$\,$ $\mathcal R_{k}(t)$. 
\label{Fig:FIG_desitter_ts5e-05_2000_100000_FR_-55_25}]{\includegraphics{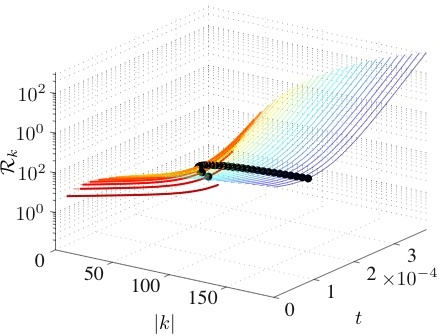}}     
}
\vspace{3mm}
\mbox{
\subfigure[$\,$  $N_{k}(t)$ projected onto the $t$-$N_{k}$ plane.
\label{Fig:FIG_desitter_ts5e-05_2000_100000_QP_0_0}]{\includegraphics{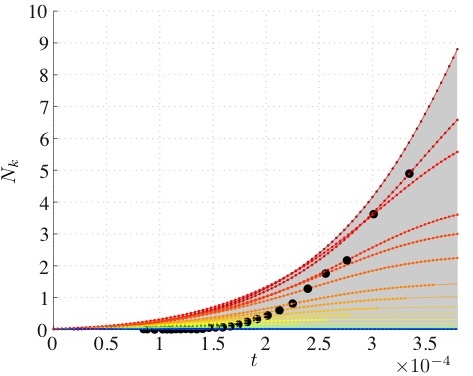}}
\hspace{3mm}
\subfigure[$\,$  $\mathcal R_{k}(t)$ projected onto the $t$-$R_{k}$ plane.
\label{Fig:FIG_desitter_ts5e-05_2000_100000_FR_0_0}]{\includegraphics{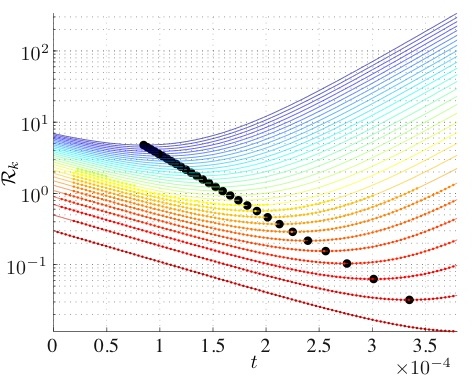}}     
}
\vspace{3mm}
\mbox{
\subfigure[$\,$  $N_{k}(t)$ projected onto the $k$-$N_{k}$ plane. 
\label{Fig:FIG_desitter_ts5e-05_2000_100000_QP_-90_0}]{\includegraphics{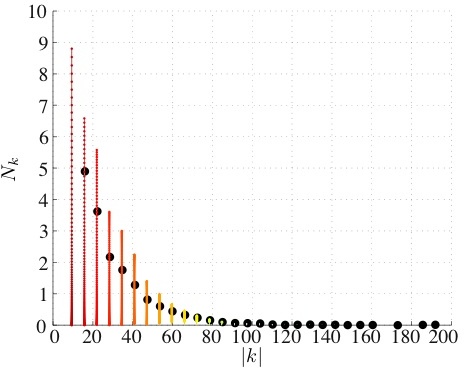}}
\hspace{3mm}
\subfigure[$\,$  $\mathcal R_{k}(t)$ projected onto the $k$-$R_{k}$ plane.
\label{Fig:FIG_desitter_ts5e-05_2000_100000_FR_-90_0}]{\includegraphics{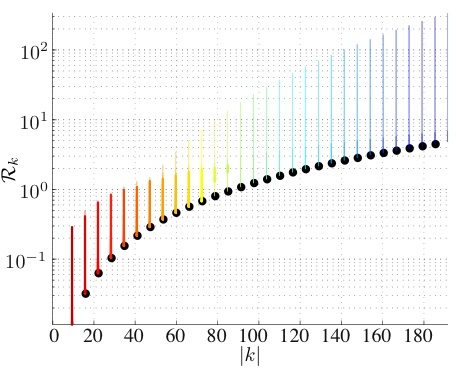}}     
}
\vspace{3mm}
\caption[Relationship between quasiparticle production and frequency ration for quantum modes.]{(Colors online only.) In this figure we compare the quasiparticle production per quantum mode (left column) with its frequency ratio (right column), for $t_{s}=5 \times10^{-5}$. Parameters are $C_{NL}(\bar{t}=0)=1 \times 10^{5}$, $N_{0}=10^{7}$ and $X=2 \times 10^{3}$. The bold plotted dots on the left hand side indicate that the frequency ratio is below one, hence the quantum mode corresponds to a super-Hubble horizon mode. While on the right hand side we indicated with the bold dots when a change in the mode occupation number is above a certain threshold --- here $\Delta N_k \geqslant 0.004$ --- to filter out quantum noise fluctuations.}
\label{Fig:desitter_ts5e-05_2000_100000_QP_FR}
\end{center}
\end{figure*}
%_figure__figure__figure__figure__figure__figure__figure__figure__figure__figure__figure__figure_
%
%
%_figure__figure__figure__figure__figure__figure__figure__figure__figure__figure__figure__figure_
\begin{figure*}[!htb]
\begin{center}
\mbox{
\subfigure[$\,$ $N_{k}(t)$. 
\label{Fig:FIG_desitter_ts1e-05_2000_100000_QP_-55_25}]{\includegraphics{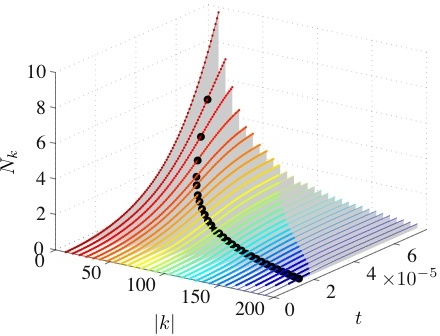}}
\hspace{3mm}
\subfigure[$\,$ $\mathcal R_{k}(t)$. 
\label{Fig:FIG_desitter_ts1e-05_2000_100000_FR_-55_25}]{\includegraphics{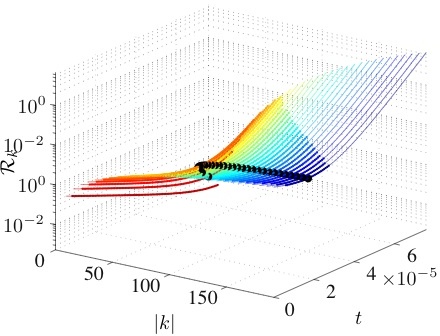}}     
}
\vspace{3mm}
\mbox{
\subfigure[$\,$  $N_{k}(t)$ projected onto the $t$-$N_{k}$ plane.
\label{Fig:FIG_desitter_ts1e-05_2000_100000_QP_0_0}]{\includegraphics{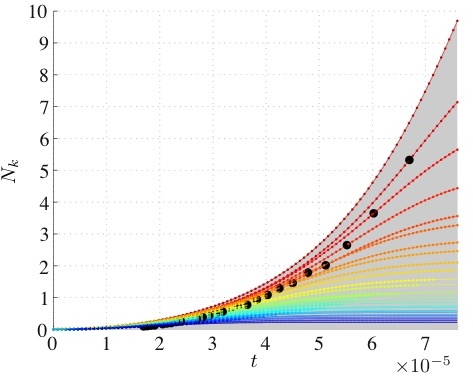}}
\hspace{3mm}
\subfigure[$\,$  $\mathcal R_{k}(t)$ projected onto the $t$-$R_{k}$ plane.
\label{Fig:FIG_desitter_ts1e-05_2000_100000_FR_0_0}]{\includegraphics{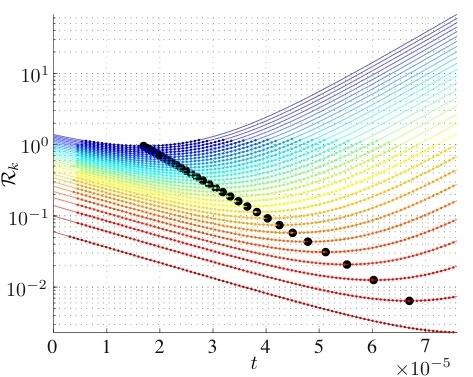}}     
}
\vspace{3mm}
\mbox{
\subfigure[$\,$  $N_{k}(t)$ projected onto the $k$-$N_{k}$ plane. 
\label{Fig:FIG_desitter_ts1e-05_2000_100000_QP_-90_0}]{\includegraphics{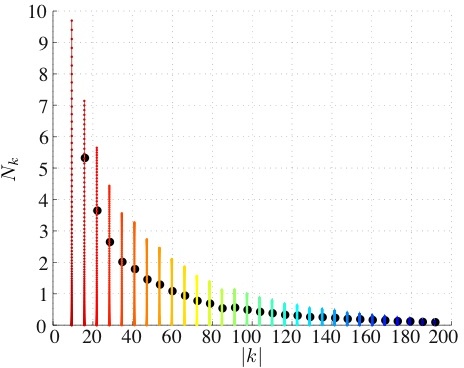}}
\hspace{3mm}
\subfigure[$\,$  $\mathcal R_{k}(t)$ projected onto the $k$-$R_{k}$ plane.
\label{Fig:FIG_desitter_ts1e-05_2000_100000_FR_-90_0}]{\includegraphics{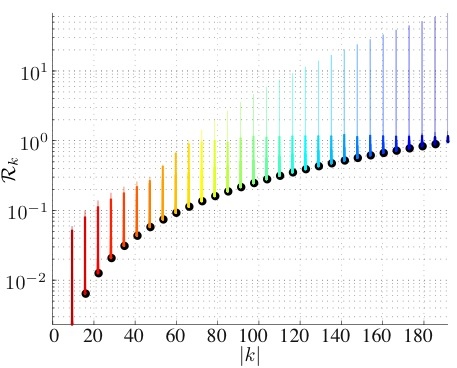}}     
}
\vspace{3mm}
\caption[Relationship between quasiparticle production and frequency ration for quantum modes.]{(Colors online only.) In this figure we compare the quasiparticle production per quantum mode (left column) with its frequency ratio (right column), for $t_{s}=1\times10^{-5}$. Parameters are $C_{NL}(\bar{t}=0)=1 \times 10^{5}$, $N_{0}=10^{7}$ and $X=2 \times 10^{3}$. The bold plotted dots on the left hand side indicate that the frequency ratio is below one, hence the quantum mode corresponds to a super-Hubble horizon mode. While on the right hand side we indicated with the bold dots when a change in the mode occupation number is above a certain threshold --- here $\Delta N_k \geqslant 0.004$ --- to filter out quantum noise fluctuations. }
\label{Fig:desitter_ts1e-05_2000_100000_QP_FR}
\end{center}
\end{figure*}
%_figure__figure__figure__figure__figure__figure__figure__figure__figure__figure__figure__figure_
%

The process of cosmological particle production in an expanding\,/\,collapsing universe can be qualitatively understood in terms of a single parameter, the frequency ratio $\mathcal{R}_{k}(t)$ as given in Eq.~\refb{Eq:Ratio.deSitter.eikonal}. 
First, we explain the connection between the qualitative behavior of the particle production process and this frequency ratio \refb{Eq:Ratio.deSitter.hydro} in the hydrodynamic limit, then we transfer these ideas to the emergent rainbow metrics we have introduced in~Sec.~\ref{Sec:FRW.type.geometries}. \\

%~~~~~~~~~~~~~~~~~~~~~~~~~~~~~~~~~~~~~~~~~~~~~~~~~~~~
\paragraph{Qualitative behavior of particle production in the hydrodynamic limit.}
%~~~~~~~~~~~~~~~~~~~~~~~~~~~~~~~~~~~~~~~~~~~~~~~~~~~~
There is a relatively simple way to understand the qualitative evolution of mode functions in an exponentially changing universe, by considering out the effective harmonic oscillator equation for the auxiliary field.
Within the hydrodynamic limit we get
\begin{equation}
\label{Eq:Qualitative.Hydro}
\ddot{\hat \chi}_{k}(t) + \left(\omega_{k}(t)^{2} - H^{2} \right) \hat\chi_{k}(t) = 0 \, ,
\end{equation}
where we used the equation of motion, see Eq.~\refb{Eq:Equation.Motion.Chi}, for the effective time-dependent harmonic oscillator frequency for phononic modes given in Eq.~\refb{Eq:Harmonic.Oscillator.deSitter.Hydrodynamic}. 

Above we have shown that the general solution is a linear combination of first order Hankel functions of the first and second kind, see  Eqns.~\refb{Eq:Mode.Functions.deSitter.v} and \refb{Eq:Mode.Functions.deSitter.vc}. These mode functions are a function of $R_{k}(t)$, and therefore in the limit of $R_{k}\to \infty$ the mode functions approach positive and negative frequency modes, while for $R_{k} \to 0$ the modes stop oscillating, and the modes exhibit exponentially growing or exponentially decaying kinematics.

A simpler way to come to the same answer is to investigate Eq.~\refb{Eq:Qualitative.Hydro} in its limits: These are $\omega_{k}(t) \gg H$, and $\omega_{k}(t) \ll H$, or equivalently in terms of the frequency ratio:

\begin{description}
\item[$\mathbf{R_{k}(t) \gg 1}$:] {It is then possible to write down an \emph{approximate} solutions for Eq.~\refb{Eq:Qualitative.Hydro},
\begin{eqnarray}
\label{Eq:WKB.Mode.Function.v}
v_{k}^{\mathrm{dS}} &=& \frac{\exp(i \int_{t_{0}}^{t} \omega_{k}(t') \, dt')}{\sqrt{\omega_{k}{t}}} \, , \\
\label{Eq:WKB.Mode.Function.vc}
(v_{k}^{*})^{\mathrm{dS}} &=& \frac{\exp(-i \int_{t_{0}}^{t} \omega_{k}(t') \, dt')}{\sqrt{\omega_{k}{t}}} \, .
\end{eqnarray}
These are approximately plane waves, but their amplitude and frequency change as a function of time.  This ansatz is referred to as the WKB approximation, which is valid within the \emph{adiabatic} limit,
when during one oscillation period $T=2\pi/ \omega_{k}(t)$ the relative change in the frequency is small (see \cite{Mukhanov:2007aa}),
\begin{equation}
\left\vert \frac{\omega_{k}(t+T)-\omega_{k}(t)}{\omega_{k}(t)} \right\vert \approx  
%\left\vert \frac{\dot{\omega}_{k}}{\omega_{k}} \, T \right\vert = 
2\pi \, \left\vert \frac{\dot \omega_{k}}{\omega_{k}^{2}} \right\vert \ll 1 \, .
\end{equation}
For de Sitter spacetimes equates to
\begin{equation}
\left\vert \frac{\dot \omega_{k}^{\mathrm{dS}}}{(\omega_{k}^{\mathrm{dS}})^{2}} \right\vert 
%= \left\vert \frac{H}{\omega_{k}^{\mathrm{in}} \, \exp(-Ht)}  \right\vert 
=   \left\vert \frac{1}{R_{k}(t)}  \right\vert 
\ll 1 \, ,
\end{equation}
the condition that the ratio $R_{k}(t)$ is much larger than one, which verifies the consistency of adopting the adiabatic approximation.\\ 

Hence, the particle production process in the infinite past is negligibly small, as all modes oscillate rapidly compared to the Hubble frequency.}
 
\item[$\mathbf{R_{k}(t) \ll 1}$:] {Here the differential equation \refb{Eq:Qualitative.Hydro}, reduces to
\begin{equation}
\label{Eq:Qualitative.Hydro.Large.H}
\ddot{\hat \chi}_{k}(t) - H^{2}  \hat\chi_{k}(t) = 0 \, .
\end{equation}
The solutions of this equation are exponentially growing, $\sim \exp(Ht)$, or decaying, $\sim \exp(-Ht)$, mode functions. \\

Therefore, the modes are no longer freely oscillating. Instead they get ``dragged along'' with the spacetime fabric.
}
\item[$\mathbf{R_{k}(t)=1}$:]{For any mode $k$ there is a time $t_{\mathrm{crossing}}$ when the frequency ratio is equal to one, and such that:
\begin{eqnarray}
t< t_{\mathrm{crossing}}  \quad  &:&  \quad \Omega_{k}(t)^{2} > 1 \, ,\\
t> t_{\mathrm{crossing}}  \quad  &:&  \quad \Omega_{k}(t)^{2} < 1 \, .
\end{eqnarray}
Therefore it is possible to associate a boundary, the so-called ``\emph{Hubble horizon}'', when $\omega_{k}=H$. For different modes with wavenumber $k$ the Hubble horizon occurs at different times. \emph{Subhorizon} modes, \ie, $\omega_{k}>H$ are distorted plane wave oscillations, while \emph{superhorizon} modes, \ie, $\omega_{k}<H$, approximately satisfy a harmonic oscillator equation with imaginary oscillator frequency, and thus they are exponentially growing or decaying. The ``crossing-time'' $t=t_{\mathrm{crossing}}$ is referred to as the time of ``Horizon crossing''. The \emph{superhorizon} modes are sometimes said to be frozen modes. Following this definition, if we retain the hydrodynamic approximation, then in the infinite future all modes will be frozen out.\\

Please bear in mind that the ``Hubble horizon'' is different from the cosmological horizon, see Sec.~\ref{Sec:Emergent.Cosmological.Horizons}. One easy way to see this, is to remember that the Hubble ``horizon'' is associated with a point in time, which is different for each mode, while the cosmological horizon is the maximum distance signals can propagate for any mode.}
\end{description}

The main lesson for the hydrodynamic limit is that it is not essential to explicitly solve the differential equation. Of course, it is much easier to investigate certain limits of the resulting mode functions, instead of investigating the limits of the equation of motion. In the hydrodynamic limit both methods are accessible. Beyond this limit, where the eikonal approximation is valid, we suggest a rather different strategy. \\

%~~~~~~~~~~~~~~~~~~~~~~~~~~~~~~~~~~~~~~~~~~~~~~~~~~~~
\paragraph{Qualitative behavior of particle production beyond the hydrodynamic limit:}
%~~~~~~~~~~~~~~~~~~~~~~~~~~~~~~~~~~~~~~~~~~~~~~~~~~~~
\emph{Ex ante} we would like to motivate this section by the remark that while the frequency ratio in the hydrodynamic limit $R_{k}(t)$ is a monotonically decreasing function in time, the ratio in the eikonal approximation $\mathcal{R}_{k}(t)$ is not. 
Therefore there is some freedom to obtain different results to the ``conventional'' particle production process. 
We demonstrate the correctness of this assertion by referring to the numerical simulations reported in \cite{Jain:2006ki}.\\

To obtain a rough estimate on the different qualitative regimes of particle production, we use the experience gained in the hydrodynamic limit, and simply exchange $R_{k}(t) \to \mathcal{R}_{k}(t)$. The eikonal frequency ratio has been introduced in Sec.~\ref{Sec:Characteristic.Value.Quantum.Process}, see Eq.~\refb{Eq:Ratio.deSitter.eikonal}, as the ratio between the modes (modified) frequency $\omega_{k}(t)=\omega_{0}\sqrt{b_{k}(t)}$, and the (rainbow) Hubble parameter $H_{k}(t)$, see Eqs.~\refb{Eq:Rainbow.Scale.Factor.deSitter} and \refb{Eq:Ratio.deSitter.eikonal}. \\

For early times, when $R_{k}(t)\gg 1$ the hydrodynamic and eikonal ratios are identical, therefore in both cases approach the adiabatic limit, where we expect the particle production process to be negligibly small. \\

As intimated, the overall slope of the eikonal ratio is not a monotonically decreasing function, since
\begin{equation}
\label{Eq:Derivative.Ratio.Eikonal}
\dot{\mathcal{R}}_{k} =  \frac{\omega_{0}\,\sqrt{\exp(-2Ht)+(k/K)^{2}}\, [\exp(-2Ht)-2 \, (k/K)^{2}]}{-\exp(-2Ht)} \, ,
\end{equation} 
and the time derivative of the eikonal limit changes its sign at 
\begin{equation}
t_{\mathrm{turn}} = \frac{\ln (K^{2}/(2\, k^{2}))}{2H} \, .
\end{equation}
For $t < t_{\mathrm{turn}}$ the slope of the ratio is negative, for $t=t_{\mathrm{turn}}$ the ratio is given by 
\begin{equation}
\mathcal{R}_{k}(t_{\mathrm{turn}})= \frac{3\sqrt{3}}{2} \, \frac{\gamma_{\mathrm{qp}}}{H} \, k^{2}\, ,
\end{equation}
and for $t>t_{\mathrm{turn}}$ the ratio is positive. Therefore the eikonal ratio has a minimum at $t_{\mathrm{turn}}$, with the maximal particle production around this point. After this point the ratio starts to increase again, and we shall soon see that the particle production process will slow down again. \\

To qualitatively describe the particle production process in our specific rainbow spacetime, we suggest the following terminology: 
\begin{description}
\item[$t\to - \infty$:] {At early times almost all modes are ``sub-Hubble-horizon'' modes, and the particle production process is negligible. The modes oscillate with much higher frequencies as their corresponding Hubble frequencies, that is $\mathcal{R}_{k}(t) \gg 1$.}
\item[$t \sim t_{ \mathrm{turn} } $:] {As times goes on the mode frequencies are decreasing, while at the same time the rainbow Hubble frequencies are decreasing as well, see Sec.~\ref{Sec:What.About.Hubble.Paramter}. Nevertheless, the ratio between them exhibits a minimum at $t_{\mathrm{turn}}$, where the particle production process is expected to be maximal. 
Even if the particle production process is maximal, this does not necessarily imply that the quantity of particle production is noticeable; the modes also need to be ``super-Hubble-horizon'' modes, or in more accurate terminology, we require $\mathcal{R}_{k}(t_{\mathrm{turn}})\ll 1$.}
\item[$t \sim t_{ \mathrm{crossing} } $:] {If there exists a time $t=t_{\mathrm{crossing}}$, such that $R_{k}(t_{\mathrm{crossing}})\sim 1$, where a mode $k$ crosses the ``Hubble horizon'', there will be a second time $t=t_{\mathrm{re-entering}}$, where the mode $k$ enters the ``Hubble horizon'', and  $R_{k}(t_{\mathrm{re-entering}})\sim 1$. We suggest that it is useful to adopt the following terminology to describe the behavior of the modes: ``freezing of the mode $k$'' occurs in the time period $t_{\mathrm{crossing}} < t < t_{\mathrm{turn}}$, whereas ``melting of the mode $k$'' occurs during $t_{\mathrm{turn}} < t < t_{\mathrm{re-entering}}$.}
\end{description}

%
%_figure__figure__figure__figure__figure__figure__figure__figure__figure__figure__figure__figure_
\begin{figure}[htb]
 \begin{center}
 \includegraphics{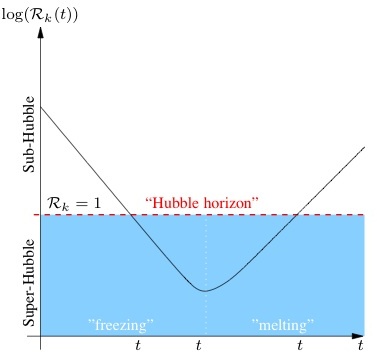}
 \caption[Freezing and melting of quantum modes.]  {\label{Frezzing.Melting.Concept} Schematic description of the freezing and melting of quantum modes.}
 \end{center} 
\index{de~Sitter}      
\end{figure}
%_figure__figure__figure__figure__figure__figure__figure__figure__figure__figure__figure__figure_
%

Note, that we have used the approximate relations $R_{k}(t_{\mathrm{crossing}})\sim 1$ and $R_{k}(t_{\mathrm{re-entering}})\sim 1$, instead of the more precise $R_{k}(t_{\mathrm{crossing}})= 1$ and $R_{k}(t_{\mathrm{re-entering}})= 1$. This purely technical complication is due to the fact that the harmonic oscillator frequency in the eikonal limit, see Eq.~\refb{Eq:Harmonic.Oscillator.Frequency.Eikanol.General}, can be written as
\begin{equation}
\Omega_{k}(t)^{2} = \omega_{k}(t)^{2} - H_{k}(t)^{2} \, [1 - 2\, \exp(2Ht) \, (k/K)^{2}] \, , \;
\end{equation}
and cannot be simplified to $\Omega_{k}(t)^{2} = \omega_{k}(t)^{2} - H_{k}(t)^{2} $.
Here the \emph{exact} times for the ``crossing'' and ``re-entering'' of horizons should be correlated with the sign-change in the harmonic oscillator frequency. Nevertheless, it can be seen that this does not change the qualitative description for the particle process. \\

Another novelty in our qualitative understanding of the particle production process in our FRW rainbow-spacetime, is the connection with the condensed matter point of view:
The minimum of the ratio $\mathcal{R}_{k}(t_{\mathrm{turn}})$ for a mode $k$, see Eqs.~\refb{Eq:Derivative.Ratio.Eikonal}, \refb{Eq.Time.Dependent.Effective.Planck.Length}, and \refb{Eq:K}, occurs at
\begin{equation}
\exp(-2Ht) - 2(k/K)^{2}=0 \quad \rightarrow \quad k=  \ell_{\mathrm{Planck}}(t)^{2} / 2 \, . 
\end{equation}
This quantity also appears in the context of conventional condensed matter physics, where it is defined as the crossover between the phonon and free-particle region. This borderline, the inverse of the healing, or coherence length $\xi$~\cite{Pethick:2001aa}, is given by
\begin{equation}
\xi^{-2} = \frac{2\, m \,n_{0} \, U(t)}{\hbar^{2}} = \frac{1}{2} \, \frac{4 m^{2}}{\hbar^{2}} \, \frac{n_{0} \, U(t)}{m} =   \ell_{\mathrm{Planck}}(t)^{2} / 2 \, ,
\end{equation}
which indicates where each mode starts to decouple from the spacetime. In other words, each mode can experience particle production, until it becomes free-particle like. 
Hence, in terms of the microscopic physics of a BEC we have a natural understanding of $\mathcal{R}_{k}(t_{\mathrm{turn}})$. For a detailed treatment of the numerical simulations please see~\cite{Jain:2006ki}.

To show the qualitative correlation between the modified frequency ratio \refb{Eq:Ratio.deSitter.eikonal} with particle production in our specific rainbow de Sitter spacetime, we have plotted the ratio for several $k$-modes as a function of time, and compared them to number occupation plots, see Figs.~\ref{Fig:desitter_ts0.001_2000_100000_QP_FR}, \ref{Fig:desitter_ts0.0001_2000_100000_QP_FR}, \ref{Fig:desitter_ts5e-05_2000_100000_QP_FR} and \ref{Fig:desitter_ts1e-05_2000_100000_QP_FR}.
This figure compares the change in the mode occupation number in each mode on the left side, with frequency ratio $\mathcal{R}_{k}(t)$ of the mode on the right side,  for a number of different scaling times $t_{s}$.
Each black line on the left hand side indicates the occupation number in the mode $k$ as a function of time. (The red dashed line on the left had side indicates the sudden limit.) On the right hand side we have plotted the frequency ratio $\mathcal{R}_{k}(t)$ for each of those modes with a different color (online only); gradually changing from infrared modes (dark red), to ultraviolet modes (dark blue). The horizontal (red dashed) line indicates where the frequency ratio $\mathcal{R}_{k}(t)$ is equal to one, while the vertical (blue dashed) line indicates the end of the expansion time in our simulations. 

The black dots in the figures to the left indicate when the modes cross over from phononic to trans-phononic behavior, \ie where each mode starts to decouple from the emergent spacetime. We can see that the black dots are located where the frequency ratio has its minimum; see Figs.~\ref{Fig:FIG_desitter_ts0.001_2000_100000_FR_-90_0}, \ref{Fig:FIG_desitter_ts0.0001_2000_100000_FR_-90_0}, \ref{Fig:FIG_desitter_ts5e-05_2000_100000_FR_-90_0}, \ref{Fig:FIG_desitter_ts1e-05_2000_100000_FR_-90_0} and \ref{Fig:FIG_desitter_ts1e-05_4000000_200000_FR_-90_0}.

Figures~\ref{Fig:desitter_ts0.001_2000_100000_QP_FR}, \ref{Fig:desitter_ts0.0001_2000_100000_QP_FR}, \ref{Fig:desitter_ts5e-05_2000_100000_QP_FR}  and~\ref{Fig:desitter_ts1e-05_2000_100000_QP_FR} correspond to four different scaling times $t_{s}$ for $t_{s}=1\times 10^{-3}$, $t_{s}=1\times 10^{-4}$, $t_{s}=5\times 10^{-5}$ and $t_{s}=1\times 10^{-5}$. In all these figures we compare quasi-particle production (left column) with the frequency ration (right column). The simulation parameters in these simulations are in correspondence to reference~\cite{Jain:2006ki}, where  the initial non-linearity was given by $C_{NL}(\bar{t}=0)=1 \times 10^{5}$, the number of atoms in the condensates correspond to $N_{0}=10^{7}$ and $X=2 \times 10^{3}$.

For example, in Fig.~\ref{Fig:FIG_desitter_ts0.001_2000_100000_QP_-55_25} we see the quasi-particle production in each mode in the de Sitter region where $t_{s}=1\times 10^{-3} $. The negligibly small population of the modes during, and after, the expansion can be explained by means of the ratio plot to its right, in Fig.~\ref{Fig:FIG_desitter_ts0.001_2000_100000_FR_-55_25}. As pointed out above, the particle production process is large only if $\mathcal{R}_{k} \ll 1$, which is impossible to achieve for such a \emph{relatively} large scaling unit $t_{s}$. We also see, that only the first two modes from the bottom of the infrared scale cross the ``Hubble horizon'', such that $\mathcal{R}_{k} <1$, and (as indicated in the figure) they eventually turn around and re-enter the ``Hubble horizon'' after a few e-foldings. Consequently, such an experimental set-up is inappropriate for mimicking cosmological particle production.

In contrast, Fig.~\ref{Fig:FIG_desitter_ts1e-05_2000_100000_QP_-55_25} shows significant quasi-particle production, for the first $100$ modes plotted in the figure. Here $t_{s}=1\times 10^{-5}$, and thus the expansion is two magnitudes faster than one we discussed before. The qualitative behavior is roughly in agreement with the mode frequency ratio plotted to the right in Fig.~\ref{Fig:FIG_desitter_ts1e-05_2000_100000_FR_-55_25}.
The particle production process slows down after each modes crosses over from the phononic to free-particle regime (at the blue point), and as we can see from the figures, this happens for the four lowest energy modes between $5-8$ e-foldings. \\

While the emergent spacetime picture is necessary to understand the time-dependent commutator relations --- in terms of the field operator and its conjugate momentum on a time-dependent classical background --- the the particle production process tends to zero during inflation is naturally explained from a condensed matter physics point of view, as it is related to the crossover between phononic and trans-phononic quasi-particle excitations.

%~~~~~~~~~~~~~~~~~~~~~~~~~~~~~~~~~~~~~~~~~~~~~~~~~~~~
\subsubsection{Emergent cosmological horizons?\label{Sec:Emergent.Cosmological.Horizons}}
%~~~~~~~~~~~~~~~~~~~~~~~~~~~~~~~~~~~~~~~~~~~~~~~~~~~~
The existence of a cosmological horizon in our specific emergent spacetime can be investigated by calculating the maximum distance, $r_{\mathrm{max}}$, travelled by small perturbations initiated at a certain time $t=t_{0}$ and certain point $\vec{r}_{0}$. We can associate a cosmological horizon to each point $(t_{0},\vec{r}_{0})$, if the maximum distance the signal --- here our excitations in the Bose--Einstein condensate --- can travel is finite. This naturally defines a region around the point of emission, and its boundary is the cosmological horizon. In the presence of cosmological horizons two points in spacetime are said to be ``causally disconnected'' if the distance between them is larger than $r_{\mathrm{max}}$.

In ``conventional'' cosmology cosmological horizons are predicted to exist for an infinitely long-lasting de Sitter universe, but what is the situation in our spacetime emerging from a Bose gas? We would like to address this question presently, and show that due to non-perturbative corrections there are, strictly speaking, no emergent cosmological horizons present in the system.

This can easily be shown, given that one only needs to integrate the \emph{group} velocity of a perturbation emitted at $(t_{0},\vec{r}_{0})$ to the infinite future,
\begin{equation}
\label{distance_deSitter_max}
r_{\mathrm{max}}
= \lim\limits_{t\rightarrow +\infty}  \int \limits_{t_{0}}^{t}  c_{\mathrm{group}} \; dt .
\end{equation}
In ``conventional'' --- that is Lorentz-invariant spacetimes --- \emph{group} and \emph{phase} velocities are identical.
This is not the case for our specific analogue model, where the \emph{phase} velocity,
\begin{equation} 
\label{phase_velocity_deSitter}
c_{\mathrm{phase}}:=\frac{\omega_{k}}{k} = c_{0} \, \sqrt{\exp(-2Ht) + (k/K)^{2}},
\end{equation} 
and the \emph{group} velocity,
\begin{equation}
\label{group_velocity_deSitter}
c_{\mathrm{group}}:=\frac{\partial \omega_{k}}{\partial k}
= \frac{c_{0}^{2} \exp(-2Ht) + 2 (k/K)^{2}}{\sqrt{c_{0}^{2} \exp(-2Ht) + (k/K)^{2}}}
\end{equation}
are clearly different.

Within the hydrodynamic limit, when $K \to \infty$, the two velocities are equivalent, $c_{\mathrm{group}}(t) \to c_{k}(t)$. Within this limit all signals travel with the same speed, and therefore wave packets ``keep their shape'', \ie~they exhibit no dispersion.
In this case a signal --- sent at $(t_{0},\vec{r}_{0})$ --- propagating forever will only travel a finite distance given by,
\begin{equation}
\label{distance_deSitter_hydro}
\lim\limits_{K \to \infty} r_{\mathrm{max}}
= \lim\limits_{t' \to +\infty}  \int \limits_{t_{0}}^{t'}  \lim\limits_{K \to \infty}  c_{\mathrm{group}} \; dt
= \frac{c_{0}}{H} \, .
\end{equation}
In the past cosmological horizons have been repeatedly mis-interpreted, and we would like to advise the interested reader to carefully read \cite{Rothman:1993aa}.

Returning to our problem of investigating the presence of cosmological horizons in the Bose gas, the maximum distance a signal can travel is given by 
\begin{equation}
\label{distance_deSitter}
r_{\mathrm{max}}
= \lim\limits_{t\rightarrow +\infty}  \int \limits_{0}^{t}   c_{\mathrm{group}} \; dt .
= \infty \, ,
\end{equation}
Thus, the emergent spacetime picture strictly does not exhibit any cosmological horizons whatsoever. This result can be generalized beyond the specific BEC based model, since many other analogue models for gravity exhibit super-sonic modifications in the dispersion relations will show similar behavior.

%
%_figure__figure__figure__figure__figure__figure__figure__figure__figure__figure__figure__figure_
\begin{figure*}[!htb]
\begin{center}
\mbox{
\subfigure[$\,$ $t_{s}=1\times 10^{-3}$ \label{Fig:deSitter.1e3}]{\includegraphics{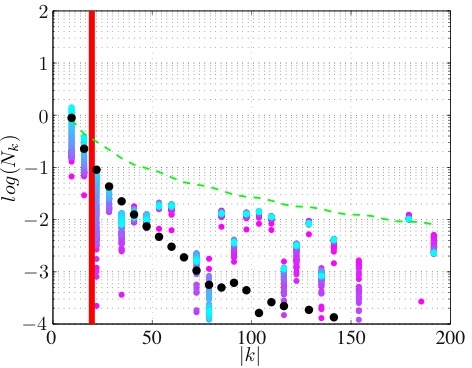}}
\hspace{5mm}
\subfigure[$\,$ $t_{s}=1\times 10^{-4}$ \label{Fig:deSitter.1e4}]{\includegraphics{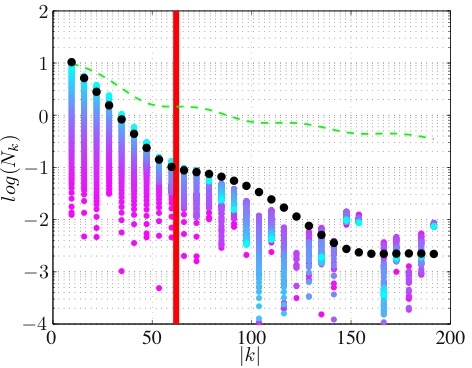}}     
}
\vspace{5mm}
\mbox{
\subfigure[$\,$ $t_{s}=5\times 10^{-5}$ \label{Fig:deSitter.5e5}]{\includegraphics{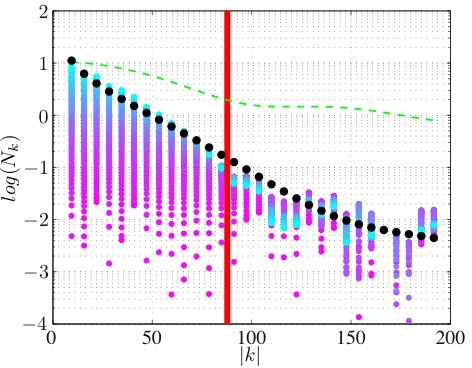}}
\hspace{5mm}
\subfigure[$\,$ $t_{s}=1\times 10^{-5}$ \label{Fig:deSitter.1e5}]{\includegraphics{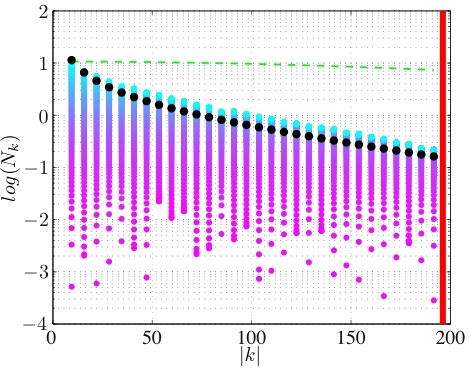}}     
}
\mbox{
\subfigure[$\,$ Duration time $T=t_{final}-t_{initial}$ in percent \label{Fig:Coloarbar}]{ \includegraphics[width=0.9\textwidth]{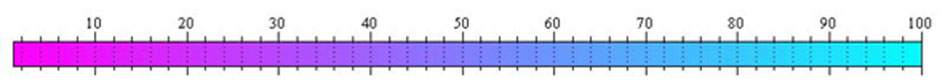}}     
}
\caption[Long lasting rainbow inflation.]{The time-evolution of the particle spectrum on a single plot, for four different scaling times $t_{s}$. As time goes on the data points linearly change their color, changing gradually from violet to turquoise. Note that at low momenta --- corresponding to those modes that twice ``cross the Hubble horizon'' by ``freezing'' and ``re-entering'' -- the final quasi-particle density is approximately log-linear, corresponding to $n_{k} \sim C \, \exp(-\sigma \, k)$. Some modes never ``cross the Hubble horizon'', if $\mathcal{R}_{k}(t_{\mathrm{turn}})\geqslant 1$. This can be seen in our plots, where we indicated 
$k_{\mathrm{static}} =  [\sqrt{2} \, 3^{(-3/4)}/ \sqrt{\gamma_{\mathrm{qp}}}] \cdot \sqrt{H}$ 
in each plot with a vertical (red) dashed line. The dashed green line shows the theoretical results for a finite de Sitter calculation obtained in the hydrodynamic limit, as presented in Sec.~\ref{Sec:Toy.Model.Conventional.Inflation}. 
\label{Fig:deSitter.Results.Free.Particle.Regime}}.
\end{center}
\end{figure*}
%_figure__figure__figure__figure__figure__figure__figure__figure__figure__figure__figure__figure_
%

%~~~~~~~~~~~~~~~~~~~~~~~~~~~~~~~~~~~~~~~~~~~~~~~~~~~~
\subsubsection{Long lasting rainbow inflation\label{Sec:Long.Lasting.Rainbow.Inflation}}
%~~~~~~~~~~~~~~~~~~~~~~~~~~~~~~~~~~~~~~~~~~~~~~~~~~~~
The present analogue model does not possess horizons in the cosmological sense. Consequently we cannot automatically assume that an infinitely long-lasting de Sitter phase would lead to a Planckian spectrum where the temperature is connected to the surface gravity at the horizon, see~\cite{Gibbons:1977aa}. Thus there is no straightforward method to calculate the overall temperature for our everlasting de Sitter rainbow spacetime. \\

At first, it seems to be a highly contrived question to ask for the final spectrum in our emergent rainbow spacetime after an infinitely long-lasting inflationary epoch. However, we demonstrate by means of numerical evidence --- from our computer simulations \cite{Jain:2006ki} --- that a \emph{sufficiently long duration expansion} is sufficient to reveal the characteristic final shape of the particle spectrum at the end of inflation. \\

Before we jump to the result, we would like to show that the final Bogoliubov coefficients are indeed time-independent, and that the mode occupation number calculated with the instantaneous Hamiltonian diagonalization method represents ``real'' quasi-particles, due to well-defined $\mathrm{in}$ states in the adiabatic regime, and \emph{also} well-defined $\mathrm{out}$ states in the free-particle regime. The mode functions in our rainbow spacetime interpolate between these two states. \\
%

%~~~~~~~~~~~~~~~~~~~~~~~~~~~~~~~~~~~~~~~~~~~~~~~~~~~~
\paragraph{Pair of coupled harmonic oscillators:}
%~~~~~~~~~~~~~~~~~~~~~~~~~~~~~~~~~~~~~~~~~~~~~~~~~~~~
Given that we are only dealing with one region, we can drop the prefactor in the equation of motion~\refb{Eq:Equation.Motion.Chi} for the auxiliary mode operator $\hat \chi_{k}$:
\begin{equation}
\label{Eq:Equation.Motion.Chi.One.Region}
\ddot{\hat{\chi}}_{k}(t) +\Omega_{k}(t)^{2} \, \hat\chi_{k}(t)  = 0 \, ,
\end{equation}
where
\begin{eqnarray}
t \to + \infty \; &:& \; \ddot{\hat{\chi}}_{k}(t) + (\omega_{k}^{\mathrm{WKB}})^{2}  \, \hat\chi_{k}(t)  = 0 \, , \\
t \to - \infty \; &:& \; \ddot{\hat{\chi}}_{k}(t) + (\omega_{k}^{\mathrm{FP}})^{2} \, \hat\chi_{k}(t)  = 0 \, .
\end{eqnarray}
In the infinite past, when the eikonal scale factor approaches the hydrodynamic (``conventional'') scale factor, the mode functions are determined within the adiabatic regime. Thus $\omega_{k}^{\mathrm{WKB}}\sim \omega_{k}(t)$, and the mode functions are approximated by Eqs.~\refb{Eq:WKB.Mode.Function.v} and \refb{Eq:WKB.Mode.Function.vc}. These mode functions represent positive and negative frequency modes at every instant of time. Moreover in the infinite past they represent the adiabatic vacuum --- for instance see \cite{Mukhanov:2007aa,Fulling:1989aa,Birrell:1984aa}.

In the infinite future all excitations behave like free particles, since
\begin{equation}
\omega_{k}^{\mathrm{FP}} = \frac{\omega_{0} \, k}{K} = \frac{\hbar \, k^{2}}{2m} \, ,
\end{equation}
represents the kinetic energy of an object with mass $m$; $E=p^{2}/(2m)=\hbar^{2} k^{2}/ (2m)$. (In terms of BEC theory this is the ``single particle energy''.) The time-independence is due to the end of the effective expansion in our emergent spacetime. The mode functions are the usual positive and negative frequency modes, as given in Eq.~\refb{Eq:Mode.Functions.Minkowski.out}, where $\omega_{k}^{\mathrm{out}}=\omega_{k}^{\mathrm{FP}}$. \\

%
% _figure__figure__figure__figure__figure__figure__figure__figure__figure_ _figure__figure__figure_
\begin{figure}[!htb]
\centering
\includegraphics{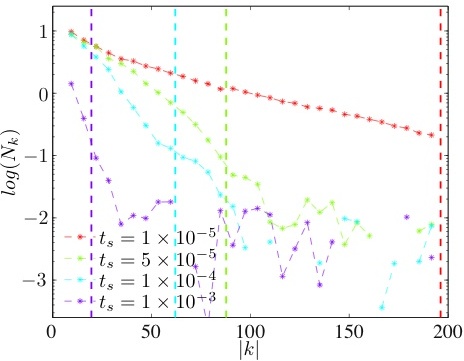}
\caption[Final particle spectra.]{
The four different particle spectra, corresponding to four different scaling times $t_s$, after a finite rainbow de Sitter phase. (Note that the ``gaps'' in the quasi-particle spectrum are a result of the logarithmic scale that is undefined for (small) negative occupation numbers, a numerical artifact arising in our simulations. For more information about the numerical method used to obtain these results consult~\cite{Jain:2006ki}.) 
}
\label{Fig:deSitter.All}
\end{figure}
% _figure__figure__figure__figure__figure__figure__figure__figure__figure_ _figure__figure__figure_
%

Here we are dealing with well-defined \emph{initial and final} vacua states. This kind of problem can be compared with the $\tanh$-expansion, studied for example in~\cite{Fulling:1989aa,Birrell:1984aa}. In this case the Bogoliubov coefficients are --- after an infinitely long-lasting expansion --- time-independent. Please note that this analytic calculation does not by itself give any details 
\emph{when and how} the mode functions change, since the Bogoliubov coefficients for a smooth scale function $b_{k}(t)$ for all $t$ are given by the globally defined time-dependent quantities 
\begin{eqnarray}
\label{Eq.Alpha.General.Continous.Normalized.Smooth}
\alpha_{k}(t) &=&\frac{1}{2i} \, W[ u_{k}, v_{k}^{*}] \, ,  \\
\label{Eq.Beta.General.Continous.Normalized.Smooth}
\beta_{k}(t) &=&\frac{1}{2i} \, W[ v_{k}^{*},u_{k}^{*}]  \, ,
\end{eqnarray}
as compared with Eqs.~\refb{Eq.Alpha.General} and \refb{Eq.Beta.General}. The Wronskian of the mode function is time-independent, therefore it can be evaluated at all times.
(Note that the general mode functions are not necessarily normalized.)
Again, $v_{k}$ and $v_{k}^{*}$ represent the $\mathrm{in}$, and $u_{k}$ and $u_{k}^{*}$ the $\mathrm{out}$ mode functions. (Note that we have changed our notation in respect to~\cite{Jain:2006ki}.) \\

%
% _figure__figure__figure__figure__figure__figure__figure__figure__figure__figure__figure__figure_
\begin{figure}[htb]
\centering
\includegraphics{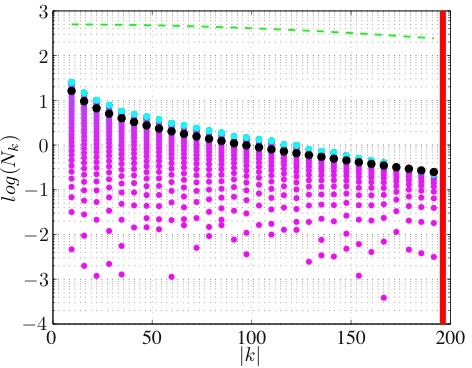}
\caption[Final particle spectra.]{The final particle spectra after a finite duration rainbow de Sitter phase for $t_{s}=1\times 10^{-5}$ after all modes have decoupled from the emergent spacetime geometry. Compare with Fig.~\ref{Fig:deSitter.1e5}. }
\label{Fig:deSitter.Longrun}
\end{figure}
% _figure__figure__figure__figure__figure__figure__figure__figure__figure_ _figure__figure__figure_
%

%
%_figure__figure__figure__figure__figure__figure__figure__figure__figure__figure__figure__figure_
\begin{figure*}[!htb]
\begin{center}
\mbox{
\subfigure[$\,$ $N_{k}(t)$. 
\label{Fig:}]{\includegraphics{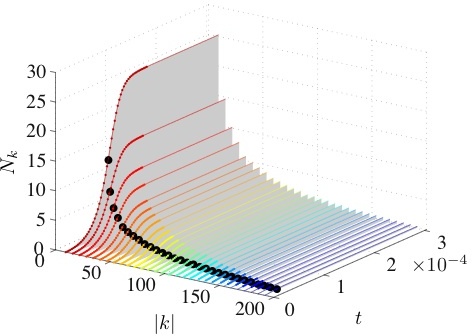}}
\hspace{3mm}
\subfigure[$\,$ $\mathcal R_{k}(t)$. 
\label{Fig:}]{\includegraphics{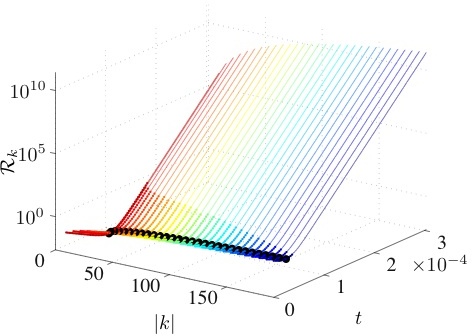}}     
}
\vspace{3mm}
\mbox{
\subfigure[$\,$  $N_{k}(t)$ projected onto the $t$-$N_{k}$ plane.
\label{Fig:FIG_desitter_ts1e-05_4000000_200000_QP_0_0}]{\includegraphics{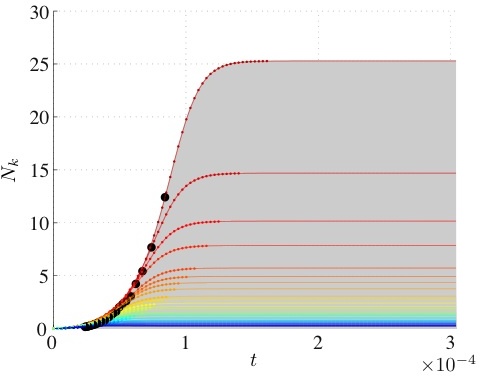}}
\hspace{3mm}
\subfigure[$\,$  $\mathcal R_{k}(t)$ projected onto the $t$-$R_{k}$ plane.
\label{Fig:FIG_desitter_ts1e-05_4000000_200000_FR_0_0}]{\includegraphics{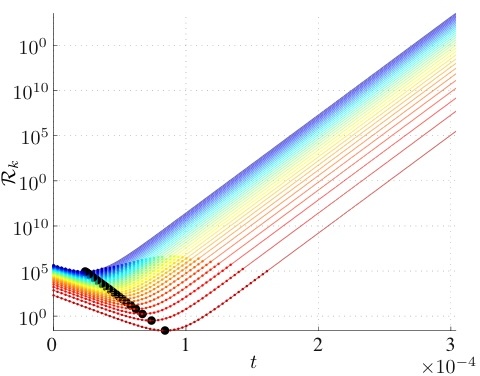}}     
}
\vspace{3mm}
\mbox{
\subfigure[$\,$  $N_{k}(t)$ projected onto the $k$-$N_{k}$ plane. 
\label{Fig:FIG_desitter_ts1e-05_4000000_200000_QP_-90_0}]{\includegraphics{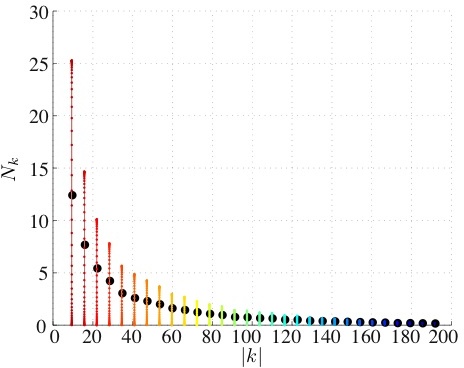}}
\hspace{3mm}
\subfigure[$\,$  $\mathcal R_{k}(t)$ projected onto the $k$-$R_{k}$ plane.
\label{Fig:FIG_desitter_ts1e-05_4000000_200000_FR_-90_0}]{\includegraphics{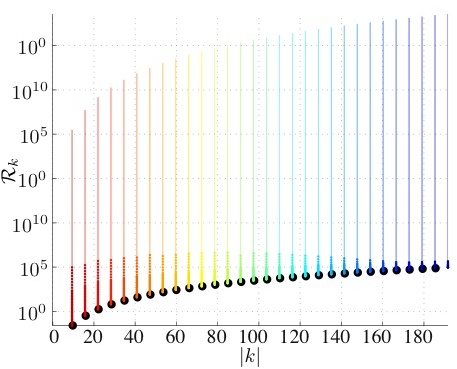}}     
}
\vspace{3mm}
\caption[Relationship between quasiparticle production and frequency ration for quantum modes.]{(Colors online only.) In this figure we compare the quasiparticle production per quantum mode (left column) with its frequency ratio (right column), for $t_{s}=1\times10^{-5}$. Parameters are $C_{NL}(\bar{t}=0)=2 \times 10^{5}$, $N_{0}=10^{7}$ and $X=4 \times 10^{6}$. The bold plotted dots on the left hand side indicate that the frequency ratio is below one, hence the quantum mode corresponds to a super-Hubble horizon mode. While on the right hand side we indicated with the bold dots when a change in the mode occupation number is above a certain threshold --- here $\Delta N_k \geqslant 0.004$ --- to filter out quantum noise fluctuations. }
\label{Fig:desitter_ts1e-05_4000000_200000_QP_FR}
\end{center}
\end{figure*}
%_figure__figure__figure__figure__figure__figure__figure__figure__figure__figure__figure__figure_
%
By means of the numerical results shown in Figs.~\refb{Fig:deSitter.Results.Free.Particle.Regime}, it appears to be rapidly approaching a final asymptotic state.
We have plotted the final spectra in Fig.~\ref{Fig:deSitter.All}, so that we easily see the relationship between the slope of the line and the inverse scaling time $t_{s}^{-1}$. 
However the temporal duration in our previous numerical simulation for $t_{s}=1\times 10^{-5}$ has not been sufficiently long, as it is not obvious that the particle production process has yet came to an end. In Fig.~\ref{Fig:desitter_ts1e-05_2000_100000_QP_FR} we can see that at the end of the numerical simulation a good fraction of the quantum field modes (\ie, roughly $\vert k \vert < 40$) are frozen. We expect particle production in those modes to contribute significantly to the infrared end of the final spectrum. We therefore repeated the numerical simulation for $t_{s}=1 \times 10^{-5}$ (so that  initially all modes are sub-Hubble horizon modes) with two times the previous duration.
In addition, the initial nonlinearity is now $C = 2 \times 10^5$ instead of $C = 1 \times 10^5$ so that all modes are ``phononic'' at the start of the simulation. We also increased our expansion rate from $X=2 \times 10^{3}$ to $X=4 \times 10^{6}$.
 As shown in Fig.~\ref{Fig:deSitter.Longrun} and in greater detail in Fig.~\ref{Fig:desitter_ts1e-05_4000000_200000_QP_FR}, at the end of the simulation all modes are trans-phononic, and the particle production process ceases. 
From our numerical simulations the final particle spectrum does not seem to nicely fit a straight line, but it seems conceivable to employ standard line-fitting tools to study the final particle spectrum as a function of $t_{s}$ and $k$. (We are currently investigating this issue.) 

%%%%%%%%%%%%%%%%%%%%%%%%%%%%%%%%%%%%%%%%%%%%%%%%%%%
%
\section{Conclusions and outlook \label{Sec:Conclusions.Outlook}}
%
%%%%%%%%%%%%%%%%%%%%%%%%%%%%%%%%%%%%%%%%%%%%%%%%%%%
In this article we put the analogy between a parametrically excited Bose--Einstein condensate and cosmological particle production to the test. Knowing that the analogy for mimicking a specific quantum effect in ``conventional'' curved-spacetime quantum-field-theory hinges on the robustness of the effect against model-specific deviations, we derived the ``whole'' model-dependent emergent rainbow spacetime. Similar work on the \emph{acoustic} Hawking effect in subsonic and supersonic (super)-fluids has been carried out in~\cite{Jacobson:1991sd}.\\

There were two main lessons learnt for the analogue model community. First, the specific model we presented --- a uniform gas of atoms with time-dependent atomic interactions --- is in general \emph{not} robust against the non-perturbative ultraviolet corrections. Secondly, we also showed that the analogy is sufficiently good for mimicking some aspects of cosmological particle production for finite changes in the size of the effective universe. We said ``some aspects'', because the analogy only holds for the low-energy part of the spectrum, and therefore the analogy is associated with a certain $k$-range. These correspond to phononic excitations, bounded by a time-dependent parameter $\vert k \vert <  1/ \ell_{\mathrm{Planck}}(t)$.

Given that we expect significant deviation from the desired quantum effect, one might ask the question
``Is the analogue model we have presented a suitable candidate for laboratory experiments?''
Previously in~\cite{Jain:2006ki}, as well as briefly in the current paper, we have presented numerical results for cosmological particle production in a ``realistic'' Bose gas. As a matter  of fact, despite many possible sources of difficulties, for example back-reaction-effects\,/\,mode-mixing, the phononic regime shows excellent agreement with the theoretical predictions\footnote{
In fact, for our numerical simulations the parameters we choose (number of atoms and nonlinearity) were such that we were working in a regime where back-reaction and mode-mixing were negligible or very small effects. But in principle these effects are included by the methodology.}.
The Bose--Einstein condensate enables us to prepare and control a quantum field to such an extent, that within a few years time the technology should be able to drive ``inflation'' between two ``natural'' vacua and --- that is the \emph{outstanding} problem~\cite{Schutzhold:2006pv,Weinfurtner:2007aa,Barcelo:2003yk} --- directly measure the resulting spectra. Of course, there are also other models involving a freely expanding condensate cloud, but we suspect similar problems --- a growing ``Planck-length'' and the lack of a (strict) cosmological horizon~\cite{Fedichev:2004on} --- to appear, and would like to stress that those models destroy the condensate during the expansion process. We leave it as an open problem to simulate and check the ``robustness'' of those models. \\

So much for the analogy, but what did we learn from the deviations occurring in our specific emergent spacetime? Are there conclusions to be drawn that are of relevance for the cosmology, or even quantum gravity programme? We leave this to our readers, and merely summarize our experience regarding ``trans-Planckian'' physics in our emergent spacetime:

The emergent spacetime we have presented is an example of emergent Lorentz-symmetry. At the infrared end of the excitation spectrum we exhibit Lorentz invariance is exhibited, while non-perturbative corrections from the microscopic substructure ``naturally'' break the Lorentz-invariance in the ultraviolet regime~\cite{Liberati:2006sj,Weinfurtner:2006nl,Weinfurtner:2006eq,Weinfurtner:2006iv,Liberati:2006kw}. These corrections also alter the spacetime picture from ``ordinary'' spacetimes to the --- more ``unusual'' but conceivable~\cite{Magueijo:2004aa} --- concept of \emph{rainbow} spacetimes. These are momentum-dependent spacetimes, where the $k$-dependence is suppressed in the infrared regime. Further, a time-dependence in the inter-atomic potential yields a FRW-type universe for phononic modes, and \emph{rainbow FRW-type spacetimes} for higher-energy modes. The borderline that divides the energy scale  into phononic and trans-phononic intervals, may be viewed as an \emph{analogue Planck-length}. (The physical behavior changes somewhere in between the phononic and free-particle regime, in the same sense that the Planck-scale is expected to exhibit new laws of physics.) The ``Planck-length'' in our model is correlated with the contact potential (scattering length), and thus we are dealing with a time-dependent ``Planck-length''. In our expanding universe the ``Planck-length'' is growing as well~\cite{Visser:2001ix}, such that more and more modes are ``trans-Planckian'' as time goes on. Consequently they gradually decouple from the emergent spacetime picture, and their behavior becomes like free particles. The \emph{rainbow} scale factor
\begin{equation}
a_{k}(t) = a(t)/ \sqrt{1+k^{2} \, \ell_{\mathrm{Planck}}(t)^{2}} \, ,
\end{equation}
and the \emph{rainbow} Hubble parameter
\begin{equation} 
H_{k}(t)= H/ (1 + k^{2} \ell_{\mathrm{Planck}}(t)^{2}) \, ,
\end{equation}
are both momentum-dependent. The growth in the ``Planck-length'' forces the \emph{rainbow} Hubble parameter to approach zero, and therefore the universe gradually --- mode by mode --- effectively stops expanding~\cite{Jacobson:1999aa,Niemeyer:2001aa}. 

This leads to interesting physics for \emph{rainbow} inflation: At early times all modes are ``sub-Hubble-horizon'' modes. As the expansion goes on some --- but not all --- modes cross the ``Hubble-horizon'' and become ``superhorizon'' modes, these modes are ``frozen modes'' and get dragged along with the spacetime fabric. Eventually the effective expansion starts to slow down --- due to the growing ``Planck-length'' --- and a process starts that we call ``\emph{the melting of the modes}''. After a while these modes ``re-enter'' the ``Hubble-horizon'' and the particle production process is finished. \\

However useful these results are to the general relativity and cosmology community, we would like to end our conclusion by commenting on the importance of BEC based analogue models to the condensed matter physics community. There are many aspects, \eg, the time-dependent commutation relations, that really seem to require the emergent spacetime picture to fully understand the physics in our parametrically excited condensate. The emergent spacetime and the Bose--Einstein condensate are two aspects of one and the same effect, and one needs to know both points of view to appreciate the full complexity of this novel state of matter.
%%%%%%%%%%%%%%%%%%%%%%%%%%%%%%%%%%%%%%%%%%%%%%%%%%
%%%%%%%%%%%%%%%%%%%%%%%%%%%%%%%%%%%%%%%%%%%%%%%%%%
\clearpage

\end{document}